\documentclass[10pt,preprint]{aastex}

\newcommand{\av}{$A_V$}

\newcommand{\etal}{et~al.}

\newcommand{\ks}{$K_{\rm s}$}

\newcommand{\lsun}{$L_{\sun}$}

\newcommand{\teff}{$T_{\rm eff}$}

\newcommand{\mum}{$\mu$m}

\begin{document}

\title{The Taurus Spitzer Survey: New Candidate Taurus Members
Selected Using Sensitive Mid-Infrared Photometry}

\slugcomment{Version from \today}

\author{
L.\ M.\ Rebull\altaffilmark{1}, 
D.\ L.\ Padgett\altaffilmark{1},
C.-E.\ McCabe\altaffilmark{1},
L.\ A.\ Hillenbrand\altaffilmark{2},
K.\ R.\ Stapelfeldt\altaffilmark{3},
A.\ Noriega-Crespo\altaffilmark{1},
S.\ J.\ Carey\altaffilmark{1},
T.\ Brooke\altaffilmark{1},
T.\ Huard\altaffilmark{4},
S.\ Terebey\altaffilmark{5},
M.\ Audard\altaffilmark{6,7},
J.-L.\ Monin\altaffilmark{8},
M.\ Fukagawa\altaffilmark{9},
M.\ G\"udel\altaffilmark{10},
G.\ R.\ Knapp\altaffilmark{11},
F.\ Menard\altaffilmark{8},
L.\ E.\ Allen\altaffilmark{12}
J.\ R.\ Angione\altaffilmark{3,5},
C.\ Baldovin-Saavedra\altaffilmark{6,7},
J.\ Bouvier\altaffilmark{8}
K.\ Briggs\altaffilmark{10},
C.\ Dougados\altaffilmark{8},
N.\ J.\ Evans\altaffilmark{13},
N.\ Flagey\altaffilmark{1}, 
S.\ Guieu\altaffilmark{1},
N.\ Grosso\altaffilmark{14}, 
A.\ M.\ Glauser\altaffilmark{10,15},
P.\ Harvey\altaffilmark{13},
D.\ Hines\altaffilmark{16},
W.\ B.\ Latter\altaffilmark{17},
S.\ L.\ Skinner\altaffilmark{18},
S.\ Strom\altaffilmark{12},
J.\ Tromp\altaffilmark{5},
S.\ Wolf\altaffilmark{19}}

\altaffiltext{1}{Spitzer Science Center/Caltech, M/S 220-6, 1200
E.\ California Blvd., Pasadena, CA  91125
(luisa.rebull@jpl.nasa.gov)}
\altaffiltext{2}{Department of Astronomy, California Institute of
Technology}
\altaffiltext{3}{Jet Propulsion Laboratory}
\altaffiltext{4}{University of Maryland, College Park}
\altaffiltext{5}{California State University, Los Angeles}
\altaffiltext{6}{ISDC Data Center for Astrophysics, University of
Geneva, Ch. d'Ecogia 16, CH-1290 Versoix,  Switzerland}
\altaffiltext{7}{Observatoire de Gen\`eve, University of Geneva, Ch.
des Maillettes 51, 1290 Versoix, Switzerland}
\altaffiltext{8}{Laboratoire d'Astrophysique de Grenoble, Universit\'e
de Grenoble -- CNRS, UMR 5571, Grenoble, France}
\altaffiltext{9}{Department of Earth and Space Science, Graduate School
of Science, Osaka University, 1-1 Machikaneyama, Toyonaka, Osaka
560-0043 Japan Nagoya University, Japan}
\altaffiltext{10}{ETH Zurich, Institute of Astronomy, 8093 Zurich, Switzerland}
\altaffiltext{11}{Princeton University}
\altaffiltext{12}{NOAO, Tucson, AZ}
\altaffiltext{13}{University of Texas, Austin}
\altaffiltext{14}{Observatoire astronomique de Strasbourg,
Universit{\'e} de Strasbourg, CNRS, INSU, 11 rue de l'Universit{\'e},
67000 Strasbourg, France}
\altaffiltext{15}{UK Astronomy Technology Centre, Royal Observatory,
Edinburgh EH9 3HJ, UK}
\altaffiltext{16}{Space Science Institute}
\altaffiltext{17}{NASA Herschel Science Center, IPAC, Pasadena, CA
91125}
\altaffiltext{18}{CASA, University of Colorado, Boulder, CO 80309-0389}
\altaffiltext{19}{University of Kiel, Institute of Theoretical
Physics and Astrophysics, Leibnizstrasse 15, 24098 Kiel, Germany}

\begin{abstract}

We report on the properties of pre-main-sequence objects in the Taurus
molecular clouds as observed in 7 mid- and far-infrared bands with the
Spitzer Space Telescope.  There are 215 previously-identified members
of the Taurus star-forming region in our $\sim$44 square degree map;
these members exhibit a range of Spitzer colors that we take to define
young stars still surrounded by circumstellar dust (noting that
$\sim$20\% of the bonafide Taurus members exhibit no detectable dust
excesses). We looked for new objects in the survey field with similar
Spitzer properties, aided by extensive optical, X-ray, and ultraviolet
imaging, and found 148 candidate new members of Taurus.  We have
obtained follow-up spectroscopy for about half the candidate sample,
thus far confirming 34 new members, 3 probable new members, and 10
possible new members, an increase of 15-20\% in Taurus members. Of the
objects for which we have spectroscopy, 7 are now confirmed
extragalactic objects, and one is a background Be star. The remaining
93 candidate objects await additional analysis and/or data to be
confirmed or rejected as Taurus members.  Most of the new members  are
Class II M stars and are located along the same cloud filaments as the
previously-identified Taurus members. Among non-members with Spitzer
colors similar to young, dusty stars are evolved Be stars, planetary
nebulae, carbon stars, galaxies, and AGN.

\end{abstract}


\section{Introduction}
\label{sec:intro}

A complete inventory of all the coeval stars in a young stellar
association, cluster, or group (hereafter ``association") enables
studies of the initial mass function (IMF), disk fraction, and stellar
rotational properties, among other pursuits.  Information from 
associations with a range of ages enables understanding of the overall
formation and evolution of young stars, including the change with time
of disk fraction and stellar rotation rate.  However, identifying all
of the member stars of a given young association can be quite
difficult.  (By ``member,'' we mean objects that are clearly young, 
close to the same age, and often still associated with their natal
cloud.)  Finding all such members requires that one employ multiple
observational techniques.  These methods include but are not limited
to X-ray surveys (e.g., Alcal\'a \etal\ 1996, Wolk \etal\ 2006),
H$\alpha$ surveys (e.g., Ogura \etal\ 2002), variability surveys
(e.g., Carpenter \etal\ 2001, Rebull 2001), ultraviolet (UV) surveys
(e.g., Rebull \etal\ 2000), and infrared (IR) surveys (e.g.,
J{\o}rgensen \etal\ 2006, Rebull \etal\ 2007). At each wavelength, we
can use the fact that stellar youth implies more flux at a given
radiometric band (X-rays, H$\alpha$, UV, IR), or more flux
variability, than older stars of comparable mass, allowing separation
of association members from field contaminants.  When we combine the
information from surveys in  multiple wavelengths, we must remember
that  the influence of extinction due to circumstellar matter and/or
the molecular cloud is vastly different at different wavelengths;
extinction affects UV and optical wavelengths much more strongly than
IR. For young associations, many, but not all, legitimate members are
identifiable using just one or a few survey methods. 

There are advantages and disadvantages to studying nearby young
associations. Identification of young members in an association is
made easier if the objects are located at smaller heliocentric
distances and therefore brighter on the whole.  This is especially the
case for low-mass members with very low luminosities; a complete
census of these stars and brown dwarfs can be made only for very
nearby star-forming regions. On the other hand, nearby star-forming
molecular clouds cover large areas of sky and therefore require large
investments of observing time. At just 137 pc (with a depth of $\sim$
20 pc; Torres \etal\ 2007, 2009), the Taurus star-forming region is
one of the closest large cloud complexes with hundreds of low-mass
stars, ongoing star formation, and objects ranging in age up to
$\sim$5 Myr.  Studies of Taurus objects have significantly influenced
our basic understanding of the star-formation process for decades
(e.g., Herbig \& Rao 1972, Kenyon \etal\ 2008). However, the Taurus
Molecular Cloud is close enough that it subtends more than 100 square
degrees of sky; surveying all or even most of it is difficult within
typical telescope time allocations.  

Infrared surveys led to the discovery that some stars have infrared
excesses, interpreted as circumstellar matter (e.g., Aumann \etal\
1984, Beichman \etal\ 1986).  Most if not all low-mass stars form with
circumstellar accretion disks, resulting in IR excesses for as long as
the dusty circumstellar material survives (e.g., Hernandez \etal\
2007). By using IR to survey a star-forming region, the stars with IR
excesses are relatively easily distinguished from stars without such
excesses. The Spitzer Space Telescope (Werner \etal\ 2004) provides an
excellent platform for surveying star-forming regions in the mid-IR
and far-IR, enabling stars with IR excesses to be identified; the
member stars which do not have IR excesses must be recovered using
different techniques such as the ones listed above.  Because Spitzer
relatively efficiently maps large regions of sky, it is a particularly
useful tool for surveying large star forming regions. 

We have conducted a large multi-wavelength imaging and spectroscopic
survey of the Taurus Molecular Cloud (TMC) in order to test if our
inventory of Taurus members with infrared excesses is, in fact,
complete.  The Spitzer imaging component is referred to as the Taurus
Spitzer Survey, and is described by Padgett \etal\ (2008; hereafter
P08) and Padgett \etal\ (2009; hereafter P09).   It covers $\sim$44
square degrees from 3.6 to 160 \mum\ and is a Spitzer Legacy Project,
so enhanced data products have been delivered back to the Spitzer
Science Center (SSC), including the catalogs on which this present
paper is based.  In addition to the Spitzer component, there are four
other major components to our Taurus survey.  XMM-Newton was used by
the XMM-Newton Extended Survey of the Taurus Molecular Cloud (XEST)
program (e.g., G\"udel \etal\ 2007 and references therein), which
mapped $\sim$5 square degrees, most of which was also mapped by the
Spitzer observations; the XEST data include X-ray imaging but also
include ultraviolet data from the XMM-Newton Optical Monitor (Audard
\etal\ 2007). XEST was deliberately pointed towards aggregates of
previously identified Taurus members. In the optical, the
Canada-France-Hawaii Telescope (CFHT) survey (Monin \etal\ in
preparation; G\"udel, Padgett, \& Dougados 2007) mapped $\sim$28
square degrees (all of which are encompassed by the Spitzer area), and
the Sloan Digital Sky Survey (SDSS) (Finkbeiner \etal\ 2004;
Padmanabhan \etal\ 2008) mapped $\sim$48 square degrees in two
perpendicular strips, about half of which overlaps the Spitzer area. 
Finally, the Five College Radio Astronomy Observatory (FCRAO)
millimeter wavelength survey (Goldsmith \etal\ 2008) mapped $\sim$100
square degrees in the CO(1-0) line, covering the Spitzer survey area
entirely.  The relative coverages of these surveys is shown in P09. 
Our extended collaboration has already begun to use this rich dataset
to search for new members of Taurus; Scelsi \etal\ (2007,2008)
identified new candidate members using the XEST data, and Guieu \etal\
(2006, 2007) identified new brown dwarf members using the CFHT data to
study their disk properties using the Spitzer data. Ongoing
investigations include searches for members via emission line spectra
(Knapp \etal, in prep), Herbig-Haro (HH) objects (Stapelfeldt \etal\
in prep), and transition disks (McCabe \etal\ in prep).  

In this paper, we select new candidate Taurus members with infrared
excesses using Spitzer and Two-Micron All-Sky Survey (2MASS; Skrutskie
\etal\ 2006) data.  We construct color-magnitude and color-color
diagrams for point sources, then use the locations of
previously-identified young stars in these diagrams to select new
candidate members with infrared excesses.  To discard obvious
extragalactic sources, we examine the source  morphology in all
available bands of the multiwavelength Taurus survey. We construct
spectral energy distributions (SEDs) from the photometry over all
available bands, again discarding objects we believe to be galaxies.  
Follow-up spectroscopy has been obtained to assess
whether or not our new candidate Taurus objects are likely Taurus
members.  We also present Spitzer flux densities for the 215
previously-identified members found in the region covered by our
Spitzer survey.  Note that (a) the 215 previously-identified objects
include those members without infrared excesses identified via other
mechanisms; (b) the 215 previously-identified objects are those
covered by our Spitzer map -- there are other legitimate Taurus
members outside the region we observed, such as in L1551; (c) our new
candidate member list is necessarily just those with IR excesses and
exclusively within the regions covered by our Spitzer observations. 
Some objects are resolved in one or more of the Spitzer images, and
extended source photometry may be a better representation of the
complete flux from the object; many of the extended sources are
discussed individually in  other papers (e.g., Tobin \etal\ 2008,
Stapelfeldt \etal\ in prep)

The observations, data reduction, and ancillary data are described in 
\S\ref{sec:obs}.  Section~\ref{sec:pickysos} describes our young
stellar object (YSO) selection process which is based on the colors of
previously-identified Taurus members, also presented here.   We
describe 34 objects that we have, thus far, identified as new members
of Taurus (plus 3 probable new members and 10 possible new members) in
\S\ref{sec:discussion} and discuss the properties of the new objects
in conjunction with (and comparison to) the previously-identified
members of Taurus.  Finally, we summarize our main points in
\S\ref{sec:concl}.  The Appendix contains spectral energy
distributions and discussion of some specific objects.

\section{Observations, Data Reduction, and Ancillary Data}
\label{sec:obs}

\subsection{Spitzer data}
\label{sec:spitzerdata}

P08 and P09 present a comprehensive discussion of the Spitzer data
acquisition, reduction, and bandmerging to the 2MASS data.  In summary,
we conducted the observations using IRAC (Infrared Array Camera; 3.6,
4.5, 5.8, \& 8 \mum; Fazio \etal\ 2004) and MIPS (Multi-band Imaging
Photometer for Spitzer; 24, 70, \& 160 \mum; Rieke \etal\ 2004) in two
epochs to enable removal of asteroids from the final point-source
catalog.  The observations were spread over three observing programs
and three years, 2005 to 2007.

For IRAC, we used MOPEX (MOsaicking and Point source EXtractor;
Makovoz \& Marleau 2005) to find the sources, and IDL to perform
aperture photometry at those locations.  With our IRAC observations,
we sacrificed redundancy for spatial coverage, and obtained just 2
IRAC frames per position (total integration time of 25.2 seconds), so
instrumental artifacts are abundant. We discarded single-band
(apparent) detections as likely artifacts.  Objects that we measured
to be brighter than the 0.6 sec saturation limits (630, 630, 4600, and
2500 mJy at the four IRAC channels) we took to be saturated, and these
appear as lower limits in our catalog.  The zero-points we used to
convert between flux densities and magnitudes were as found in the
IRAC Data Handbook on the SSC website: 280.9, 179.7, 115.0, and 64.13
Jy for IRAC's four channels, respectively. 

For MIPS, we used MOPEX point-response-function (PRF) fitting
photometry at 24 and 70 \mum.  The total integration time at 24 \mum\
was 30 seconds per position; the MIPS scan legs were interleaved to
provide complete coverage at 70 \mum, and a total integration time of
15 seconds per position.  We took objects brighter than 4.7 Jy at 24
\mum\ and 6.5 Jy at 70 \mum\ to be saturated (or at least
non-linear).  These objects appear in our catalog as having lower flux
density limits.  Following the MIPS Data Handbook on the SSC website,
the zero-points that convert between flux densities and magnitudes are
7.14 and 0.775 Jy for MIPS-24 and 70. 

The observations at 160 \mum\ present special challenges.  All of the
things that can affect the other bands (such as saturation, extended
emission, confusion with the cloud background and nearby objects --
whether they be Taurus or background objects -- and instrumental
artifacts), also affect this band.  Because the Taurus cloud emission
and instrumental effects are both very strong, and because the
resolution is the poorest of all the Spitzer bands, these items are of
particular concern at this bandpass. Additionally, due to scattered
data gaps in our 160 \mum\ map (see P08,P09),  measurements (or
limits) at 160 \mum\ are missing or compromised for some sources. 
MOPEX does not detect any point sources automatically at 160 \mum,
because all detected point-like objects appear to be slightly
resolved.  For $\sim$100 objects that were apparent by eye in the 160
\mum\ image (or for which limits were of interest for this paper), we
performed aperture photometry on the cleaned image which was smoothed
by a 4 pixel median filter to minimize the influence of holes in the
map and image artifacts.   We used a 32$\arcsec$ aperture, an annulus
from 64-128$\arcsec$, and an aperture correction of 1.97 (valid for
temperatures between 500 and 2000 K\footnote{The dust we see at 160
\mum\ is likely to be cooler than this, but the aperture correction
changes only by 0.7\% between this temperature range and 50 K, well
within the uncertainties.}). Based on a  comparison of the flux
densities obtained from the filtered image and the unfiltered image,
we took the flux density uncertainties to be 20\% below 5~Jy and 30\%
for higher flux densities.  Some objects that fell in regions with too
many missing flux density values (from saturation or gaps in the map)
are not retrievable and thus do not have a measurement (or limits); in
essence, they are off the edge of the map.   Visual inspection of each
160 micron source was used in order to  determine whether the object
was clearly detected as a single point source,  confused with another
nearby source, or contaminated by data dropouts and/or  saturation
issues which would cause the reported flux density to be a lower
limit; these are indicated in the data tables below.  The zero-point
for MIPS-160 is 0.159 Jy, again from the MIPS Data Handbook on the SSC
website. 

We extracted $JH$\ks\ data from the 2MASS point source catalog for our
region, retaining flux densities only for those objects with
high-quality  2MASS data flags.  There were a handful of objects of
interest for this paper for which we made the following manual
modifications, and therefore these modified values were used for the
color-magnitude diagrams (CMDs) and spectral energy distributions
(SEDs) below.  Measurements for five objects did not meet the data
flag criteria we imposed on the 2MASS catalog, but their as-reported
flux densities were completely consistent with the rest of their SED,
so we adopted their measurements as good detections; these objects are
SST Tau 041542.7+290959, 041542.7+290959, 042517.6+261750,
043835.4+261041, and 043354.7+261327. Flux densities for one or more
bands for the following previously-identified Taurus members do not
exist in the point source catalog, and thus the flux densities used
below were taken from the extended source catalog: SST Tau
042757.3+261918, 044112.6+254635, 043535.3+240819, and
043316.5+225320.  

We bandmerged all the available point sources between 2 and 70 microns
using position matching alone, with a wavelength-dependent maximum
matching radius. Sources between 2 and 8 \mum\ were matched over a
radius of  1$\arcsec$, and at 24 \mum, the matching radius to the rest
of the catalog was 2$\arcsec$.  All matching radii are values
empirically determined by inspection of histograms of nearest
neighbors between bands, and spot-checking individual sources;
histograms of positional offsets and additional discussion appears in
P09.  We pre-merged the 24 and 70 \mum\ catalogs before merging to the
rest of the catalog using an empirically-determined radius of
10$\arcsec$.  Because the spatial resolution of the 70 \mum\ images is
so much worse than the 2 \mum\ images, often more than one NIR (or
optical) source can be matched to the 70 \mum\ source; however, it is
extremely likely that if we detect a source at 70 \mum, it will also
appear at 24 \mum, so by implementing the pre-merge of 24 and 70 \mum,
we are preferentially matching the 70 \mum\ sources to a likely
physical match.  

We include here a brief aside on the accuracy of blind merging by
position. In the generic case of surveys across wavelengths, relative
astrometric accuracy and spatial resolution is paramount. In our case
of a catalog primarily driven by Spitzer+2MASS sources, astrometric
accuracy is not the dominant source of error, since each instrument is
internally consistent and calibrated to the 2MASS coordinate system.
For many star-forming regions studied with Spitzer, e.g., many of
those observed by the Cores-to-Disks (c2d; Evans \etal\ 2003) and
Gould's Belt (Allen \etal\ in preparation) projects, the source
surface density is high enough that blind merging by position causes
an unacceptable rate of false matches, and multiple short-wavelength
sources should in reality be assigned to a single long-wavelength
source.  However, the source surface density in Taurus is low enough
($\sim$4 sources per square arcminute, compared to $\sim$20 in some
star-forming regions) that source confusion in general is not as much
of a concern.  On the other hand, in Taurus, many objects are known to
be close binaries, and we do not apportion the flux density we observe
between the two objects; for close binaries, we have treated the
object as single here.  In any case, in three of the objects
investigated in detail for this paper ($<$1\% of the objects
investigated, $<$0.5\% of the total number of 70 \mum\ sources in the
entire catalog), we determined by individual inspection that the 24/70
flux densities were incorrectly matched to a nearby faint
short-wavelength source, rather than the correct, bright, slightly
farther away, short-wavelength source.  In these cases, we manually
tweaked the flux assignment; the position change is well within the
expected uncertainties between the long- and short-wavelength
catalogs. We are confident that for the large majority of the sources
in our entire catalog, our blind merging by position works, but for
any particular source not discussed here, the images and the flux
assignment in our delivered catalogs should be carefully scrutinized.

In creating our Spitzer-centric catalog, we dropped any object without
a Spitzer detection (e.g., sources off the edges of our maps) before
proceeding and after each additional merging step below (e.g., when
combining the SDSS catalog and the Spitzer catalog, we did not retain
SDSS-only sources from off the edges of our map or whose SED falls so
rapidly that they are too faint for our shallow Spitzer survey). 
There are nearly 700,000 sources in the catalog with Spitzer flux
densities in at least one band.  This does not include the asteroids,
which will be discussed by Hines \etal, in preparation. It also does
not include sources that are substantially extended (except at 160
$\mu$m, see below); many of those will be covered by Stapelfeldt
\etal, in preparation.  The vast majority of the $\sim$700,000
sources are well behind the Taurus Molecular Cloud.  Sources which are
members are generally expected to be bright in the shorter bands and
detected in multiple bands.

P09 discusses the survey sensitivity in detail.  In summary, the 3.6
and 4.5 \mum\ sensitivities are quite comparable, and 98\% of the
objects detected in 3.6 \mum\ are also detected at 4.5 \mum. The 5.8
and 8 \mum\ channels are much less sensitive; only 22\% of the objects
detected at 3.6 \mum\ are also detected at 5.8 \mum, and just 17\% of
the objects detected at 3.6 \mum\ are also detected at 8 \mum\ (where
nebulosity can also be a factor in point source identification and
extraction).  Our 24, 70, and 160 \mum\ sensitivity is a strong
function of position in the image because of nebular and high Zodiacal
dust emission contributions to the background. In part because of the
varying background but also because of the effective sensitivity of
the instrument to photospheres at the distance of Taurus given our
exposure times, just 1.4\% of the objects detected at 3.6 \mum\ are
also detected at 24 \mum, and just 0.15\% are detected at 70 \mum. 
The faintest independent 24 \mum\ detection is $\sim$10th mag
($\sim$M2 spectral type photosphere at Taurus age and distance), and
the histogram of detections turns over at $\sim$9.3 mags (e.g., there
is a steep fall-off between [24]$\sim$9.3 and 10, where the bracket
notation denotes the magnitude at that band). Similarly, at 70 \mum,
the faintest object is [70]$\sim$3.5, and a more typical value is
2.5-2.7 (note that 70 \mum\ is much more strongly affected by nebular
emission and instrumental effects than 24 \mum).  Only O or B
photospheres would be that bright at the distance of Taurus, so we can
only detect legitimate Taurus objects at 70 (or 160) \mum\ which have
substantial IR excesses.  

A full completeness analysis is given by P09, but as another
independent way of assessing our completeness, we examined by hand
each of the images at each of the Spitzer bands for each of the
previously identified Taurus members and new candidate Taurus members
discussed here (see \S\ref{sec:known} and
\S\ref{sec:selectionoverview}).  If the object could be seen in the
image but a flux density was not initially reported at that band, we
made a manual assessment of the flux density or upper/lower limit, as
appropriate. For IRAC, $\sim$1-2\% (depending on the band) of the
previously identified Taurus members and $\sim$4-11\% of the new
candidate members were missing photometry and were filled in manually
(having lower signal-to-noise).  The most common band in which flux
densities were erroneously missing (e.g., not in the automatically
generated catalog but visible in the images) was 5.8 $\mu$m, which is not
particularly surprising, as this is the least sensitive of the IRAC
bands. For MIPS, $\sim$9\% of the previously identified Taurus members
were missing photometry in either 24 or 70 $\mu$m, and  $\sim$6\% of
the new candidate objects were missing photometry at 24 or 70 \mum.
The lower fraction of missing photometry in the new candidate objects
as compared to the previously identified objects is a reflection of
the fact that our selection mechanism is somewhat biased towards
objects with MIPS detections; see \S\ref{sec:newsamplesummary}
below.   

Many sources detected at shorter wavelengths are undetected at longer
wavelengths, and it is important for our science analysis to obtain
upper limits for our sources in the Spitzer photometric bands.  For
the list of coordinates of previously identified and new candidate
Taurus members, at 24 $\mu$m, we used MOPEX to look for a source whose
photometry could be obtained via PRF-fitting at that location.  If an
object was detected, we took a weighted average flux density based on
all detections (e.g., between epochs and tiles). If the final
signal-to-noise ratio (SNR) was $<$3, we took the error to be the
1$\sigma$ upper limit, and multiplied by 3 to get the 3$\sigma$ limits
found in the data tables below and SEDs in \S\ref{sec:seds}. If the
PRF fitting failed,  we performed aperture photometry at the location
of the object (at each tile and epoch available), and took a weighted
average of all the resultant positive aperture flux densities. If that
average had a SNR $>$3, we took that weighted average to be a
detection. For all objects that had SNR$<$3, we took the upper limit
to be the 1$\sigma$ limit (and multiplied by 3 to get the 3$\sigma$
limits found in the data tables and SEDs below).  If all the measured
aperture flux densities were negative, we took the 1$\sigma$ limit to
be the straight average of the errors (and multiplied by 3 to get the
3$\sigma$ limits found in the tables and SEDs below).  Finally, still
at 24 \mum, the objects that cannot be resolved from a companion are
reported as simply unknown, where the presence of a companion is known
from shorter-wavelength higher-spatial-resolution observations.  At 70
$\mu$m, some objects are unresolved from a nearby object and are
impossible to estimate (again, where the presence of a companion is
known from shorter-wavelength higher-spatial-resolution
observations).  Upper limits at 70 \mum\ for all remaining undetected
sources were obtained by performing aperture photometry at the
expected source location, using a 35$\arcsec$ aperture and a
multiplicative aperture correction of 1.22, as discussed in the MIPS
Data Handbook, available at the SSC website.  These 1-$\sigma$ errors
were multiplied by 3 to get the 3-$\sigma$ errors shown in the Tables
and Figures here. Upper limits at 160 \mum\ were individually assessed
using aperture photometry on the uncertainty image using the same
parameters as for detected sources above (32$\arcsec$ aperture, an
annulus from 64-128$\arcsec$, and an aperture correction of 1.97) to
obtain 1-$\sigma$ errors, and then multiplied by 3 to get the
3-$\sigma$ errors shown in the Tables and Figures here.

\subsection{Complementary survey photometric data}

In \S\ref{sec:spitzerdata}, above, we discussed merging the 2MASS and
Spitzer data. We also need to match our catalog to the other Taurus
surveys, listed in \S\ref{sec:intro} above.  Again, we match by
position, with a radial offset tolerance customized empirically to
each band or catalog.  

The extracted CFHT point sources (see Monin \etal\ in preparation or
Guieu \etal\ 2006 for more details on the extraction process; also
see Monin \etal\ 2007 or Guieu 2008) are merged first to 2MASS. 
Sources that are CFHT-only are then dropped in order to remove objects
that are very statistically likely to be
instrumental artifacts.  The CFHT sources are then merged to the
Spitzer catalog.  The CFHT $I$ bandpass is converted to Cousins $I$
via the following equation:
\begin{equation}
I_C = I_{CFHT} - 0.531\times (I_{CFHT}-z_{CFHT}) - 0.278
\end{equation}
and then converted to flux densities using the Cousins $I$ zero-point from
Bessell (1979), $2.55\times10^{-23}$ W m$^{-2}$ Hz$^{-1}$.  For these
points, we use the $I_C$ effective wavelength of 0.79 \mum. 
About 138,000 of the point sources in our catalog have CFHT $I_C$ magnitudes
(20\% of the entire catalog).

The SDSS photometry arrives from the SDSS pipeline with photometric
measurements in $ugriz$ in flux density units of nanomaggies\footnote{In
SDSS, a ``maggy'' is the ratio of the flux density of the object to a
standard flux density.  The Sloan magnitudes are AB magnitudes, as
opposed to Vega magnitudes.  In the AB system, a flat spectrum object
with 3631 Jy at each band should have every magnitude equal to zero,
and all maggies equal to one. Flux densities returned by the Sloan pipeline
are nanomaggies, and can be converted to $\mu$Jy.  For more
discussion, see Padmanabhan \etal\ (2008).}, which can be converted
to the same flux density units as the rest of the catalog data.  We retained
only those flux density measurements with good quality flags, and we merged
the source lists for each of the SDSS tiles to each other by position
to remove duplicates before merging to the master Spitzer catalog.
The effective wavelengths are 3590, 4810, 6230, 7640, and 9060 \AA\
for $ugriz$, respectively.  We also made note of whether the object
appearing in the SDSS images was flagged by the pipeline as extended
or not.  There are SDSS $z$-band flux densities for about 300,000 objects in
our catalog (45\% of the entire catalog).  There are 6400 spectra
available from SDSS, about 3400 of which overlap the Spitzer survey
region; for each object, we matched by position to our Spitzer
catalog, and accepted the spectral classification produced by the
SDSS pipeline as a spectral classification of the object, unless
another spectral classification was available in the literature (see
below). 

There are $\sim$1000 objects in our catalog (0.1\% of the entire
catalog) with X-ray measurements from the XEST survey (G\"udel \etal\
2007). The XEST catalog includes flux densities from the XMM-Newton
Optical Monitor (OM).  The OM has a field of view comparable to, but
not exactly identical to, the main X-ray field of view; see Audard
\etal\ (2007) for more discussion of the XEST-OM sample.  These data
are in one of three ultraviolet bandpasses ($U$, UVW1, or UVW2).  To
convert these values to flux densities, we used the following
equation, found in the XMM-Newton OM calibration document (Chen \etal\
2004):
\begin{equation}
F_{\nu} = 10^{(0.4(Z-m))} \times f \times \lambda^2/c \times 10^{23}
\end{equation} 
where $m$ is the reported magnitude (and $F_{\nu}$ the flux density) for
a given object, $Z$ = 18.259, 17.204, and 14.837, and  $f$ =
1.94$\times10^{-16}$, 4.76$\times10^{-16}$, and
5.71$\times10^{-15}$ ergs cm$^{-2}$ s$^{-1}$ \AA$^{-1}$
counts$^{-1}$ sec for $U$, UVW1, and UVW2 (respectively). In
the equation, $\lambda$ is in units of \AA, and $c$ is
$3\times10^{18}$ \AA\ s$^{-1}$    The effective wavelengths are
0.344, 0.291, and 0.212 \mum\ for $U$, UVW1, and UVW2.  There
are $\sim$1600 objects with XMM-Newton OM flux densities in our catalog
(0.2\% of the entire catalog).  

We note that many of the X-ray detected XEST sources are likely
background galaxies (see G\"udel \etal\ 2007) and that XEST included
regions not covered by our map, such as L1551.

The XEST team assembled a catalog of supporting data from the
literature, such as optical photometric measurements, for all of the
previously-identified Taurus members (see \S\ref{sec:known} below);
we have included these photometric points in our database, converting
Johnson magnitudes to flux densities using zero-points available in
the literature (e.g., Cox 2001 and references therein).  

The SEDs presented in this paper use all of these supporting data
where available (except for the X-ray fluxes), and are presented as
$\lambda F_{\lambda}$ in cgs units (erg s$^{-1}$ cm$^{-2}$), against
$\lambda$ in microns. 

\subsection{Spectroscopy}
\label{sec:spectra}

We obtained follow-up spectroscopy for $\sim$75\% of the candidate
YSOs discussed in this paper.  Some previously identified Taurus
members missing spectral types in our database (as discussed in
\S\ref{sec:known} below) were also observed. Data were obtained over
six runs between 2007 and 2009 at Keck and the Palomar
200$^{\prime\prime}$.  Most of the $\gtrsim$200 spectra are
low-resolution optical, obtained with the Double Spectrograph at
Palomar (30 Nov - 3 Dec 2008) or LRIS at Keck (Feb 2007). Many spectra
were taken in the infrared with Triplespec at Palomar (21-24 Nov 2008
and 21 Dec 2008) or NIRSPEC at Keck (Dec 2007 and Feb 2008).  The
infrared spectra will be discussed in a forthcoming paper; the optical
spectra are discussed here. 

%3712-5657, 916 px
%6231-8724, 1023 px

The optical spectra from Palomar are taken in two segments, blue and
red; the blue covered $\sim$3710-5660 \AA\ at $\sim$2 \AA/px and the
red, $\sim$6230-8720 \AA\ at $\sim$2.5 \AA/px.  Instrument settings
were a 316 lines mm$^{-1}$ grating blazed at 7500 \AA\ and used at
grating angle 24.75$\arcdeg$ for the red side, and a 300 lines
mm$^{-1}$  grating blazed at 3900 \AA\ and used at grating angle
23.12$\arcdeg$ for the  blue side.  LRIS at Keck is also a
double-barreled spectrograph which we used with a 400 lines mm$^{-1}$
grism blazed at 3400 \AA\ in the blue  and a 400 lines mm$^{-1}$
grating blazed at 8500 \AA\ and positioned at grating angle
23.49$\arcdeg$ in the red. We obtained continuous wavelength coverage
from the blue atmospheric cutoff to $\sim$ 9400 \AA\ at 1.86 \AA\
px$^{-1}$. The Double Spectrograph and LRIS data were both reduced
using the Image Reduction and Analysis Facility (IRAF)\footnote{IRAF
is distributed by the National Optical Astronomy Observatories, which
are operated by the Association of Universities for Research in
Astronomy, Inc., under cooperative agreement with the National Science
Foundation.} ccdred and onedspec packages. Images were trimmed,
bias-subtracted, and flattened prior to spectral extraction with the
IRAF task apall. Wavelength calibration was performed using a Fe-Th-Ar
lamp in the blue and Th-Ar lamp in the red. The wavelength solution
was applied using the IRAF task dispcor. The white dwarf Feige 34 was
observed each night, providing an approximate flux calibration
reference for all the scientific targets.

Spectral classification was obtained via visual examination of each
spectrum and comparison to a standard grid composed of $>$60 stars
ranging in types from B8 to M9. Four authors performed the
classification independently to achieve an estimated accuracy of
roughly a subclass. 

The red spectra included H$\alpha$. For each spectrum, we used IRAF to
measure an equivalent width for H$\alpha$, following the usual
convention where negative values indicate emission (see
\S\ref{sec:newsamplesummary}).  We also noted if the Ca IR triplet was
in emission at the time of observation; this is indicated in the data
tables below. 

The spectroscopic data are sufficient to rule out redshifted galaxies,
to classify stars, and to find stars with  H$\alpha$ in emission. 
However, our data are of insufficient resolution to, e.g., detect the
presence of lithium, or determine surface gravities for most types.  
Section~\ref{sec:newsamplesummary} discusses an analysis similar to
that presented by Slesnick \etal\ (2008) which uses the TiO 8465 \AA\
index and Na 8190 \AA\ index to determine an estimate of the surface
gravity of the star. Additional follow-up data will be required to
assess membership for stars with no H$\alpha$ in emission and a small
IR excess at Spitzer bands, and/or types earlier than M1 where the
gravity analysis is not applicable.

\section{YSO candidate selection}
\label{sec:pickysos}

\subsection{Overview of YSO selection process}
\label{sec:selectionoverview}

First, we establish two comparison samples from the literature, and
then we discuss the process by which we selected new candidate YSOs in
Taurus.

\subsubsection{Template sample of previously-identified Taurus members}
\label{sec:known}

We first informed our search for new Taurus members by determining the
regions of color space occupied by previously identified Taurus
members.   By ``Taurus member,'' we mean an object that is confirmed
via multiple mechanisms to be young and associated with the Taurus
Molecular Cloud and the other Taurus members, e.g., sharing communal
properties such as stellar activity. By ``previously identified,'' we
mean identified as a member by other authors in the literature using
data sets other than the Taurus Spitzer Legacy Survey.  However, in
order to appear in our catalog, the object must be within the region
we mapped with Spitzer. There are legitimate Taurus members outside
our region, including, e.g., those in the L1551 region.

The core of our sample of previously-identified Taurus members is the
list assembled by the XEST team for their analysis (see G\"udel \etal\
2007 and references therein).  We have updated this list with more
recently confirmed objects (e.g., Scelsi \etal\ 2008), as well as
scattered additional previously-identified Taurus members found in the
literature.  Kenyon \etal\ (2008) also report previously identified
Taurus objects, with $\sim$30 more objects in our Spitzer field of
view, which we have also included (but see Appendix \ref{sec:popups}
for one object from Kenyon \etal\ (2008) which we rejected).  We have
thus defined our sample of previously identified Taurus members as
basically an updated XEST+Kenyon \etal\ list convolved with our survey
coverage. There are 215 previously-identified Taurus members in our
Spitzer maps.  There are $\sim$100 more objects from Kenyon \etal\
which are outside our mapped region. Binary objects that are
unresolved in any of our Spitzer maps are regarded here as a single
object (e.g., FS Tau Aab, whose separation is $<1\arcsec$).  We
discuss the overall Spitzer properties of this list in some detail
below.  We note here that all of these 215 previously-identified
Taurus members were detected by Spitzer, but not all of them have IR
excesses; this list includes the young stars without IR excesses
(e.g., mostly weak-lined T~Tauri stars, WTTS) discovered by other
means.

Hartmann \etal\ (2005) report IRAC observations of a set of previously
identified members covered by IRAC guaranteed time observation (GTO)
team observations.  This region of the sky is also covered by our
shallower map, to the same depth as the rest of our survey.  Our
photometry agrees within the errors expected from photometry
methodology and from the intrinsic variability of the stars.  

We do not include the objects reported by Luhman \etal\ (2006,
2009a,b) as previously known objects because they were found with an
independent analysis of the same data used here -- our Taurus Legacy
data, in part along with the XEST data -- and the derived values
agree.

%Luhman \etal\ (2006, 2009) use the publicly-available data from the
%Taurus-1 part of our survey, as reduced by the GLIMPSE pipelines, to
%search for new members of Taurus, and report the photometry for a list
%of previously identified members.  Since the same data are used, there
%is no variation in the derived photometry that comes from intrinsic
%stellar variability.  Again, within the errors expected from our
%photometry and those reported by the other team, our photometry
%agrees. Since the new Taurus members discussed in Luhman \etal\ (2006,
%2009) were selected using the data discussed here, and since we have
%independently identified every one of them here, we do not treat these
%stars as previously identified members; they are noted in our lists of
%new Taurus members.  

Scelsi \etal\ (2008) present spectroscopic follow-up on potential new
Taurus members discovered by the XEST survey (Scelsi \etal\ 2007). 
Three confirmed new Taurus members from Scelsi \etal\ (2008) were
independently discovered and confirmed by us using Spitzer data (SST
Tau 043456.9+225835 = XEST 08-003, SST Tau 043542.0+225222 = XEST
08-033, and SST Tau 042215.6+265706 = XEST 11-078).  Since we report
these in the list of new Taurus members, these do not appear in the
list of previously identified Taurus members; they are noted in the
tables below.  The new members reported there that we did not
rediscover are included in our list of previously identified Taurus
members.  A complete discussion of the Spitzer properties of all of
the candidate members found using XEST X-rays that were presented by
Scelsi \etal\ (2007) will appear in Audard \etal, in preparation.

\subsubsection{Template sample of non-members}
\label{sec:nonmembers}

Aside from previously-identified Taurus members, there are a
large variety of other previously-identified objects in our survey
region.  Many of these are clearly not Taurus members, but some
are more ambiguous. The previously-identified objects include known
extragalactic objects, named objects of unknown nature, confirmed
non-members, and potential (unconfirmed) members of Taurus.  

To construct this list, we first searched in SIMBAD over our entire
field to obtain a list of $\sim$8000 known objects.  For objects that
did not already have high-precision coordinates, we went back to the
original article reporting the discovery of the object and attempted
re-identification of the object using finding charts and 2MASS
images.  If no finding charts were available, the brightest close
object from 2MASS was assigned to the object's name.  Some objects are
not recoverable, but most were identified; nearly 90\% of the entire
$\sim$8000-object list has high-accuracy coordinates in the end.  New
coordinates were reported back to the SIMBAD team for inclusion in
their database. 

We also included the results from several papers from the literature
reporting specifically confirmed non-members.  These confirmed
non-members can be candidate member objects from other Taurus surveys
such as Luhman \etal\ (2006) or Scelsi \etal\ (2008), that failed a
spectroscopic test for membership. They can also be spectroscopically
confirmed background giants from studies of the ISM (e.g., studies of
the Taurus dark cloud). These confirmed non-members have not
necessarily been ingested into SIMBAD, since they were not the primary
scientific result of the paper. Note that we did not list as
non-members those objects merely assumed but not confirmed (via
spectroscopy) to be background giants.

We merged this list by position with our master catalog to identify
objects seen in our survey.  For each object for which a match was
found in our catalog, we went back to the original literature in an
attempt to identify it as a known extragalactic object, a named
object of unknown nature (e.g., objects from an all-sky survey where
no specific follow-up has been done), a confirmed non-member (as
defined immediately above), or a potential (unconfirmed) member of
Taurus.  In the case of objects from the literature listed as
potential but unconfirmed Taurus members, we noted and bookkept these
objects separately, and we mention them where relevant below; some
are indeed recovered here by our Spitzer-based searches for YSOs.

Thus, the sample of Taurus non-members is certainly biased and far
from comprehensive and is {\em defined} to include mixtures of stars,
other Galactic objects (such as planetary nebulae), and extragalactic
objects.  This sample can be indicative of some typical colors to
expect from a variety of types of infrared-bright non-member objects.

As a further diagnostic for non-members (including extragalactic
objects), we merged by position to the 2MASS extended object catalog. 
Objects in this catalog are likely but not guaranteed to be all
extragalactic objects -- 11 (out of 215) previously identified Taurus
members are also 2MASS extended objects (due to, e.g., scattered light
from extended dust structures), but 107/148 previously-known galaxies
are 2MASS extended objects.  2MASS extended object identifications, if
relevant, are noted in the data tables and SEDs below.

\subsubsection{The Process}
\label{sec:process}

In order to find new candidate Taurus members, we first examined
various color-color and color-magnitude spaces using our entire Taurus
catalog, highlighting the locations of the previously identified
objects (both members and non-members).  We compared these diagrams to
discussions in the literature also seeking to identify YSOs from
Spitzer photometric measurements (e.g., Allen \etal\ 2004, Padgett
\etal\ 2008b, Rebull \etal\ 2007, Harvey \etal\ 2007, Gutermuth \etal\
2008).  There is no single color selection criterion that is 100\%
reliable in separating members from non-member contaminants.  Exactly
which color selection criteria work best can be a strong function of
the relative bandpass sensitivities and saturations, since 2MASS,
IRAC, and MIPS do not all detect the same faintest objects (due not
only to sensitivities but also degree of interstellar reddening and
embeddedness of the young protostellar objects), or saturate for the
same brightest objects.  After extensive empirical investigation using
diagrams from the literature as well as new diagrams, we selected four
color-magnitude diagrams (CMDs) and one color-color diagram (CCD)
which provided the best diagnostics for YSOs, and we used them to
construct an initial list of new candidate YSOs.  In each diagram, we
define regions most likely to harbor YSO candidates, and regions most
likely to contain galaxies or other non-members; these are listed in
detail in \S\ref{sec:samplecriteria}.  

By imposing these color selections, we are selecting objects that have
infrared excesses (e.g., flux densities above that expected for a
photosphere) and whose overall brightness is consistent with objects
at the Taurus distance. We interpret these excess objects as dusty
objects, with circumstellar disks and/or envelopes.  We do not select
objects without infrared excesses.

One aspect of our survey which makes it different from many of the
Spitzer-based surveys in the literature is our extensive optical
imaging.  While the SDSS and CFHT imaging data do not cover every
square arcminute of the Spitzer maps, they cover most of it.  The
SDSS spatial resolution is only slightly better than the IRAC
resolution at $\sim$1.4$\arcsec$, but the CFHT data has much better
spatial resolution at $\sim0.6-0.8\arcsec$.  We examined the images
at all available bands for each of the nearly 900 objects meeting the
color-color or color-magnitude criteria (plus many more objects in
the process of establishing these color spaces). At any band, if the
object is a resolved galaxy, or projected in the vicinity of a
galaxy cluster, we dropped it from further consideration. Some of the
objects that are resolved are actually previously-identified YSOs.
Many of the objects that have Spitzer colors similar to YSOs turn out
to be resolved galaxies when examined with SDSS or CFHT.  These
optical imaging data have been crucial to our ability to distinguish
galaxies from YSO candidates.  

If the candidate object meets the color criteria in any one of the
color-magnitude spaces we investigated and passes the imaging/spatial
resolution test, we regard it as a provisional YSO candidate, pending
additional scrutiny discussed below.  Objects meeting the color
criteria but failing the imaging/spatial resolution test are
``candidate non-members'' and appear separately in the Figures below.

\subsubsection{Gradations of Confidence for YSO candidates}
\label{sec:gradations}

Previously identified Taurus members tend to be bright, because
previous infrared (and optical) surveys were shallower than our
surveys.  True new Taurus members are also likely to be generally
bright.  Very red (embedded or cool) objects could also be members,
especially since this survey goes fainter in the infrared than any
prior survey of the region (excepting the Spitzer GTO observations in
the various core regions, Hartmann \etal\ 2005).  However, the fainter
objects are also statistically more likely to be galaxies, especially
over our survey area of more than 44 square degrees at $-15\arcdeg$
galactic latitude.  Thus, we specifically focused our attention on
bright and/or red objects meeting our color selection criteria. Faint
red objects meeting our color selection criteria were also considered
but  are statistically more likely to be galaxies than YSO Taurus
members.

In addition to the easily quantifiable Spitzer magnitude and color
criteria, we also individually assessed each candidate YSO using
qualitative judgments.  These include but are not limited to:
morphology in imaging data in each available band; relative brightness
at all bands from $U$ to 160 $\mu$m (e.g., infrared excess, but
optical too bright to be a Taurus member); amplitude of excess; shape
of SED; apparent (projected) proximity to other previously identified
Taurus members; apparent (projected) proximity to clearly-identifiable
galaxies (e.g., appearing to be part of a galaxy cluster); resolvable
spiral arms or tidal tails; previous identifications (e.g., with the
2MASS extended source catalog); estimated \av\ from the 160 \mum\ map
(e.g., objects seen in high extinction regions are likely Taurus
members); and star counts (a similar criterion to proximity to
galaxies or estimated \av).  These assessments were done over several
weeks by groups of co-authors and resulted in increased appreciation
of the range of contaminants, and more objects being identified as new
likely galaxies.  (See the Appendix discussion on ``8 \mum\ pop-up
objects," \S\ref{sec:popups}, for an example of a class of objects we
rejected.)  We looked critically at the shape of the excess above the
photosphere, and if the excess appeared only at one band (8 or 24
\mum), we retained the object as a YSO candidate only if it was more
than 4$\sigma$ above the photosphere (see \S\ref{sec:insignifirx} for
many of the rejected low-significance sources).  For the surviving YSO
candidates, based on all of the available information from any
wavelength (spectroscopic as well as photometric, plus derived
information such as placement in a theoretical Hertzprung-Russell
Diagram -- see \S\ref{sec:ldlstar}) as well as all the criteria listed
above, we assigned a letter grade, A/B/C, with grades of ``A''  as
more likely members than those with grades of ``C.''

These qualitative criteria can fail to recover some of the
previously-identified Taurus members.  Several of the
previously-identified Taurus members do not have IR excesses and are
therefore not recoverable by our search.  Seven of the
previously-identified Taurus members (e.g., 042146.3+265929 or
042307.7+280557) would probably have been rejected because in the
optical images, these objects are in front of a field of galaxies,
i.e., they appear to be part of a galaxy cluster, and are not in a
high \av\ region. One additional previously identified member, the
well-known edge-on disk IRAS 04302+2247 (=SST Tau 043316.5+225320),
was temporarily identified as a galaxy because its appearance was so
unusual in the optical image.  While our process is clearly imperfect,
we are confident that, working as a group and using all of the
available multi-wavelength information, we have identified a reasonably
high-confidence sample of candidate Taurus members present in our
photometric catalog. 

Our approach for finding YSOs is customized to our data set. This
labor-intensive process is not one that can be blindly applied to
other regions, even regions where similar extensive supporting
optical data are available. While our color selection can be easily
applied to any Spitzer+2MASS catalog, the manual examination of each
object is not necessarily easily duplicated and certainly automating
this process is not currently possible.  However, because our survey
is wide-area, and the contamination rate is high, this process is
unavoidable and has been crucial to our YSO selection.  The time we
spent in vetting the candidate list enabled more efficient use of
our follow-up spectroscopic telescope time, e.g., there was little
time wasted in taking spectra of contaminants.

\subsubsection{The Figures and Sample Selection Criteria}
\label{sec:samplecriteria}

\begin{deluxetable}{lllllll}
\tablecaption{Sample Selection Criteria \label{tab:samplecriteria}}
\tabletypesize{\footnotesize}
\tablewidth{0pt}
\tablehead{
\colhead{sample selected via...} & \colhead{YSO selection} &
\colhead{faint flag} }
\startdata
24/70 CMD  & either [24]$<$7 OR [24]$-$[70]$>$6 & [24]$>$7 \\
\ks/24 CMD & \ks$<$14 AND \ks$-$[24]$>$1 & \ks$>$13.5\\
8/24 CMD   &  [8]$-$[24]$>$0.5 AND  & [8]$>$9.5\\
           & ( ([8]$-$[24]$<$4 and [8]$<$10) or 
([8]$-$[24]$\geq$ 4 and [8]$<$2.5$\times$([8]$-$[24])) )  \\
4.5/8 CMD  &          [4.5] $<$ 6 AND   & [4.5]$>$11\\
          &   ( [4.5] $\ge$ 6 and [4.5] $\leq$ 11.5 \\
          &      and [4.5]$-$[8] $>$ 0.4) 
           OR ( [4.5] $>$ 11.5 and 
                [4.5] $<$ 0.6944$\times$([4.5]$-$[8])+11.22) 
\\
IRAC CCD   &   [3.6]$-$[4.5]$>$ 0.15 and [5.8]$-$[8]$>$ 0.3 
        and [3.6] $<$ 13.5
& \ldots \\
\enddata
\end{deluxetable}

The color-color and color-magnitude spaces we have chosen to use (see
\S\ref{sec:process} for overview) are the following:  [24] vs.\
[24]$-$[70] (\S\ref{sec:cmd1}), \ks\ vs.\ \ks$-$[24]
(\S\ref{sec:cmd2}), [8] vs.\ [8]$-$[24] (\S\ref{sec:cmd3}), [4.5] vs.\
[4.5]$-$[8] (\S\ref{sec:cmd4}), and finally,  [3.6]$-$[4.5] vs.\
[5.8]$-$[8] with an additional [3.6] brightness cutoff 
(\S\ref{sec:cmd5}).  Table~\ref{tab:samplecriteria} summarizes the
details of the sample selection criteria for each parameter space. Our
final selection includes objects selected in any of these parameter
spaces (not just objects selected in all of them); this will be
discussed in more detail in \S\ref{sec:newsamplesummary}. (We
explicitly compare this selection method to others from the literature
in \S\ref{sec:othermeth} below.) We discuss each of these parameter
spaces, in the order given above and in
Table~\ref{tab:samplecriteria}, in \S\ref{sec:2470}-\ref{sec:irac};
for each, there is a figure (Figures~\ref{fig:2470}-\ref{fig:irac})
consisting of 6 panels.  Each of the panels contains either a
subsample or a comparison sample to clearly demonstrate our selection
techniques.  In the remainder of this section, we discuss each of the
panels in introductory terms only.

In the upper left of each Figure is the SWIRE (Spitzer Wide-area
Infrared Extragalactic Survey; Lonsdale \etal\ 2003) ELAIS N1
extragalactic field\footnote{VizieR Online Data Catalog, II/255 (J.
Surace et al., 2004)} (the c2d reduction -- see Evans \etal\ 2007 --
is used here, as in Rebull \etal\ 2007 and Padgett \etal\ 2008b).  The
ELAIS N1 field is a $\sim$6 square degree field centered on 16h08m44s
+56d26m30s (J2000), or galactic coordinates ($l,b$) of 86.95, +44.48
(to be compared with the Taurus map center of $l,b\sim$173,$-$15). The
SWIRE sample is expected to be essentially entirely galaxies and
foreground stars, and as such provides a visual guide to the locations
where such objects appear in the corresponding diagram.  Note that
this is just the $\sim$6 square degree field, as observed; it has {\em
not} been scaled up to represent $\sim$44 square degrees of Taurus
data, because in this case we are primarily interested in the range of
colors sampled by the galaxies, not the overall numbers.  As we will
see below, many newly discovered extragalactic objects in our survey
have colors very similar to many certifiable YSOs, and different than
the colors of objects found in SWIRE. Note also the Galactic latitude
difference; this difference in Galactic latitude is likely to dominate
the source counts in IRAC bands 1 and 2.  More discussion of relative
source counts will appear in P09.  

In the upper right panel of each Figure, our entire Taurus
catalog is represented, so that various sub-samples can be
seen in the context of the larger catalog.  The Taurus catalog
is expected to consist of YSOs, foreground/background stars (and
other non-stellar galactic objects such as planetary nebulae),
and background galaxies (recall that the asteroids have already
been removed).  To first order, then, the objects in the Taurus
catalog that do not resemble the objects found in SWIRE are the
YSOs. However, the populations are not necessarily
well-separated, as can be seen in the remaining panels of the
Figures.  

The remaining 4 panels in each Figure are subsets of the Taurus
catalog. The second row of plots contains YSOs, both the previously
identified Taurus members (left) and new candidate Taurus members
(right), those selected by that particular diagram ($+$) as well as
others selected from other diagrams (grey dots).  The distribution of
previously identified Taurus members includes those selected based on
infrared excess and those selected via other mechanisms, and thus
includes objects without IR excesses; of course, we will not find
objects like the latter using Spitzer. Note also that the distribution
of previously identified Taurus members often includes objects that
have colors resembling galaxies.  This is not surprising, since the
galaxies are indeed undergoing star formation; thus, these color
selection mechanisms are successfully finding star formation, just not
necessarily in Taurus.  The set of new candidate Taurus members is
constructed from a color and magnitude cut on the entire sample, and
then examining all of the available data for each of the candidates,
dropping the likely galaxies (\S\ref{sec:gradations}).  

The third and final row contains the distributions for non-members,
both previously identified and newly identified here.  The left panel
is the sample of previously identified non-members which, as discussed
in \S\ref{sec:nonmembers}, includes stars identified via proper
motions, background giants, and galaxies identified in the
literature.  The last panel is the sample of all objects identified as
possibly YSOs based on Spitzer colors but then rejected as such, based
primarily on inspection of the optical images and SEDs; see
\S\ref{sec:gradations}.   Objects which passed all the other tests to
be YSO candidates but failed the spectroscopic test (see
\S\ref{sec:spectra}) are indicated in the last two panels of the
Figures by grey stars. (Note that, having been selected by the other
tests, they {\em also} appear in the 4th panel as candidate YSOs.)

Thus, for the color-magnitude or color-color space represented by
each Figure, one can examine and compare the distribution of galaxies
(SWIRE, previously identified non-members, new non-members), the
distribution of YSOs (previously identified Taurus members), the
distribution of foreground/background stars (SWIRE, previously
identified non-members), and the distribution of new candidate Taurus
members.

\begin{deluxetable}{lcccccccccccccc}
\tablecaption{Sample properties I.\tablenotemark{a} \label{tab:sampleproperties2new}}
\tabletypesize{\tiny}
\rotate
\tablewidth{0pt}
\tablehead{
\colhead{sample} & \multicolumn{2}{c}{24/70} & \multicolumn{2}{c}{\ks/24} &
\multicolumn{2}{c}{8/24} & \multicolumn{2}{c}{4.5/8} & \multicolumn{2}{c}{IRAC} &
\multicolumn{2}{c}{{\em ANY}} &
\multicolumn{2}{c}{objects identified} \\
 & \multicolumn{2}{c}{CMD}  &\multicolumn{2}{c}{CMD}  &\multicolumn{2}{c}{CMD}  &
\multicolumn{2}{c}{CMD}  &\multicolumn{2}{c}{CCD} 
&\multicolumn{2}{c}{objects identified} &
\multicolumn{2}{c}{in {\em ALL } diag.} }
\startdata
initial sample size & 447 && 357 && 381&& 334 && 266&& 883 && 103\\
%\hline
\cutinhead{{\bf Entire sample}}
%\hline
\# previously identified YSOs (all) & 89  && 135  &&  124 && 102  && 102  && 144  && 65 \\
\hspace{0.2in} (\% out of CMD selection ) && 19 && 37 && 32 && 30 && 38 && 16 && 63\\
\hspace{0.2in} (\% out of 215 previously identified YSO sample) && 41 && 63 && 58 && 47 && 47 && 67 && 30  \\
\hline
\# new candidate YSOs (all) & 34 && 85 && 81 && 65 && 57 && 148  && 16 \\
\hspace{0.2in} (\% out of CMD selection ) && 7 && 23 && 21 && 19 && 21 && 16 && 15\\
\hspace{0.2in} (\% out of 148 new candidate YSO sample) && 22 && 57 && 55 && 44 && 39 && 100 && 11  \\
\hline
\# previously identified NM (all) &  47 &&  51  &&  33 &&  38 &&  30 && 98  && 10  \\
\hspace{0.2in} (\% out of CMD selection ) && 10 && 14 && 8 && 11 && 11 && 11 && 9 \\
\hspace{0.2in} (\% out of 821 previously identified NM sample) && 6 && 6 && 4 && 5 && 4 && 12 && 1\\ 
\hline
\# new NM (all) & 270 &&  92 &&  147 &&  121 &&  72 && 487 && 12 \\
\hspace{0.2in} (\% out of CMD selection ) && 60 && 25 && 38 && 36 && 27 && 55 && 11\\
\hspace{0.2in} (\% out of 489 new NM sample) && 55 && 19 && 30 && 25 && 15 && 100 && 2 \\
\hline
\hline
\# SWIRE (all; for comparison) & 57 &&  29 &&  20 &&  42 &&  7 && 109 && 1 \\
\hspace{0.2in} (\% out of entire SWIRE sample ) && 3 && 1 && 0.9 &&2&& 0.3 && 5 && 0.04\\
\hline
\cutinhead{{\bf Just the faint sample}}
%\hline
\# previously identified YSOs (faint) & 0 && 3 && 15  && 5 && \ldots  && 19\tablenotemark{b} && 2\tablenotemark{b} \\
\hspace{0.2in} (\% out of CMD selection ) && 0 && 0.8 && 3 && 1 && \ldots && 2\tablenotemark{b} && 1\tablenotemark{b}\\
\hspace{0.2in} (\% out of 215 previously identified YSO sample) &&  0 && 1 && 7 && 2 && \ldots && 9\tablenotemark{b} && 1\tablenotemark{b}  \\
\hline
\# new candidate YSOs (faint) & 6 &&  9 &&  21 &&  32 &&  \ldots   && 56\tablenotemark{b} && 4\tablenotemark{b}\\
\hspace{0.2in} (\% out of CMD selection ) && 4 && 3 && 6 && 10 && \ldots && 8\tablenotemark{b} && 5\tablenotemark{b}\\
\hspace{0.2in} (\% out of 148 new candidate YSO sample) && 4 && 6 && 14 && 22 && \ldots  && 38\tablenotemark{b} && 3\tablenotemark{b}  \\
\hline
\# previously identified NM (faint) & 8 &&  16 &&  6 &&  32 &&  \ldots   && 49\tablenotemark{b} && 8\tablenotemark{b} \\
\hspace{0.2in} (\% out of CMD selection ) && 1 && 4 && 1 && 9 && \ldots && 5\tablenotemark{b} && 7\tablenotemark{b} \\
\hspace{0.2in} (\% out of 821 previously identified NM sample) &&  1 && 2 && 0.7 && 4 && \ldots && 5\tablenotemark{b} && 1\tablenotemark{b} \\
\hline
\# new NM (faint) & 156 && 43 && 110 && 111&& \ldots    && 371\tablenotemark{b} && 11\tablenotemark{b} \\
\hspace{0.2in} (\% out of CMD selection ) && 34 && 12 && 28 && 33 && \ldots && 42\tablenotemark{b} && 10\tablenotemark{b}\\
\hspace{0.2in} (\% out of 489 new NM sample) && 32 && 9 && 22 && 23 && \ldots && 76\tablenotemark{b} && 2\tablenotemark{b} \\
\hline
\hline
\# SWIRE (all; for comparison) & 40 &&  15 &&  13 &&  24 &&  \ldots &&107 && 1 \\
\hspace{0.2in} (\% out of entire SWIRE sample ) && 2 && 0.7 && 0.6
&&1&& \ldots && 5\tablenotemark{b} && 0.04\tablenotemark{b}\\
\enddata
\tablenotetext{a}{As an example for how to read this table, in the
case of the 24/70 CMD, 447 objects are selected by our color cuts in
this diagram.  Of these, 19\% are previously identified YSOs, 7\% are
new candidate YSOs, 10\% are previously identified non-members, and
60\% are new non-members. And, out of the 215 stars that compose our
previously identified Taurus member sample, 41\% are recovered in the
24/70 diagram, of the 148 objects in our candidate YSO sample, 22\%
are found here, of the 821 previously identified non-members, 6\% are
found in this diagram, and of the 489 new non-members, 55\% are found
here.}
\tablenotetext{b}{Objects listed here were picked in {\em any} diagram as
being faint.}
\end{deluxetable}

\begin{deluxetable}{lrrrr}
\tablecaption{Sample properties II. \label{tab:sampleproperties1new}}
\tabletypesize{\footnotesize}
\tablewidth{0pt}
\tablehead{
\colhead{property} & \colhead{prev.\ ident.\ YSOs\tablenotemark{a}} & \colhead{prev.\
ident.\ YSOs,
bright\tablenotemark{c}} &
\colhead{new candidate YSOs} & \colhead{new candidate YSOs,
bright} \\
\colhead{} & \colhead{total (fraction)\tablenotemark{b}} & \colhead{total (fraction)} & 
\colhead{total (fraction)} & \colhead{total (fraction)} }
\startdata
total sample size  &    215 (1.00) & 196 (1.00) & 148 (1.00)& 92 (1.00) \\
having IRAC 3.6 \mum\   &    193 (0.90) & 174 (0.89) & 142 (0.96)& 86 (0.93)  \\
having IRAC 4.5 \mum\   &    189 (0.88) & 170 (0.87) & 145 (0.98)& 90 (0.98) \\
having IRAC 5.8 \mum\   &    212 (0.99) & 193 (0.98) & 147 (0.99)& 91 (0.99)  \\
having IRAC 8.0 \mum\   &    206 (0.96) & 187 (0.95) & 147 (0.99)& 92 (1.00) \\
having MIPS 24 \mum\    &    173 (0.80) & 154 (0.79) & 135 (0.91)& 82 (0.89)  \\
having MIPS 70 \mum\    &     95 (0.44) &  89 (0.45) &  35 (0.24)& 17 (0.18)  \\
having all 4 IRAC  &    187 (0.87) & 168 (0.86) & 141 (0.95)& 86 (0.93)  \\
having all 4 IRAC+MIPS 24&   154 (0.72) & 135 (0.69) & 128 (0.86)& 76 (0.83)  \\
having both MIPS   &     89 (0.41) &  83 (0.42) &  35 (0.24)& 17 (0.18)  \\
having \ks\        &    209 (0.97) & 194 (0.99) & 140 (0.95)& 92 (1.00)  \\
having prior name  &    215 (1.00) & 196 (1.00) &  65 (0.44)& 46 (0.50)  \\
\enddata
\tablenotetext{a}{Previously identified YSO Taurus members}
\tablenotetext{b}{Notation indicates in each case that the
first number is the total number in that (sub-)sample, and the second number
(in parentheses) is the sample fraction, e.g., 192 out of 215
previously identified Taurus members have 3.6 \mum\ measurements, or
90\% of the 215 star sample.}
\tablenotetext{a}{Previously identified YSO Taurus members, bright
sub-sample, e.g., those not flagged as red and faint; see
\S\ref{sec:gradations}.  The purpose of extracting the bright sample
as distinct from the entire sample is to demonstrate the properties of
the brightest objects, e.g., in the case of the new candidate Taurus
members, those least statistically likely to be galaxies.}
\end{deluxetable}

\clearpage

\subsection{Implementation of the Spitzer Selection Criteria}

\subsubsection{Selection via [24] vs.\ [24]$-$[70]}
\label{sec:2470}
\label{sec:cmd1}

\begin{figure*}[tbp]
\epsscale{0.9}
\plotone{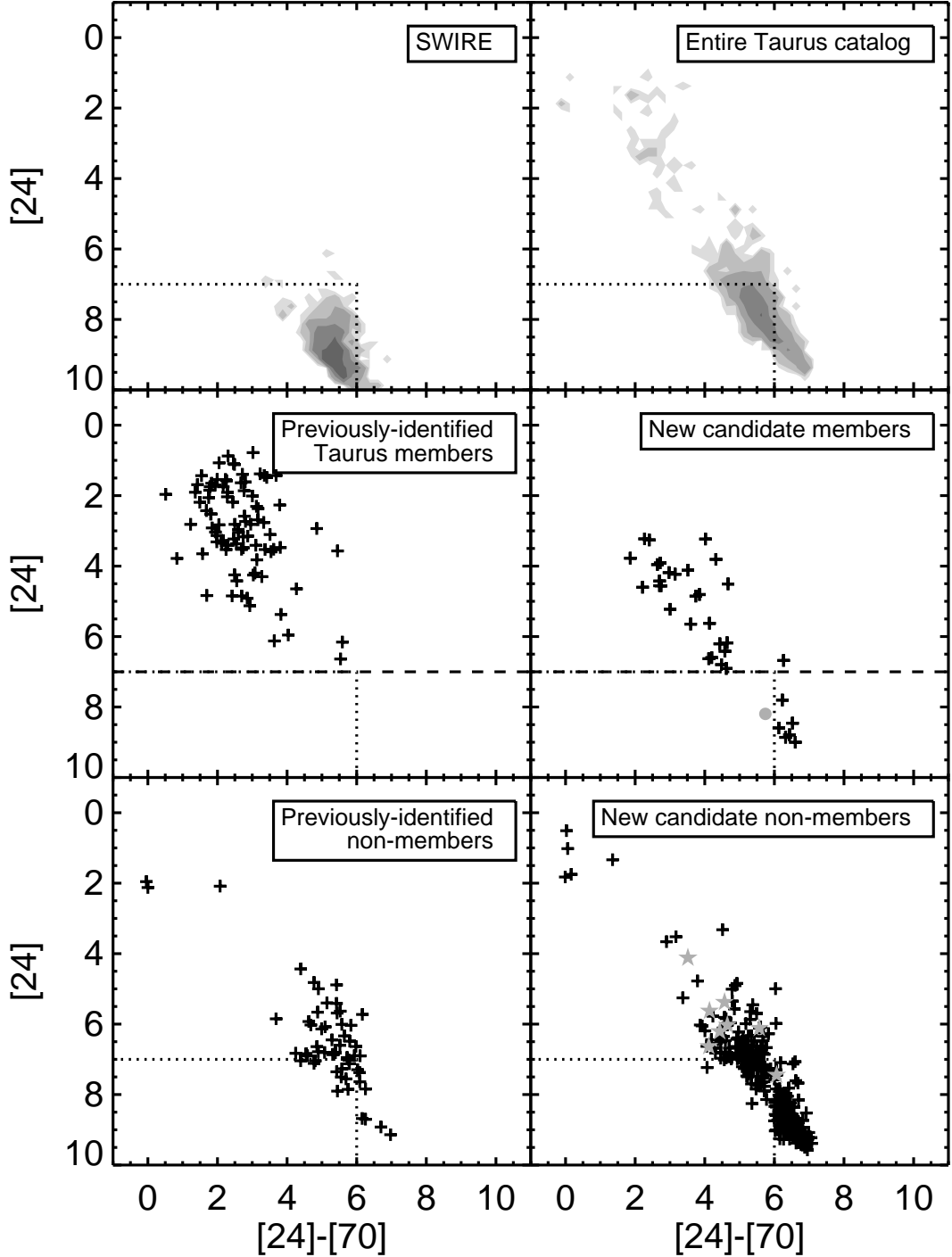}
\caption{[24] vs.\ [24]$-$[70] for (TOP) the SWIRE sample (essentially
all galaxies; contours at 1,2,4,8,16 objects), the entire Taurus
sample (YSOs+contaminants; contours at 1,2,5,15,35; dotted line
indicates region considered for YSO candidacy), (MIDDLE) the sample of
previously-identified Taurus members, and the sample of all new
candidate members ($+$ = objects selected in this color-magnitude
space; grey dots=objects selected based on other color-magnitude
spaces; dotted line indicates region considered for YSO candidacy;
dashed line indicates cutoff for ``faint'' flag), and (BOTTOM) the
sample of previously-identified non-members (stars and galaxies) and
the sample of new candidate non-members(stars and galaxies) ($+$ =
objects selected in this color-magnitude space; grey stars=objects
selected as YSOs but spectroscopically confirmed to be non-members;
dotted line indicates region considered for YSO candidacy).}
\label{fig:2470}
\end{figure*}

The [24] vs.\ [24]$-$[70] diagram has been used before to find
new candidate YSOs (e.g., Padgett \etal\ 2008b, Rebull \etal\ 2007). 
Figure~\ref{fig:2470} shows this color-magnitude diagram for the 6
samples mentioned in \S\ref{sec:samplecriteria} above (left to
right, top to bottom): SWIRE (expected to be essentially entirely
galaxies), the entire Taurus sample, previously identified Taurus
members, new candidate Taurus members, previously identified
non-members (stars identified via proper motions, background giants,
and galaxies identified in the literature), and new candidate
non-members identified here.

By inspection of Figure~\ref{fig:2470}, we find that objects with
[24]$>$7 and [24]$-$[70] between about 4 and 7 are statistically
likely to be galaxies.  Unadorned photospheres (e.g., old foreground
stars) will be bright and have [24]$-$[70]$\sim$0; an A3 ZAMS
photosphere has [24]$\sim$7 at the distance of Taurus, and for a
median Taurus-age member, [24]$\sim$7 corresponds to mid-K.  Compared
to the SWIRE catalog, the entire Taurus catalog contains many objects
with similar colors, but also many objects that are similarly red and
much brighter at [24], and therefore are candidate dusty young
stars.  

Based on the properties of the previously identified member and
non-member samples, the properties of the SWIRE sample, and
discussions in the literature, the selection we impose to search for
new candidate YSOs is either [24]$<$7 or [24]$-$[70]$>$6.  Statistics
on this sample are given in Table~\ref{tab:sampleproperties2new}
(along with statistics from the SWIRE sample for comparison); in
summary, this cut yields $\sim$450 objects, each of which we
investigated at all our available imaging bands; $\sim$20\% of them
are previously identified Taurus members, $\sim$7\% of them survive
the tests to be potential new YSOs, and $\sim$70\% are previously
identified or new non-member objects.  Nearly all of the previously
identified Taurus members that appear in this plot have [24]$<$7. 
This is a likely bias in that all of the previous infrared surveys
searching for Taurus members were much shallower than this Spitzer
survey -- the IRAS sensitivity limit was about 0.3 Jy, so we are going
7 magnitudes fainter than IRAS. However, in order to appear in this
plot, the objects have to have been detected at 70 \mum\ as well, so
the sensitivity of the 70 \mum\ survey is usually the limiting
factor. Of the entire sample of previously identified Taurus members
within our survey, 80\% are detected at 24 \mum\ (see
Table~\ref{tab:sampleproperties1new}), and just 45\% are detected at
70 \mum; $\sim$5\% of the previously identified Taurus members are
saturated in MIPS-24 and $\sim$3\% are saturated in MIPS-70. 

In this diagram, the set of previously-identified Taurus objects is
generally distinguished from the distribution of faint objects found
in the SWIRE sample. Fainter Taurus objects ([24]$>$7) could exist,
but objects that faint are statistically likely to be galaxies; their
properties at other bands could suggest otherwise.  Objects surviving
the imaging test (and other qualitative criteria -- see
\S\ref{sec:gradations}) but with [24]$>$7 are therefore further
identified as ``faint.''  About 17\% of the $\sim$35 potential new
YSOs in this parameter space are faint.  Many of the other faint
objects selected by our color/magnitude cut indeed resolve into
galaxies when examined using the CFHT or SDSS imaging -- $\sim$60\% of
all of the objects selected in this space are new galaxy candidates. 
About 100 of the brighter objects selected by this color cut resolve
into galaxies, so faintness alone is insufficient for locating and
identifying galaxies. As can be seen in Figure~\ref{fig:2470}, most of
the previously-identified non-members and new candidate non-members
resemble the colors of objects found in SWIRE.   Figure~\ref{fig:2470}
and the statistics in Table~\ref{tab:sampleproperties1new} also
demonstrate that our sample of candidate new YSOs is on average redder
and fainter than the sample of previously identified YSOs.

As discussed above (\S\ref{sec:spectra}), we have obtained Palomar
and/or Keck spectroscopy of many of our candidate objects. Because
this color selection uses bandpasses far from optical, these objects
are often very faint indeed at optical or NIR bands. We have
spectroscopy for about 70\% of the 34 new candidate YSOs selected in
this color space.  So far, almost 90\% of those are stellar (e.g.,
YSOs or stars that could still be shown to be foreground stars or
background giants), and just 4 are confirmed to be extragalactic
objects.

% PULL OUT AND CONSOLIDATE THESE PARAS SOMEWHERE ELSE:
% About 100 objects selected by this color cut are identified with
% previously-assigned names that are not previously identified Taurus
% members; most of these previously-identified objects are 2MASS extended
% sources (which could be galactic or extragalactic objects), and some
% of which are planetary nebulae, carbon stars, galaxies, and objects
% identified in the literature as potential (unconfirmed) members of
% Taurus.   Our sample of candidate YSOs selected from this
% color-magnitude space includes some objects regarded as candidate
% non-members in the literature, but which we have promoted to candidate
% YSO status; see \S\ref{sec:miscobj}.

\subsubsection{Selection via \ks\ vs.\ \ks$-$[24]}
\label{sec:k24}
\label{sec:cmd2}

\begin{figure*}[tbp]
\epsscale{0.9}
\plotone{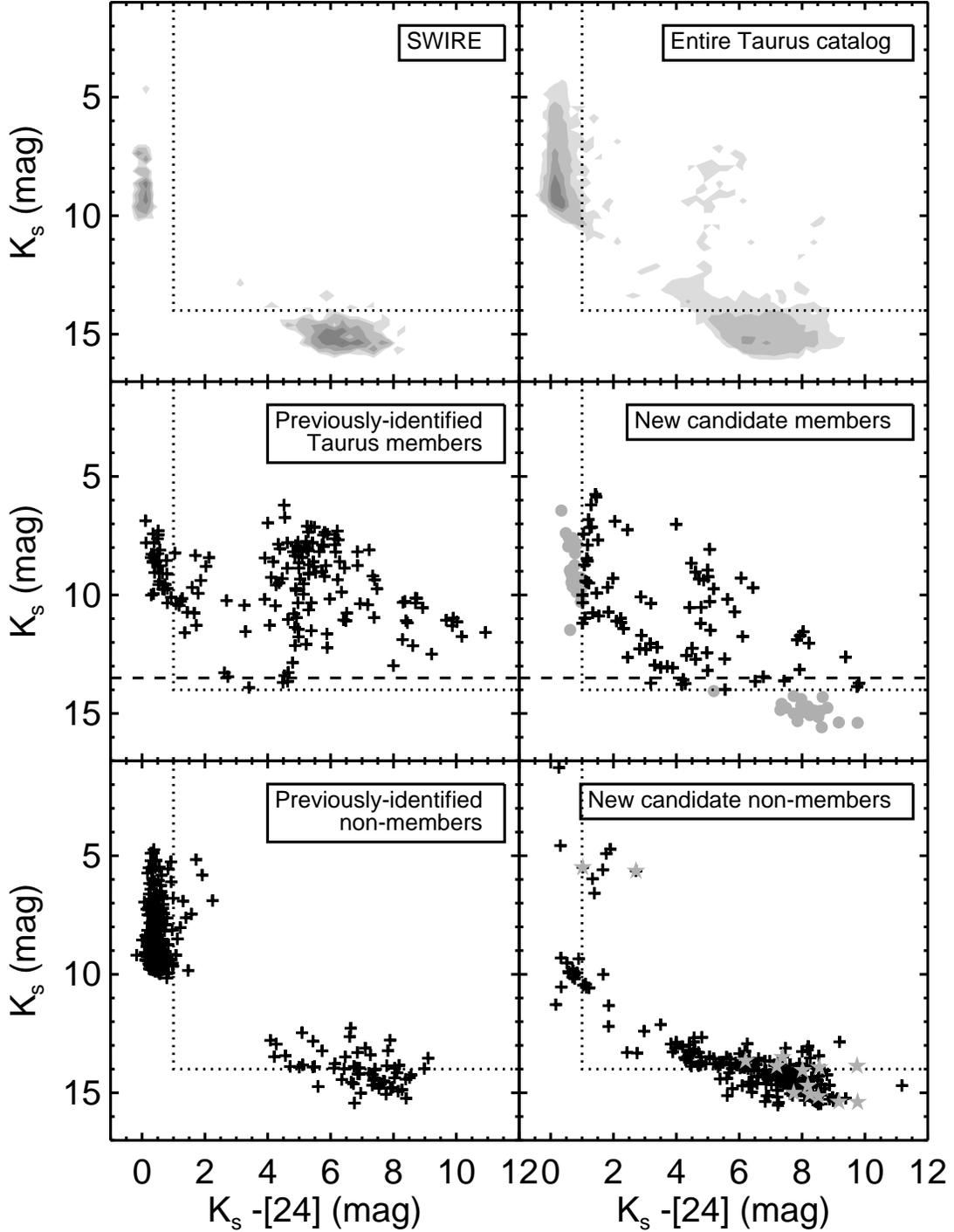}
\caption{\ks\ vs.\ \ks$-$[24] for (TOP) the SWIRE sample (galaxies \&
foreground stars; contours at 1,2,4,8,16 objects), the entire Taurus
sample (YSOs+contaminants; contours at 1,5,50,100,200 objects; dotted
line indicates region considered for YSO candidacy), (MIDDLE) the
sample of previously-identified Taurus members, and the sample of all
new candidate members ($+$ = objects selected in this color-magnitude
space; grey dots=objects selected based on other color spaces; dotted
line indicates region considered for YSO candidacy; dashed line
indicates cutoff for ``faint'' flag), and (BOTTOM) the sample of
previously-identified non-members (stars and galaxies) and the sample
of new candidate non-members(stars and galaxies) ($+$ = objects
selected in this color-magnitude space; grey stars=objects selected as
YSOs but spectroscopically confirmed to be non-members; dotted line
indicates region considered for YSO candidacy).}
\label{fig:kk24}
\end{figure*}

As for [24] vs.\ [24]$-$[70], the \ks\ vs.\ \ks$-$[24] diagram has
been used previously to find new candidate YSOs (e.g., Padgett \etal\
2008b, Rebull \etal\ 2007).  Figure~\ref{fig:kk24} shows this
color-magnitude diagram for the same samples as Fig.~\ref{fig:2470}
(see \S\ref{sec:selectionoverview}).  The SWIRE sample clearly (more
obviously than the previous diagram) consists of both galaxies
(\ks$\gtrsim$14 and \ks$-$[24] between about 4 and 8) and 
stars (\ks$\lesssim$10 and \ks$-$[24]$\sim$0).  As before, the entire
Taurus catalog has many objects with colors similar to the SWIRE
sample, but also many objects that have properties different than the
SWIRE sample, e.g., redder than \ks$-$[24]$\sim$1 and brighter than
\ks$\sim$14, as well as redder than \ks$-$[24]$\sim$8. Note that the
lack of sources in the lower left of each panel is an artifact of the
sensitivity limits of the survey.

The sample of previously identified Taurus members, for the most part,
have \ks$<$14, which generally avoids the region populated by galaxies
in the SWIRE sample, but there are legitimate YSOs mixed in with the
galaxies in this parameter space.  Here too, the historical bias
towards brighter objects in prior surveys can be seen, and faint red
objects could be legitimate YSOs. Essentially all of the previously
identified YSO sample has a \ks\ measurement in our database, but as
mentioned above, just $\sim$80\% are cleanly detected at 24 \mum\ (see
Table~\ref{tab:sampleproperties1new}).  As above, fainter objects are
statistically likely to be galaxies.

The selection we impose on this parameter space to search for new
candidate YSOs is that \ks$<$14 and \ks$-$[24]$>$ 1. Again,
Table~\ref{tab:sampleproperties2new} summarizes the sample sizes;
$\sim$360 objects meet these criteria, including most of the
(detected) previously identified Taurus members; note that there are
some stars without apparent IR excess (e.g., likely WTTS) with
\ks$-$[24]$\sim$0, and there are YSOs that have colors resembling
galaxies.  Note also that late-type stars do not have \ks$-$[24]=0
(Gautier \etal\ 2007).  Objects with \ks$>$13.5 are further identified
as ``faint'' and thus statistically likely to be galaxies. Previously
identified Taurus members compose 135 of the objects meeting the basic
color criteria; about 50 are previously identified non-member objects,
most of which are 2MASS extended sources (which could be galactic or
extragalactic objects).  By inspection of the individual images, about
100 of the objects selected here are clearly resolved galaxies. 
Besides the previously identified Taurus objects, 85 objects are
indistinguishable from point sources, or have morphologies consistent
with YSO candidates, and meet all the other qualitative criteria (see
\S\ref{sec:gradations}) for potential new YSOs selected via this color
space; a quarter of these were already found via the 24/70 color
magnitude diagram above.

In this parameter space, there is still a bias (relative to the
previously identified Taurus member sample) towards finding red and
faint objects, but this appears to be not nearly as strong as it was
in the 24/70 space above.   Of the 85 candidate YSOs found in this
space, we have already obtained Palomar and/or Keck spectroscopy (see
\S\ref{sec:spectra}) for 85\% of them. All of them except for 1 are
stellar (e.g., YSOs or stars that could still be shown to be
foreground stars or background giants); just 1 is rejected outright as
a galaxy.    

A relatively high fraction of literature background giants appear as
selected in this parameter space.  Because there is a large
difference in wavelength between \ks\ and [24], this search is
particularly sensitive to objects with small excesses, which could be
interesting transition disk candidates. However, these objects could
also be subject to reddening from the Taurus cloud that is high
enough to significantly affect \ks\ but not 24 \mum, or Taurus cloud
emission affecting the 24 \mum\ but not the \ks\ photometry --
background giants are therefore potentially  selected in this space. 
Several objects presented in the literature as {\em candidate}
background objects based on shorter-wavelength photometric
observations (i.e., without confirming spectroscopy) appear here as
objects with potential excesses only at the longer wavelengths. With
the information we have, we are unable to distinguish currently
between transition disk candidates (i.e., Taurus objects with
excesses only at 24 \mum) and confirmed background giants.  These
objects are all identified in Table~\ref{tab:posnewyso} as candidate
non-members which we have promoted to low-grade candidate YSOs.  The
SEDs that appear in Appendix~\ref{sec:seds} reveal that several of
our candidate objects indeed have \ks\ values significantly affected
by reddening and some candidate objects with clear cloud emission at
8 and/or 24 \mum\ are indicated in Table~\ref{tab:newysot3}; also see
\S\ref{sec:insignifirx} for discussion of objects with very small
excesses, usually just at 24 \mum. We expect that several of the
objects we have identified here will turn out to be background
giants. Some candidate transition disk objects will be discussed in
McCabe \etal\ (2009).

We note here that for surveys where the IRAC and MIPS coverage is
well-matched, using [3.6] or even [4.5] in place of \ks\ for this 
color-magnitude space is likely to be a better choice for searching
for YSOs for  two reasons: (a) minimizing the influence of reddening
on \ks\ (3.6 or 4.5 \mum\ is less affected by reddening than \ks; see
Padgett \etal\ 2008b for more discussion on the influence of \av), and
(b) minimizing the intrinsic range of star colors  -- the intrinsic
\ks$-$[24] color of M stars is not zero (Gautier \etal\ 2007) whereas
[3.6]$-$[24] or [4.5]$-$[24] is zero for those stars.  Specifically
for our survey, the overall \av\ towards Taurus is low, and all young
stars at the distance of Taurus should be visible to 2MASS unless they
are edge-on substellar objects.  Moreover, using [3.6] or [4.5] in
place of \ks\ does not reveal any YSO candidates  not already selected
by the color spaces used here, and finds in total only 2 more
extragalactic objects.  Had we used either [3.6] or [4.5] in place of
\ks, however, we would have found a factor of $\sim$4 fewer objects
that we believe (based on inspection and our qualitative criteria) to
be likely reddened background giant contaminants.

\subsubsection{Selection via [8] vs.\ [8]$-$[24]}
\label{sec:824}
\label{sec:cmd3}

\begin{figure*}[tbp]
\epsscale{0.9}
\plotone{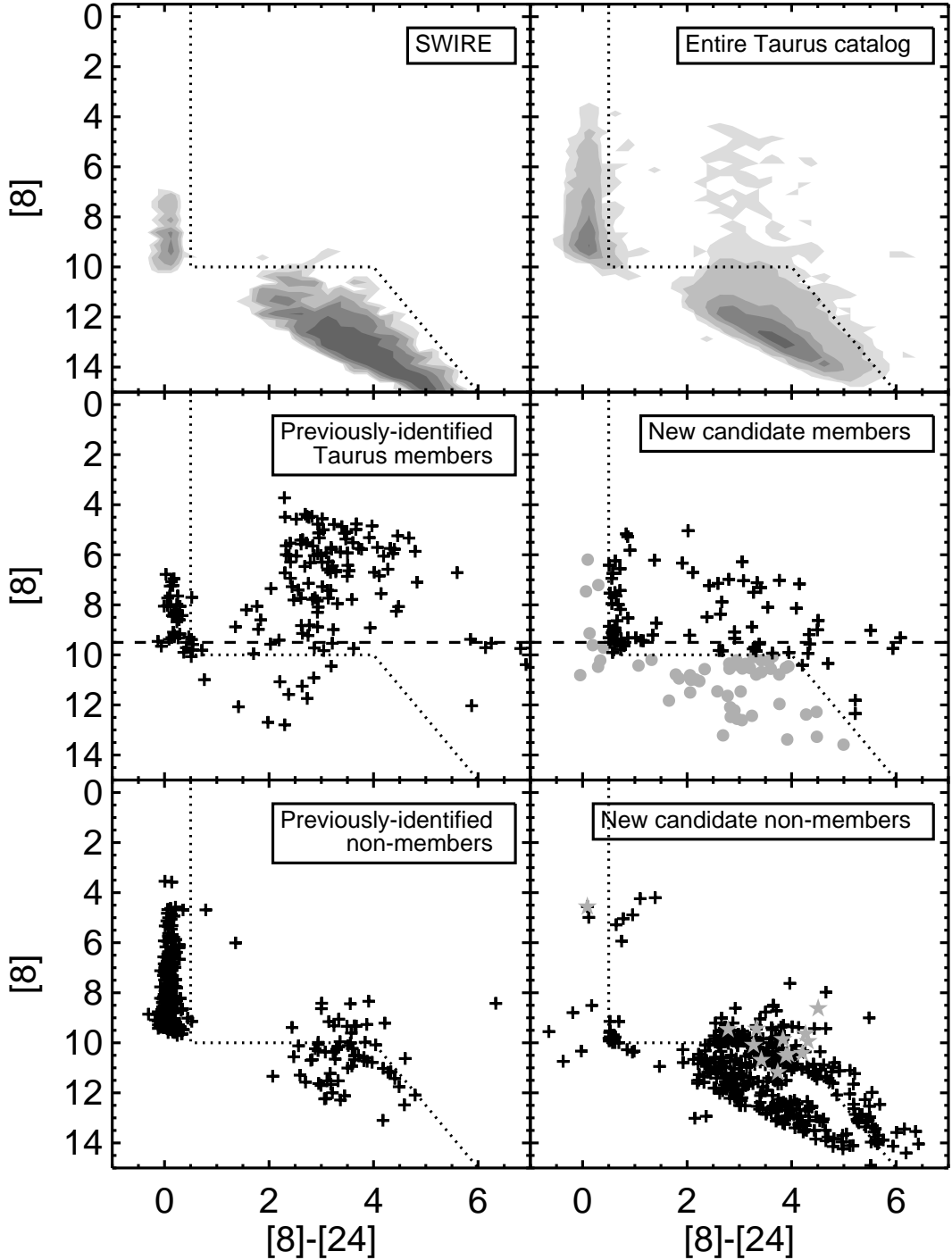}
\caption{[8] vs.\ [8]$-$[24] for (TOP) the SWIRE sample (galaxies \&
foreground stars; contours at 1,2,4,8,16 objects), the entire Taurus
sample (YSOs+contaminants; contours at 1,5,50,100,200 objects),
(MIDDLE) the sample of previously-identified Taurus members, and the
sample of all new candidate members ($+$ = objects selected in this
color-magnitude space; grey dots=objects selected based on other color
spaces; dotted line indicates region considered for YSO candidacy;
dashed line indicates cutoff for ``faint'' flag), and (BOTTOM) the
sample of previously-identified non-members (stars and galaxies) and
the sample of new candidate non-members(stars and galaxies) ($+$ =
objects selected in this color-magnitude space; grey stars=objects
selected as YSOs but spectroscopically confirmed to be non-members;
dotted line indicates region considered for YSO candidacy).}
\label{fig:8824}
\end{figure*}

While essentially all of the previously identified YSOs have 2MASS
detections at \ks, some fainter legitimate YSOs may be embedded enough
that the relatively shallow 2MASS survey will not detect the objects
at \ks, whereas they will be detected by our IRAC survey.  Thus, we
chose to investigate the [8] vs.\ [8]$-$[24] parameter space; see
Figure~\ref{fig:8824}. 

The morphology of this space is similar to the \ks\ vs.\ \ks$-$[24]
space, except the region occupied primarily by galaxies is now more
elongated in color and has a more prominent slope towards fainter and
redder objects.  In order to select specifically for objects redder
than most galaxies, the selection criteria we used consist of three
line segments:  (a) [8]$-$[24]$>$0.5 to avoid the stars without
excesses; and (b) ([8]$-$[24]$<$4 and [8]$<$10) to catch the bright
stars in the middle of the plot; OR  ([8]$-$[24]$\geq$ 4 and
[8]$<$2.5$\times$([8]$-$[24])) to obtain the reddest stars.  Objects
with [8]$>$9.5 are statistically likely to be galaxies, and thus those
are further identified as ``faint'' in Table~\ref{tab:posnewyso}. 

About 380 objects in the catalog meet these criteria, $\sim$120 of
which are previously identified YSOs, $\sim$180 of which are
non-members (previously identified or new), and $\sim$80 of which
survive the imaging test (and other qualitative criteria -- see
\S\ref{sec:gradations}) and remain potential YSOs.  Of these,
$\sim$25 were already found using the [24] and [70] selection
criteria (\S\ref{sec:2470}), $\sim$55 were found using \ks\ and [24]
(\S\ref{sec:k24}), and $\sim$20 were found in all three
color-magnitude planes. 

% Of all the objects retrieved which are any sort of
% previously-identified objects but not previously-identified YSOs, most
% again are 2MASS extended object sources, PNe, some IRAS sources, etc. 
% Of the previously-identified non-members, just one object has not
% already been retrieved by the selections above, and this object was
% identified as a galaxy based on our optical imaging, and independently
% by spectroscopic follow-up by Luhman \etal\ (2006). 

There is an apparent gap in the last panel of Figure~\ref{fig:8824}
inside the ``galaxy blob'' at [8]$-$[24]$>$4 and [8]$>$10. This is a
direct result of our selection methodology. The fainter, bluer part of
the distribution are largely those objects that were not obviously
point sources in the optical imaging, and the redder part of the
distribution closely tracks the (dotted) dividing line we used for our
selection criteria. This population is composed of objects we
rejected as candidate YSOs based on the qualitative criteria listed in
\S\ref{sec:gradations} above.

We have followup spectroscopy (see \S\ref{sec:spectra}) for $\sim$80\%
of the 81 new candidate YSOs selected in this color space.  Nearly all
(95\%) of these are stellar; just 3 are rejected as galaxies.

\subsubsection{Selection via [4.5] vs.\ [4.5]$-$[8]}
\label{sec:4.58}
\label{sec:cmd4}

\begin{figure*}[tbp]
\epsscale{0.9}
\plotone{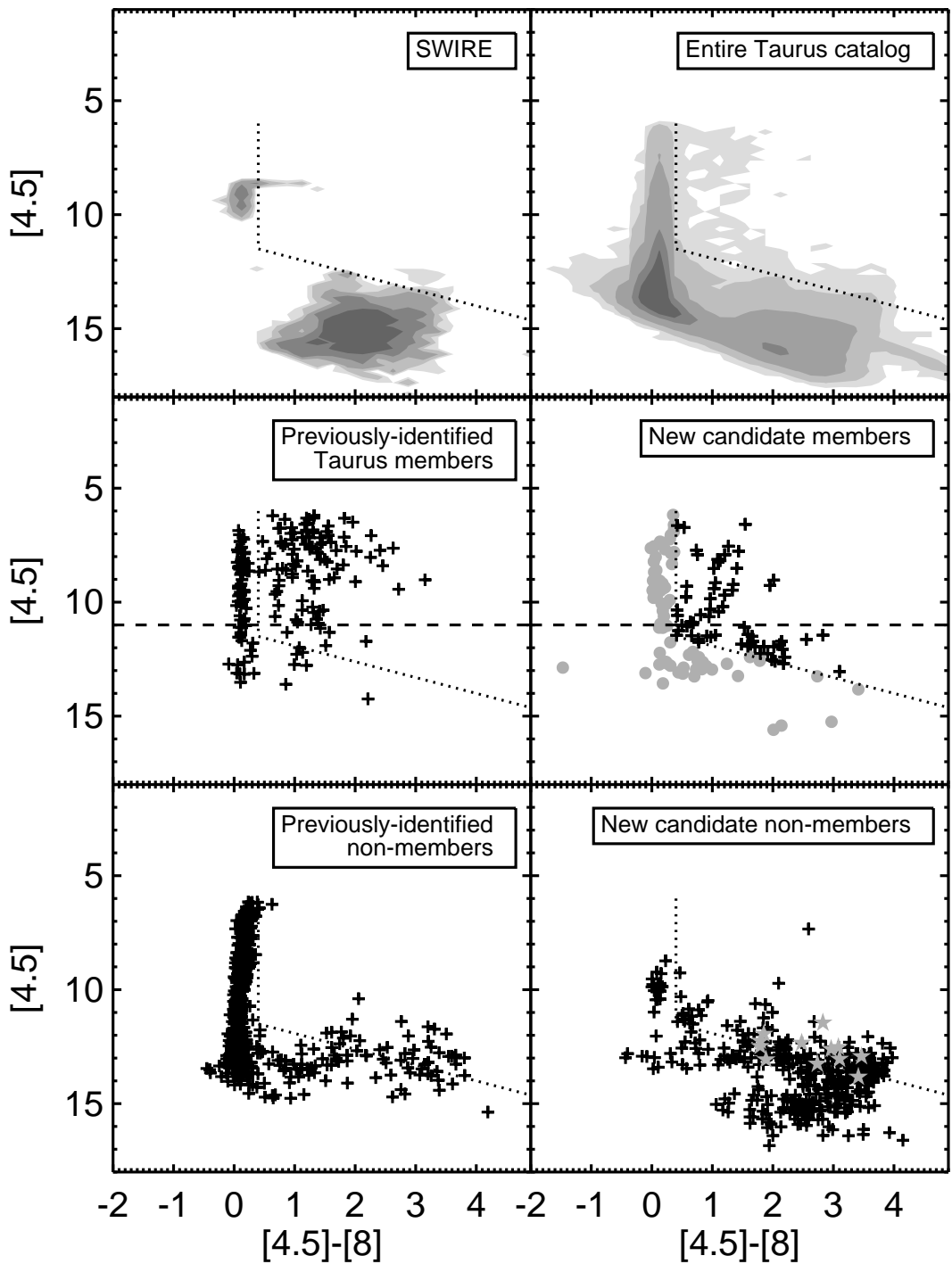}
\caption{[4.5] vs.\ [4.5]$-$[8] for (TOP) the SWIRE sample (galaxies
\& foreground stars; contours at 1,2,4,8,16 objects), the entire
Taurus sample (YSOs+contaminants; contours at 1,10,100,1000,2000
objects; dotted line indicates region considered for YSO candidacy),
(MIDDLE) the sample of previously-identified Taurus members, and the
sample of all new candidate members ($+$ = objects selected in this
color-magnitude space; grey dots=objects selected based on other color
spaces; dotted line indicates region considered for YSO candidacy;
dashed line indicates cutoff for ``faint'' flag), and (BOTTOM) the
sample of previously-identified non-members (stars and galaxies) and
the sample of new candidate non-members(stars and galaxies) ($+$ =
objects selected in this color-magnitude space; grey stars=objects
selected as YSOs but spectroscopically confirmed to be non-members;
dotted line indicates region considered for YSO candidacy).}
\label{fig:4.58}
\end{figure*}

To this point, we have required MIPS-24 detections for YSO candidate
selection, which strongly biased our sample of new potential YSOs
towards the generally brighter and/or larger excess objects (by
comparison to the rest of the catalog).   As an example of how many
YSOs we may be missing by requiring 24 \mum, just $\sim$80\% of the
previously identified YSOs are detected at 24 \mum.   This results
from a combination of intrinsic disk properties (where disk emission
makes the objects easier to detect at 24 \mum) and the Spitzer
sensitivity relative to low-mass photospheres at the Taurus distance
(for those YSOs without disks). 
By loosening
this restriction and not requiring MIPS-24, we extend the sample of
potential objects, but also the potential contamination.   We now
investigate the [4.5] vs.\ [4.5]$-$[8] parameter space; see
Figure~\ref{fig:4.58}.  This parameter space, on its own, provides the
largest possible initial sample size we have yet investigated, as 17\%
of the entire catalog is detected in these two IRAC bands (compared
with just 2\% of the entire catalog detected at MIPS-24; see P09 for
additional similar statistics). To first order, the morphology of this
space is similar to the other spaces we have investigated, with the
photospheres clustering around [4.5]$-$[8]$\sim$0 and the galaxies in
a red and faint grouping.  There are some new features apparent in
this space, however.  Saturation at 4.5 \mum\ occurs at 650 mJy (6.1
mags), so the locus of colorless objects is truncated at that level. 
In the sample of previously identified Taurus members, there is a
clear distinction between the disked and non-disked population (a gap
near [4.5]$-$[8]=0.5) which is not seen when considering the entire
catalog. 

The selection criteria we used to find candidate YSOs in this space
selected the brighter and redder objects; we did this by stitching
together several line segments.  They are: (a) [4.5] $<$ 6 AND  (b)
([4.5] $\geq$ 6 and [4.5] $\leq$ 11.5  and [4.5]$-$[8] $>$ 0.4)  OR (c)
([4.5] $>$ 11.5 and  [4.5] $<$ 0.6944$\times$([4.5]$-$[8])+11.22).   
About 335 objects meet these basic criteria, $\sim$100 of which are
previously identified YSOs, and $\sim$160 of which are new or
previously identified  non-members. About 65 objects survive the
imaging test (and other qualitative criteria -- see
\S\ref{sec:gradations}) and are new candidate YSOs. The distribution
of these objects in this color-magnitude space is very different than
that for the previously identified members; these new objects are
distinctly fainter and redder on the whole than the previously
identified sample.  Objects with [4.5]$>$11 are given the ``faint''
flag in Table~\ref{tab:posnewyso}. 

About a third of the new candidate YSO sample selected here are also
retrieved from the 24/70 space above and about two-thirds are also
retrieved from the \ks/24 and 8/24 spaces above. 

About 65\% of these 65 candidate Taurus objects have Palomar and/or
Keck spectroscopy (see \S\ref{sec:spectra}), and nearly all are
stellar; just 4 are dropped as galaxies.  

%Of the $\sim$30 objects with previous
%identifications in the literature, many are candidate brown
%dwarfs from other papers, including Luhman \etal\ (2006). 

\subsubsection{Selection via IRAC color-color diagram}
\label{sec:irac}
\label{sec:cmd5}

\begin{figure*}[tbp]
\epsscale{0.85}
\plotone{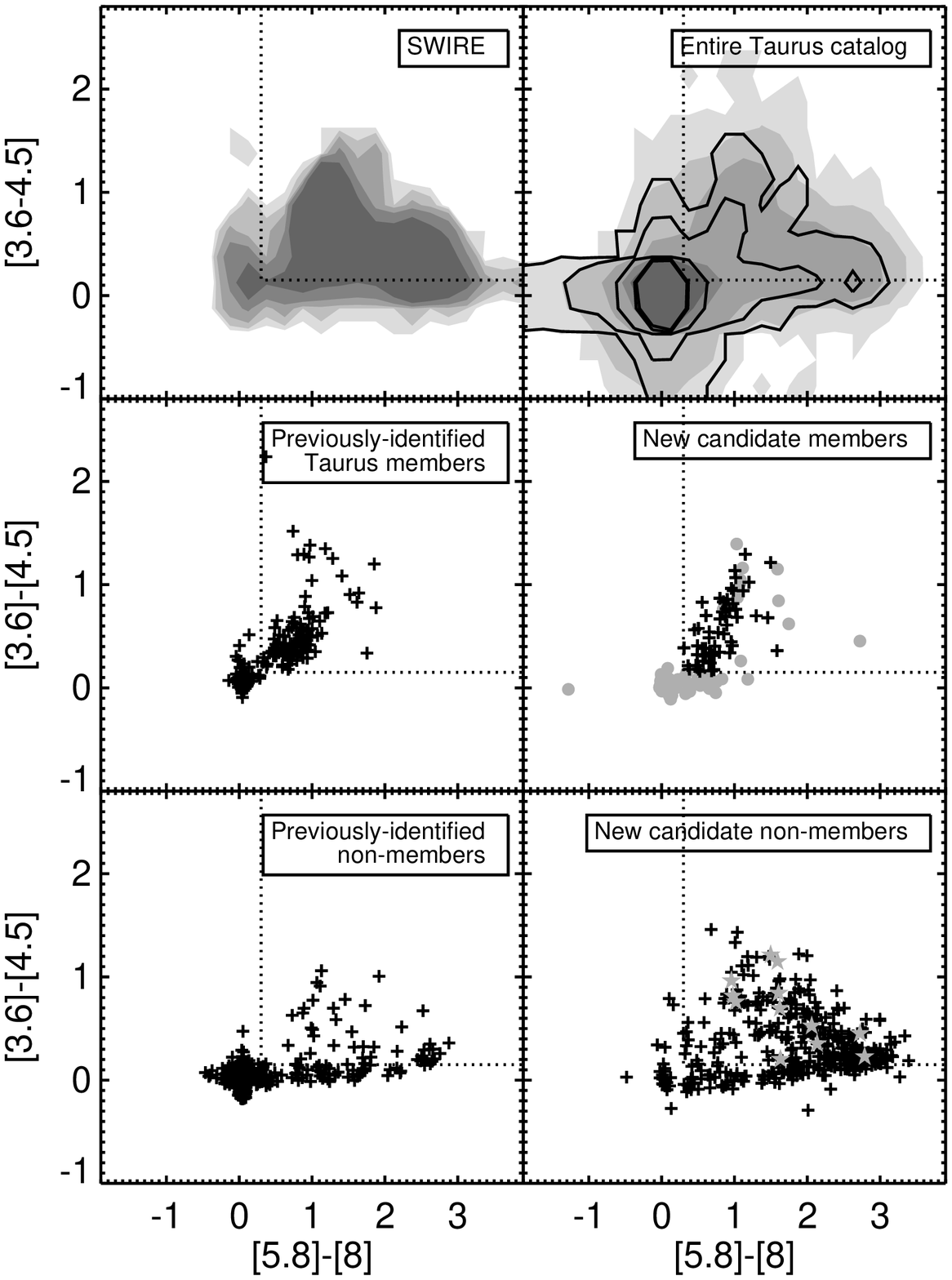}
\caption{[3.6]$-$[4.5] vs.\ [5.8]$-$[8] for (TOP) the SWIRE sample
(galaxies \& foreground stars; contours at 1,2,4,8,16 objects), the
entire Taurus sample (YSOs+contaminants; grey contours at
1,10,100,1000,2000 objects; solid line contours are for the entire
Taurus sample with an additional [3.6]$<$13.5, same contour limits as
the grey contours), (MIDDLE) the sample of previously-identified
Taurus members, and the sample of all new candidate members ($+$ =
objects selected in this color-color space; grey dots=objects selected
based on other color spaces; dotted line indicates region considered
for YSO candidacy), and (BOTTOM) the sample of previously-identified
non-members (stars and galaxies) and the sample of new candidate
non-members(stars and galaxies) ($+$ = objects selected in this
color-color space; grey stars=objects selected as YSOs but
spectroscopically confirmed to be non-members; dotted line indicates
region considered for YSO candidacy). An additional [3.6] brightness
cut was also imposed on the YSO selection in this color space; see
text.}
\label{fig:irac}
\end{figure*}

As our final selection mechanism, we use the IRAC color-color diagram
(as seen in, e.g., Allen \etal\ 2004).  This parameter space, on its
own, provides an initial sample size comparable to the previous 4.5/8
color selection, as 13\% of the entire catalog is detected in all 4
IRAC bands (to be compared with just 2\% of the entire catalog being
detected at MIPS-24, and 17\% detected at 4.5 and 8 microns; see P09
for additional similar catalog statistics).  However, by using the
IRAC color-color diagram on its own, we are blind to any luminosity
information about the sources.  This information was present
previously because we were using color-magnitude, not color-color,
diagrams.  Given the surface density of galaxies, as well as the fact
that the galaxy/YSO separation is not as vivid in this parameter
space,  the luminosity information is crucial.  We  imposed a {\em
requirement} of [3.6] $<$ 13.5, and  the cut we used on the IRAC
color-color space was (based on literature discussions)
[3.6]$-$[4.5]$>$ 0.15 and [5.8]$-$[8]$>$ 0.3.   This approach is
different than in our consideration of the above parameter spaces,
where we specifically called out YSO candidates fainter than a
specific level.  

Using these criteria, $\sim$265 objects are selected, $\sim$100 of
which are previously identified Taurus members, $\sim$100 of which
are non-members (previously identified or new), and $\sim$65 of which
are new candidate YSOs.  Of those, $\sim$35\% were also found using
24/70, $\sim$60\% were found using \ks/24 or 8/24, and $\sim$80\%
were found using 4.5/8.  (More on objects selected using all CMDs in
\S\ref{sec:newsamplesummary} below.)

As can be seen in Figure~\ref{fig:irac}, the previously identified
Taurus members roughly fall into two groups -- those with little or no
IRAC excess, and those with substantial excesses.  Among our new
candidate members, we have some objects with little or no IRAC excess
(selected from other parameter spaces), as well as objects with more
substantial excesses, but the division is not as clean, suggesting
contaminants in our YSO candidate list.  We have also selected some
objects that are very red in [5.8]$-$[8] but nearly colorless in
[3.6]$-$[4.5].  These could be disks with large inner holes, or
galaxies.

%Of the $\sim$35 objects with previous identifications in the
%literature, several are candidate brown dwarfs from other papers,
%including Luhman \etal, and several are 2MASS extended objects.  

A little more than half of these 57 candidate objects have
spectroscopy (see \S\ref{sec:spectra}) from Palomar and/or Keck;
just 4 are galaxies.

\subsection{Tables of objects}
\label{sec:finalverdicts}
\label{sec:datatables}

Now that we have used the Spitzer properties of the
previously-identified Taurus member sample to select a new candidate
YSO sample, we present the data tables with observed and derived
properties of both sets of objects.  In this section, we review the
contents of Tables~\ref{tab:knownyso}--\ref{tab:newmembers}. The
contents of Tables~\ref{tab:knownyso} and \ref{tab:posnewyso} are
similar but not identical, as are the contents of
Tables~\ref{tab:knownysot3} and \ref{tab:newysot3}.
Table~\ref{tab:newmembers} summarizes the new members, sorted by
confidence level.

Tables~\ref{tab:knownyso} and \ref{tab:posnewyso} present, for each
of the previously identified Taurus members and the new candidate
Taurus members identified here, respectively, SST Tau names (note, as
IAU-compliant names, the Right Ascension and Declination as given in
the name are truncated, not rounded), a common name from previous
studies (if applicable), and measurements in the Spitzer bands.  For
most of these objects, this is the first time that MIPS flux densities have
appeared in the literature.  Since many of the previously identified
and new candidate Taurus members appear as having YSO-like colors in
more than one of the color-magnitude and color-color diagrams of
Table~\ref{tab:samplecriteria} and
Figures~\ref{fig:2470}-\ref{fig:irac}, Tables~\ref{tab:knownyso} and
\ref{tab:posnewyso} also present the color criteria that are met by
each object individually, for each of the color spaces we use here to
identify YSO candidates.  Objects appearing faint and red (see
\S\ref{sec:gradations} above) are indicated as such.  SEDs  for each
of the previously identified members and new candidates appear
in Appendix \ref{sec:seds}; the SED properties
of the sample as whole will be discussed in \S\ref{sec:netsed}
below.  Table~\ref{tab:posnewyso} additionally contains notes about
individual objects.  These tables are sorted by SST Tau name,
effectively sorted by RA and then Dec.

Additional information about each of the previously identified Taurus
members and the new candidate Taurus members identified here appears
in Tables~\ref{tab:knownysot3} and \ref{tab:newysot3}, respectively. 
Both of these tables start by repeating the SST Tau name, and a
previous common name (if applicable).  Table~\ref{tab:newysot3} then
lists the grade ranking (see \S\ref{sec:gradations} above) that we
assigned each of the candidate objects. Note that
these tables are still sorted by position; Table~\ref{tab:newmembers}
later presents the candidate Taurus members in order of confidence.

We compare our search method to others in the literature in
\S\ref{sec:othermeth} below; in preparation for that,
Tables~\ref{tab:knownysot3} and \ref{tab:newysot3} indicate, for each
of the previously identified Taurus members,  how (or if) the c2d
(Harvey \etal\ 2007) or Gutermuth \etal\ (2008) criteria for YSOs
identified the object.  (NB: whether or not our search recovered each
previously identified object can be found in
Table~\ref{tab:knownyso}.) 

Tables~\ref{tab:knownysot3} and \ref{tab:newysot3} contain a YSO
classification.  The near- to mid-IR slope of each SED, $\alpha$, is
what we used for determining a YSO classification for these objects.
For each of the previously identified and new candidate objects in our
survey, we performed a simple ordinary least squares linear fit to all
available photometry (just detections, not including upper or lower
limits) between 2 and 24 $\mu$m, inclusive.  Note that errors on all of the
infrared points are so small as to not affect the fitted SED slope,
and that a forthcoming paper will investigate the (small) effects of
fitting a line to all available points within a different wavelength
range, e.g., 3.6 to 24 \mum. In the spirit of Wilking \etal\ (2001),
we define $\alpha = d \log \lambda F_{\lambda}/d \log  \lambda$,
where  $\alpha > 0.3$ for a Class I, 0.3 to $-$0.3 for a
flat-spectrum  source, $-$0.3 to $-$1.6 for a Class II, and $<-$1.6
for a Class III.  We realize that the precise definition of $\alpha$
can vary, resulting in different classifications for certain objects;
detailed discussion of this issue is beyond the scope of this paper.
Classification via this method is provided for all previously
identified and new candidate objects specifically to enable comparison
within this paper via internally consistent means.

Adopted spectral types appear in Tables~\ref{tab:knownysot3} and
\ref{tab:newysot3}. This spectral type comes from the literature (see
\S\ref{sec:known}) or from our spectra (\S\ref{sec:spectra}); if the
latter, it is indicated as such in the ``notes'' column.  If we
obtained a spectrum for the object, and if H$\alpha$ was measurable,
we report the equivalent width in these Tables (see
\S\ref{sec:spectra} for analysis details). If the Ca infrared triplet
was in emission at the time of our observation, it is noted in the
``notes'' column. 

Tables~\ref{tab:knownysot3} and \ref{tab:newysot3} include an estimate
of the star's luminosity ($L_{\rm *}$) and an estimate  of the ratio
of the infrared excess luminosity to the star's total
(photospheric+infrared) luminosity ($L_{\rm IR}/L_{\rm total}$, where
$L_{\rm total} = L_{\rm *} + L_{\rm IR}$).   We describe briefly our
procedure to determine the infrared luminosity.  Many faint sources
were undetected at the longest wavelengths; we extrapolated missing
data points for wavelengths longer than 8.0~$\mu$m. We used the
longest wavelength available data point as a reference and assumed
that its flux density corresponded to blackbody emission peaking at
that wavelength. We then used this blackbody function to estimate the
missing fluxes at the longer wavelengths. We compared the measured
$J-H$ colors and the expected photospheric $(J-H)_0$ colors (as
tabulated in Chapter 7 of Allen's Astrophysical Quantities; Cox 2001),
attributing any difference to extinction, \av, and we corrected the
photometric fluxes for it. We obtained a spline curve through the
corrected fluxes in $\log F_{\lambda}$, $\log \lambda$ space on a
wavelength grid from 0.1~\AA\ to 1000~\AA\ with a $\Delta\lambda =
0.02$~\AA. To determine the infrared excess, we determined the
underlying stellar photospheric emission using PHOENIX stellar
atmosphere models for $T_{\rm eff} \leq 10,000$~K and Kurucz models
for $T_{\rm eff} > 10,000$~K and matched the stellar atmosphere model
to the corrected $J$-band fluxes. The infrared luminosity was
calculated by measuring the difference between the spline curve and
the stellar atmosphere model. We emphasize that care was taken to
ensure that we included only adequate excess luminosity contributions
at each wavelength of the grid (e.g., if the stellar atmosphere model
showed a drop in flux between the \ks\ and $3.6~\mu$m fluxes, and the
spline curve was above but there was no obvious sign of infrared
excess at those wavelengths, the contribution to the infrared
luminosity was not included). We also only included contributions if
the spline curve was at least 1.5 times larger than the stellar
atmosphere contribution. Finally, we determined if the calculated
infrared luminosity was an upper limit by checking that at least one
of the corrected photometric fluxes beyond $3.6~\mu$m were real and
not extrapolated. If the latter, the infrared luminosity was indicated
as an upper limit.  Note that this method is subject to several
caveats: (a) objects that do not have spectral types (particularly
common in the list of new candidate objects) do not have a calculated
\av, and therefore no $L_*$ or $L_{\rm IR}/L_{\rm total}$; (b) this
method of dereddening using $J-H$ ignores the potential for near-IR
excess (as well as veiling and the H-opacity minimum at 1.6 \mum) and
therefore this method can overestimate $L_*$; (c) YSOs intrinsically
vary at essentially all the wavelengths used here, and the photometry
across the SED for each object was not obtained contemporaneously. Our
values for $L_{\rm IR}/L_{\rm total}$ are probably good to a factor of
2, and are sufficient to determine if a source has an envelope, is an
optically thick disk only, a highly flattened / thinning disk, or a
debris-like system.  (See \S\ref{sec:ldlstar} for more discussion.)
This approach is demonstrably inaccurate in comparison to more complex
methods, such as those employing bolometric corrections to
reddening-corrected flux densities in well-calibrated passbands.
However, more detailed modeling is beyond the scope of this paper.

For the 148 new candidate Taurus members, Table~\ref{tab:newysot3}
also contains an adopted membership classification indicating whether
or not we regard the object as a Taurus member. This classification
combines information from all of the available photometric and optical
spectroscopic data (see \S\ref{sec:newsamplesummary} and
\S\ref{sec:gradations}).  Extragalactic objects are noted as
``xgal.''   Objects that we believe to be reliable new members are
indicated as ``new member.''  Objects that we think are likely to be
new members but there is still some doubt are listed as ``probable new
member.''  Objects with the next gradation of confidence are
``possible new members.''  Objects for which we have spectra but
cannot determine clear membership are listed as ``needs additional
followup'' and, finally, objects with no spectroscopic data yet are
``pending followup."  There are 34 new members, 3 probable new
members, 10 possible new members, 7 extragalactic objects, 2 other
objects, 60 stars needing additional follow-up observations, and 33
pending any follow-up observations. Note that while all the ``new
members'' are also grade A objects, not all grade A objects are ``new
members'', because of the need for additional data in many cases. 

A final data table, Table~\ref{tab:newmembers}, lists the new members
in order of quality, sorted by their category (new member, probable
new member, possible new member, needs additional followup, or pending
followup), and then by our rank (and then by catalog number), such
that the objects we grade as most likely to be new members appear at
the top of the list. Table~\ref{tab:newmembers} also contains the
projected angular separation from to the nearest previously-identified
Taurus object.

\begin{deluxetable}{rrrrrrrrrrrrrrl}
\tablecaption{Spitzer measurements for sample of previously identified Taurus members\tablenotemark{a}\label{tab:knownyso}}
\tabletypesize{\tiny}
\rotate
\tablewidth{0pt}
\tablehead{
\colhead{SST Tau name} & \colhead{common name} & \colhead{[3.6]} &
\colhead{[4.5]} & \colhead{[5.8]} 
&\colhead{[8]} & \colhead{[24]} & \colhead{[70]} & \colhead{[160]} & 
\colhead{ident.\ in} & \colhead{ident.\ in} & 
\colhead{ident.\ in} & \colhead{ident.\ in} & 
\colhead{ident.\ in}  & \colhead{notes}
\\
\colhead{ } & \colhead{} & \colhead{(mag)} & \colhead{(mag)} &
\colhead{(mag)} 
&\colhead{(mag)} & \colhead{(mag)} & \colhead{(mag)} & \colhead{(mag)} &
\colhead{24/70} & \colhead{\ks/24} & 
\colhead{8/24}  & \colhead{4.5/8} & 
\colhead{IRAC} 
\\
\colhead{ } & \colhead{} & \colhead{} & \colhead{} &
\colhead{} 
&\colhead{} & \colhead{} & \colhead{} & \colhead{} &
\colhead{CMD} & \colhead{CMD} & 
\colhead{CMD}  & \colhead{CMD} & 
\colhead{CCD} 
}
\startdata
   041314.1+281910&                   LkCa 1&       8.54$\pm$       0.05&       8.50$\pm$       0.05&       8.41$\pm$       0.05&       8.43$\pm$       0.05&       8.28$\pm$       0.08&$>$       1.30             &     \nodata               &    \ldots&        no&        no&        no&        no & \nodata \\
   041327.2+281624&                   Anon 1&       7.23$\pm$       0.05&       7.24$\pm$       0.05&       7.17$\pm$       0.05&       7.08$\pm$       0.05&       6.95$\pm$       0.05&$>$       1.27             &     \nodata               &    \ldots&        no&        no&        no&        no & \nodata \\
   041353.2+281123&         IRAS04108+2803 A&       9.02$\pm$       0.05&       8.37$\pm$       0.05&       7.67$\pm$       0.05&       6.57$\pm$       0.05&       3.44$\pm$       0.04&$>$       0.19             &$>$      -2.05             &    \ldots&       yes&       yes&       yes&       yes &  b \\
   041354.7+281132&         IRAS04108+2803 B&       9.38$\pm$       0.05&       8.03$\pm$       0.05&       6.96$\pm$       0.05&       5.78$\pm$       0.05&       1.38$\pm$       0.04&      -1.84$\pm$       0.22&$>$      -1.94             &       yes&       yes&       yes&       yes&       yes &  b \\
   041357.3+291819&           IRAS04108+2910&       7.48$\pm$       0.05&       6.84$\pm$       0.05&       6.26$\pm$       0.05&       5.54$\pm$       0.05&       3.13$\pm$       0.04&       1.15$\pm$       0.22&$>$      -3.39             &       yes&       yes&       yes&       yes&       yes & \nodata \\
   041411.8+281153&        J04141188+2811535&      10.93$\pm$       0.06&      10.40$\pm$       0.05&      10.12$\pm$       0.07&       8.99$\pm$       0.06&       5.76$\pm$       0.01&$>$       1.03             &     \nodata               &    \ldots&       yes&       yes&       yes&       yes & \nodata \\
\enddata
\tablenotetext{a}{Table will be presented both in the print and in the
electronic version.  To convert between magnitudes and flux densities,
we use $M=2.5 \log
(F_{\rm zeropt}/F)$ where the zero-point flux densities for the seven Spitzer
bands are 280.9, 179.7, 115.0, and 64.13 Jy for IRAC and 7.14, 0.775,
and 0.159 Jy for MIPS.  IRAC effective wavelengths are 3.6, 4.5, 5.8,
and 8.0 \mum; MIPS effective wavelengths are 24, 70, and 160 \mum.}
\tablenotetext{b}{MIPS-160 flux density for this object is subject
to confusion with a nearby source or sources.}
\tablenotetext{c}{MIPS-160 flux density for this object is
compromised by missing and/or saturated data.}
\tablenotetext{d}{MIPS-160 flux density for this object is hard
saturated.}
\tablenotetext{e}{IRAC flux densities for 043835.4+261041=HV Tau C do not
appear in our automatically-extracted catalog. Flux densities here are
those from Hartmann \etal\ (2005); since their observations have more
redundancy at IRAC bands, they are able to obtain reliable flux
densities 
for this object at IRAC bands.  MIPS flux densities are determined from
our data.}
\tablenotetext{f}{The image morphology around 041426.2+280603 is
complex; careful PSF subtraction and modeling will be required to
apportion flux densities among the three local maxima seen in close proximity
in the IRAC images, which may or may not be three physically distinct
sources.}
\end{deluxetable}

\begin{deluxetable}{rrrrrrrrrrrrrrr}
\tablecaption{Spitzer measurements for sample of new candidate 
Taurus members\tablenotemark{a}\label{tab:posnewyso}}
\tabletypesize{\tiny}
\rotate
\tablewidth{0pt}
\tablehead{
\colhead{SST Tau name} & \colhead{common name} & \colhead{[3.6]} &
\colhead{[4.5]} & \colhead{[5.8]} 
&\colhead{[8]} & \colhead{[24]} & \colhead{[70]} & \colhead{[160]} & 
\colhead{ident.\ in} & \colhead{ident.\ in} & 
\colhead{ident.\ in} & \colhead{ident.\ in} & 
\colhead{ident.\ in} & \colhead{notes}
\\
\colhead{ } & \colhead{} & \colhead{(mag)} & \colhead{(mag)} &
\colhead{(mag)} 
&\colhead{(mag)} & \colhead{(mag)} & \colhead{(mag)} & \colhead{(mag)} &
\colhead{24/70} & \colhead{\ks/24} & 
\colhead{8/24}  & \colhead{4.5/8} & 
\colhead{IRAC} 
\\
\colhead{ } & \colhead{} & \colhead{} & \colhead{} &
\colhead{} 
&\colhead{} & \colhead{} & \colhead{} & \colhead{} &
\colhead{CMD} & \colhead{CMD} & 
\colhead{CMD}  & \colhead{CMD} & 
\colhead{CCD} 
}
\startdata
   041159.7+294236&                  \nodata&     \nodata               &      10.91$\pm$       0.05&     \nodata               &      10.22$\pm$       0.06&       9.88$\pm$       0.35&$>$       1.16             &$>$       1.33             &    \ldots&       yes&        no&       yes&    \ldots & \nodata \\
   041332.3+291726&                  \nodata&      13.04$\pm$       0.06&      12.96$\pm$       0.06&      12.79$\pm$       0.07&      11.96$\pm$       0.07&       8.19$\pm$       0.09&       2.45$\pm$       0.22&$>$       0.08             &        no&       yes&        no&        no&        no & \nodata \\
   041339.4+292114&                  \nodata&      12.80$\pm$       0.06&      11.94$\pm$       0.05&      11.15$\pm$       0.06&      10.35$\pm$       0.06&       7.54$\pm$       0.06&$>$       1.20             &$>$       1.46             &    \ldots&        no&        no&yes--faint&       yes & \nodata \\
   041427.3+255130&                  \nodata&      13.01$\pm$       0.06&      12.95$\pm$       0.06&      12.97$\pm$       0.07&      12.22$\pm$       0.08&       9.31$\pm$       0.21&$>$       1.33             &$>$      -0.25             &    \ldots&       yes&        no&        no&        no & \nodata \\
   041535.6+284741&                  \nodata&      13.20$\pm$       0.06&      12.26$\pm$       0.06&      11.33$\pm$       0.06&      10.34$\pm$       0.06&       5.65$\pm$       0.04&       2.05$\pm$       0.22&$>$      -1.53             &       yes&        no&yes--faint&yes--faint&       yes & \nodata \\
   041539.1+281858&                  \nodata&       8.77$\pm$       0.05&       8.52$\pm$       0.05&       8.17$\pm$       0.05&       7.50$\pm$       0.05&       4.23$\pm$       0.04&       1.08$\pm$       0.22&$>$      -2.26             &       yes&       yes&       yes&       yes&       yes & \nodata \\
\enddata
\tablenotetext{a}{Table will be presented both in the print and in the
electronic version.  To convert between magnitudes and flux densities,
we use $M=2.5 \log
(F_{\rm zeropt}/F)$ where the zero-point flux densities for the seven Spitzer
bands are 280.9, 179.7, 115.0, and 64.13 Jy for IRAC and 7.14, 0.775,
and 0.159 Jy for MIPS.  IRAC effective wavelengths are 3.6, 4.5, 5.8,
and 8.0 \mum; MIPS effective wavelengths are 24, 70, and 160 \mum.}
\tablenotetext{b}{MIPS-160 flux density for this object is subject
to confusion with a nearby source or sources.}
\tablenotetext{c}{MIPS-160 flux density for this object is
compromised by missing and/or saturated data.}
\tablenotetext{d}{MIPS-160 flux density for this object is hard
saturated.}
\tablenotetext{e}{Object previously identified in the literature as a
potential YSO.}
\tablenotetext{f}{Object previously identified in the literature as a
potential non-member.}
\tablenotetext{g}{Object identified as a Taurus member by other authors
using portions of this Spitzer Legacy data set. } 
\end{deluxetable}

\begin{deluxetable}{lllllllllllll}
\tablecaption{Derived properties of previously identified Taurus members\tablenotemark{a}\label{tab:knownysot3}}
\tabletypesize{\tiny} 
\rotate
\tablewidth{0pt}
\tablehead{
\colhead{SST Tau name} & \colhead{common name}  &
\colhead{c2d} & \colhead{Gutermuth} & \colhead{our} 
&\colhead{spectral type} & \colhead{our H$\alpha$ }
&\colhead{\av}
&\colhead{log $L_*$}
&\colhead{$L_{\rm IR}/L_{\rm total}$} 
&\colhead{notes}\\
\colhead{ } & \colhead{}  &
\colhead{category} & \colhead{category} & \colhead{class} 
&\colhead{} & \colhead{EQW (\AA)}
&\colhead{(mags)} 
&\colhead{ (\lsun)} 
&\colhead{} 
&\colhead{}}
\startdata
   041314.1+281910&                   LkCa 1&            yso&           star&            III&             M4&        \nodata&            1.3&           -0.3&    $<$ 2.0e-05&        \nodata\\
   041327.2+281624&                   Anon 1&            yso&           star&            III&             M0&        \nodata&            3.2&            0.4&    $<$ 3.8e-05&        \nodata\\
   041353.2+281123&         IRAS04108+2803 A&            yso&       class II&           flat&        \nodata&        \nodata&        \nodata&        \nodata&        \nodata&              d\\
   041354.7+281132&         IRAS04108+2803 B&            yso&       embedded&              I&        \nodata&        \nodata&        \nodata&        \nodata&        \nodata&              d\\
   041357.3+291819&           IRAS04108+2910&            yso&       class II&             II&            K5e&         -201.4&            4.8&           -0.4&           0.37&            b c\\
   041411.8+281153&        J04141188+2811535&            yso&       class II&             II&          M6.25&        \nodata&            1.7&           -1.6&           0.25&        \nodata\\
   041412.2+280837&          IRAS04111+2800G&            yso&       embedded&              I&        \nodata&        \nodata&        \nodata&        \nodata&        \nodata&              d\\
   041412.9+281212&             V773 Tau ABC&        \nodata&        \nodata&             II&             K3&        \nodata&            2.8&            0.9&          0.047&        \nodata\\
\enddata
\tablenotetext{a}{Table will be presented both in the print and in the
electronic version. }
\tablenotetext{b}{Ca IR triplet in emission in our spectrum.}
\tablenotetext{c}{New spectral type, reported here.}
\tablenotetext{d}{Unreduced IR spectrum in-hand, to be discussed in a
later paper.}
\tablenotetext{e}{043552.7+225423=HP Tau AB
was observed twice in the same night (3 Dec 2008), and within 7 hrs
the H$\alpha$ equivalent width changed from 1.95 \AA\ in absorption to
11.24 \AA\ in emission. }
\end{deluxetable}

\begin{deluxetable}{llllllllllllll}
\tablecaption{Derived properties of new potential Taurus members\tablenotemark{a}\label{tab:newysot3}}
\tabletypesize{\tiny}
\rotate
\tablewidth{0pt}
\tablehead{
\colhead{SST Tau name} & \colhead{common name} & \colhead{our} &
\colhead{c2d} & \colhead{Gutermuth} & \colhead{our} 
&\colhead{spectral type} & \colhead{our H$\alpha$}
&\colhead{\av} 
&\colhead{log $L_*$} 
&\colhead{$L_{\rm IR}/L_{\rm total}$} 
&\colhead{adopted memb.\ } &\colhead{notes}\\
\colhead{ } & \colhead{ } & \colhead{rank} &
\colhead{category} & \colhead{category} & \colhead{class} 
&\colhead{} & \colhead{EQW (\AA) }
&\colhead{(mags)}
&\colhead{ (\lsun)} 
&\colhead{ } 
&\colhead{classification} &\colhead{}
}
\startdata
   041159.7+294236&                  \nodata&              C&        \nodata&        \nodata&            III&          G8-K1&            0.9&            2.8&           -0.8&         0.0020&     needs add'l followup&              c\\
   041332.3+291726&                  \nodata&              A&        not yso&           star&             II&        \nodata&        \nodata&        \nodata&        \nodata&        \nodata&         pending followup&              d\\
   041339.4+292114&                  \nodata&             B-&            yso&       embedded&           flat&        \nodata&        \nodata&        \nodata&        \nodata&        \nodata&         pending followup&              k\\
   041427.3+255130&                  \nodata&             B+&        not yso&           star&            III&             G5&            2.5&            0.4&           -1.7&         0.0095&      possible new member&            c o\\
   041535.6+284741&                  \nodata&             A-&            yso&       embedded&              I&        \nodata&        \nodata&        \nodata&        \nodata&        \nodata&         pending followup&              d\\
   041539.1+281858&                  \nodata&             A+&            yso&       class II&             II&             M5&           -7.2&            2.8&           -0.5&          0.075&               new member&          c g m\\
   041542.7+290959&                  \nodata&             A+&            yso&           star&             II&       M0 w/ Av&           -1.9&            2.5&           -0.5&          0.099&               new member&            c g\\
   041557.9+274617&                  \nodata&             A+&            yso&       class II&             II&             M6&          -32.2&            1.1&           -1.1&           0.12&               new member&          c g m\\
\enddata
\end{deluxetable}

\begin{deluxetable}{llllll}
\tablecaption{List of new confirmed, probable, possible, and pending Taurus members\label{tab:newmembers}}
\tabletypesize{\footnotesize}
%\rotate
\tablewidth{0pt}
\tablehead{
\colhead{SST Tau name} & \colhead{common name} & 
\colhead{RA (J2000)} & \colhead{Dec (J2000)} & \colhead{our rank} &
\colhead{distance\tablenotemark{a} (arcmin)}  }
\startdata
\cutinhead{new members}
   041539.1+281858&                  \nodata&   63.913178&   28.316282&  A+&        12.66\\
   041542.7+290959&                  \nodata&   63.928283&   29.166586&  A+&         4.14\\
   041557.9+274617&                  \nodata&   63.991642&   27.771547&  A+&         7.54\\
   041810.7+251957&          [GBM90] L1506 1&   64.544954&   25.332630&  A+&        62.64\\
   042016.1+282132&                  \nodata&   65.067129&   28.359041&  A+&        10.67\\
   042021.4+281349&                  \nodata&   65.089345&   28.230326&  A+&        16.91\\
   042025.8+281923&                  \nodata&   65.107642&   28.323265&  A+&        13.65\\
   042026.0+280408&                  \nodata&   65.108609&   28.069153&  A+&        17.42\\
\enddata
\tablenotetext{a}{Angular distance to nearest previously-identified Taurus
member.}
\end{deluxetable}
\clearpage

\section{Discussion}
\label{sec:discussion}

\subsection{Overall Sample Properties}
%\label{sec:samplepropsintro}
\label{sec:newsamplesummary}

\subsubsection{The sample as a whole}

There are 215 previously identified Taurus members and 148 candidate
new Taurus members discussed here.  
Table~\ref{tab:sampleproperties2new} summarizes the samples selected
from each color-magnitude or color-color space, along with relevant
numbers from the SWIRE sample for comparison. 
Table~\ref{tab:sampleproperties1new} summarizes the fraction of the
previously identified and new candidate sources detected at each
Spitzer band.  

Table~\ref{tab:sampleproperties2new} captures the fraction of objects
found in each CMD or CCD that fall into the categories of previously
identified YSOs, new candidate YSOs, previously identified
non-members, and new non-members, and also the fraction of each of
these categories that is found in each of the CMDs or CCD.  For
example, in the case of the 24/70 CMD, 447 objects are selected by our
color cuts in this diagram.  Of these, 19\% are previously identified
YSOs, 7\% are new candidate YSOs, 10\% are previously identified
non-members, and 60\% are new non-members. And, out of the 215 stars
that compose our previously identified Taurus member sample, 41\% are
recovered in the 24/70 diagram, of the 148 objects in our candidate
YSO sample, 22\% are found here, of the 821 previously identified
non-members, 6\% are found in this diagram, and of the 489 new
non-members, 55\% are found here.  This table reveals that the \ks/24
diagram recovers the highest fraction of previously identified objects
and of new candidate objects. The 24/70 CMD finds the highest fraction
of new non-members (by fraction of objects found in this diagram as
well as by fraction of the entire new non-member list); this is not
particularly surprising, as the objects that are detected at 70 \mum\
tend to be either YSOs or extragalactic objects. This table also shows
that the $\sim$6$\arcdeg$ ELAIS N1 SWIRE sample is largely found
outside of our color selection criteria; few galaxies of the sort
found in this SWIRE sample are likely to be selected by our criteria.

Table~\ref{tab:sampleproperties1new} shows, of the entire sample of
previously-identified Taurus members, essentially all are seen in at
least one band of IRAC, most ($\sim$80\%) are detected in MIPS-24, and
just $\sim$45\% are seen at MIPS-70.  However, $\sim$10\% of these
famous, bright objects are saturated in at least one Spitzer band, and
many objects do not have infrared excesses (or have excesses too weak
to be measured at, e.g., 70 \mum).  Those that are missing flux
densities (e.g., without even limits) in Table~\ref{tab:knownyso} (or
Table~\ref{tab:posnewyso}) are missing because they are off the edge
of the covered area, or there is a cosmic ray cluster near the
location of the source, corrupting the photometry. 

As can be seen in Table~\ref{tab:sampleproperties2new}, out of the
215 previously identified Taurus members, 144 (67\%) are selected in
at least one of our color spaces as having a Spitzer infrared
excess.  This is roughly consistent with the 2/3rds disk fraction in
Taurus.  Just 65 (30\%) are selected in all of our color spaces
simultaneously. Out of our 148 new candidate Taurus members, 17
(16\%) are selected in all of our color spaces simultaneously (two of
which are likely galaxies, 043349.5+291528 and 044554.8+240843, based
on spectroscopy). Interestingly, there are a comparable number of
non-members (new or previously identified), 21 objects, that are
selected in all of our color spaces together.

For the new candidate members, we have an obvious bias in that we
cannot find stars without Spitzer IR excesses; note that the sample of
previously identified Taurus members includes Taurus members without
IR excesses.  We also have a bias in that three of these color
selections use 24 \mum\ in some fashion (either as overall brightness
or as part of the color).  This selection mechanism biases our sample
of YSO candidates towards a high fraction with 24 \mum\ detections;
see Table~\ref{tab:sampleproperties1new}. Just $\sim$9\% of these
objects are not seen at 24 \mum.  Moreover, because previous surveys
on the whole were using less sensitive instruments, the new potential
objects that we have discovered here are on average fainter in the
optical and NIR than the sample of previously identified Taurus
members (see Table~\ref{tab:sampleproperties1new}).

Many objects newly identified here have large IR excesses, but several
of them have small excesses at 8 \mum\ and either a low excess or no
detection at all at longer wavelengths. These objects are the
unselected objects (grey dots) in the \ks\ vs.\ \ks$-$[24] diagram
(Fig.~\ref{fig:kk24}) that have \ks$-$[24] near 0. These objects
are included in our list of new candidate Taurus members only if they
have more than 4$\sigma$ excess (see \S\ref{sec:gradations}); we have
dropped objects whose apparent 8 \mum\ excess is completely
inconsistent based on Planck function considerations with a
photospheric 24 \mum\ measurement (or limit). These 4$\sigma$ 8 \mum\
points seem to be real, in that the distribution of, e.g., [3.6]$-$[8]
colors are near zero for the overwhelming majority of stars in the
catalog, and these objects are clearly redder than average. However,
without detailed modeling beyond the scope of this paper, it is
puzzling how objects could have legitimate, real 4$\sigma$ excesses at
8 \mum\ and small 24 \mum\ excesses.  These objects have low grades of
confidence in Table~\ref{tab:newysot3}, and generally will
require additional observations to resolve.

\subsubsection{Additional information from the spectroscopy}

\begin{figure*}[tbp]
\epsscale{0.85}
\plotone{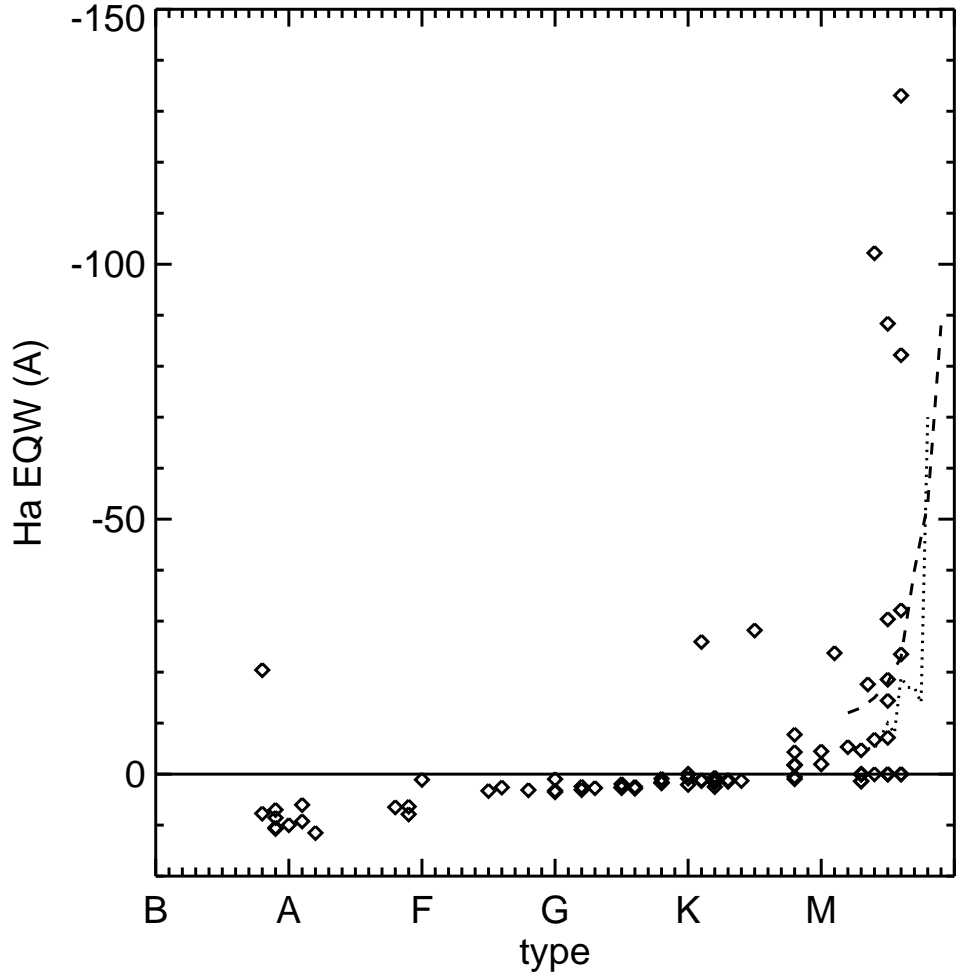}
\caption{H$\alpha$ equivalent widths (where values $<$0 indicate
emission) as a function of spectral type, for all objects with
spectral types reported here (including stars listed solely in the
Appendix and dropped as non-members).  Note that there is an
additional M0 star, not shown, with an H$\alpha$ equivalent
width of $-$498\AA. The dotted and dashed lines are,
respectively,  lines from Slesnick \etal\ (2008) and Barrado y
Navascu\'es \& Mart\'in (2003), and indicate the expected
quiescent H$\alpha$ emission from normal stellar activity. We
take stars with emission beyond these levels as clear YSOs.}
\label{fig:halpha}
\end{figure*}

\begin{figure*}[tbp]
\epsscale{0.85}
\plotone{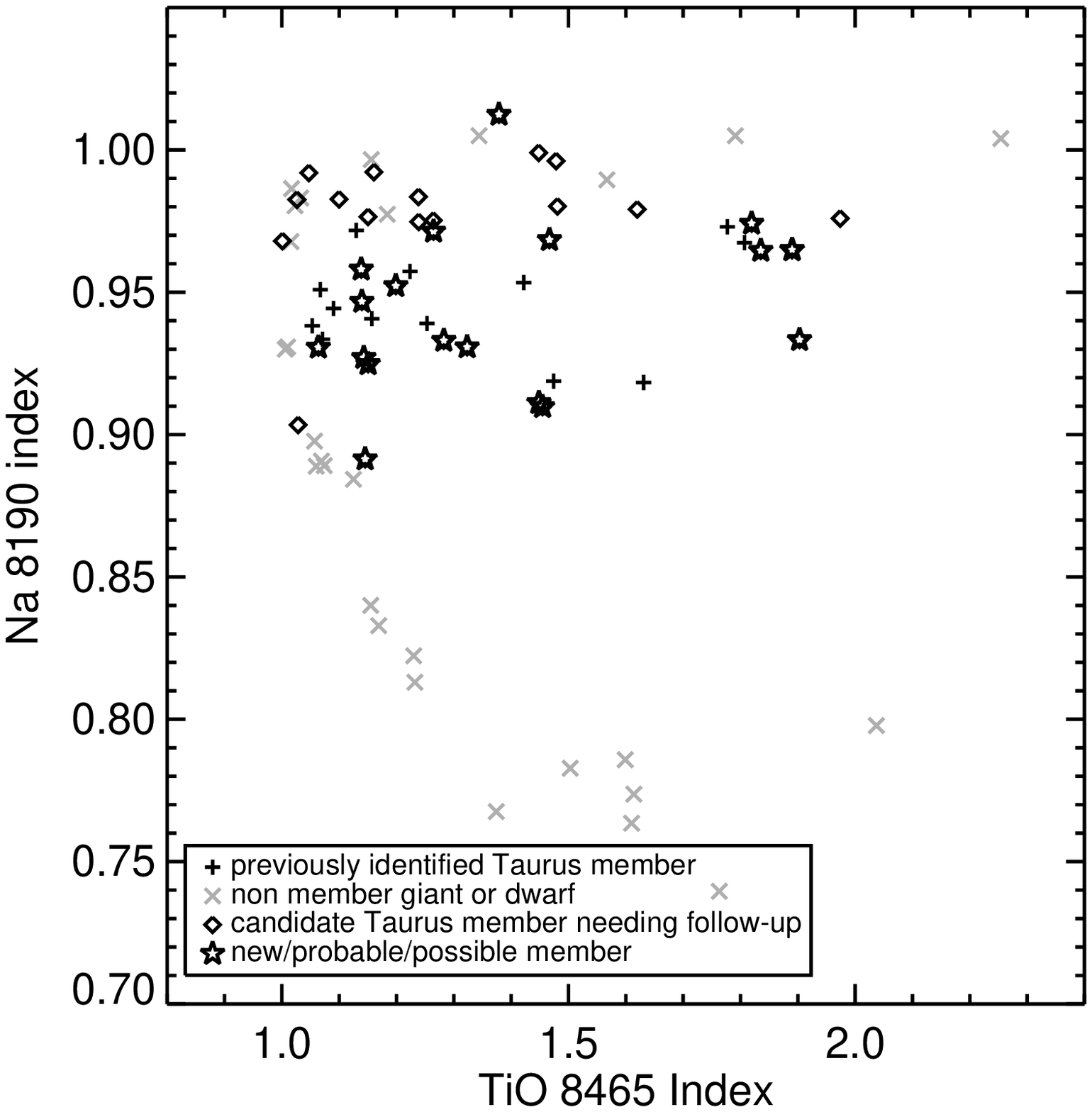}
\caption{Na 8190 and TiO 8465 indices for the M1 and later stars in
our optical spectroscopic sample.  Grey $\times$ symbols are objects
from Slesnick \etal\ (2008) that are known giants (large Na 8190
indices) or dwarfs (the remaining points, with smaller Na 8190
indices). The $+$ signs are previously-identified Taurus members.  The
star symbols are new members, probable new members, and possible new
members. The diamonds are remaining objects with spectra from the
148-object list of new candidate YSOs, e.g., those objects needing
more follow-up.   All but one of our new members have gravity
measurements consistent with YSO gravities, and even that one may be
YSO-like (see text).
}
\label{fig:gravities}
\end{figure*}

Of the 83 stars with optical spectroscopy, not all of them are
securely identified members of Taurus.   Additional information, such
as emission lines which are typical of 1-5 Myr T~Tauri stars, can help
inform our membership assessment. Twenty-six have H$\alpha$ in
emission at any level.  M stars that are not members of Taurus but
possess typical levels of stellar activity can also have H$\alpha$ in
emission.  Figure~\ref{fig:halpha} shows the H$\alpha$ equivalent
width as a function of spectral type for all stars reported in this
work (including those stars listed solely in the Appendix which we
dropped as non-members). The lines shown are from Slesnick \etal\
(2008) and Barrado y Navascu\'es \& Mart\'in (2003), dividing accretion
from ``normal'' activity levels of H$\alpha$ at the late M types. 
Stars for which we have detected a strong IR excess plus an H$\alpha$
equivalent width in emission larger than the cutoff as a function of
spectral type are objects we have placed in the secure ``new member''
bin -- see \S\ref{sec:finalverdicts} and the data tables -- except for
the Be star which cannot be a member of Taurus (see
\S\ref{sec:backgroundBs}).  We also list objects that have a more
moderate H$\alpha$ emission but still a very strong IR excess (not
Class III but Class I, II, or flat) as secure new members.  Objects
with moderate or no H$\alpha$ emission and small IR excess (Class III)
need additional indications of membership, such as lithium, radial
velocities, proper motions, or high-spectral-resolution observations
of gravity-sensitive lines.

%Note that the H$\alpha$ equivalent widths are not reported in Luhman
%\etal\ (2006) for the objects discussed there (and many of them appear
%in the figures to be reasonably low), so those objects cannot be
%similarly analyzed. 

One way to identify intermediate-gravity objects, e.g., neither giants
nor dwarfs but YSO-like gravities, is found in Slesnick \etal\ (2008)
and applies to stars of type M1 and later. It uses the TiO 8465 \AA\
index and Na 8190 \AA\ index; this analysis for our qualifying stars
is presented in Figure~\ref{fig:gravities}.  All of our new members
except one have gravities in the region of this diagram occupied by
previously identified Taurus members; all of the rest of our candidate
members shown here are also roughly consistent with YSO gravities,
given the scatter in this measurement.  These objects are indicated in
the notes column of Table~\ref{tab:newysot3}. The one new member that
may have more giant-like gravity is SST Tau 042920.8+274207, which may
appear in that location in the diagram due to reddening; reddening
will push points up and to the left in this figure.  This object is
particularly perplexing, in that it has many characteristics of youth
and also those of the  post-main sequence. Our calculation of \av\
(\S\ref{sec:finalverdicts}) suggests that \av$\sim$4. Additional
modeling (beyond the scope of this paper) is needed to further
investigate this object. We note here that several of the objects
discarded as giants (based on visual examination; see
\S\ref{sec:giants}) would appear in this diagram as having gravities
consistent with giant gravities, if they were included in this Figure.

\subsubsection{Summary of properties of all the new members}

Out of the 148 candidate new Taurus members, according to the letter
grades we assigned (\S\ref{sec:gradations}), there are 50 with letter
``A+'', ``A'', or ``A-'' (most believable), 66 with grade ``B+'',
``B'', or ``B-'', and 25 with grade ``C+'', ``C'', or ``C-'' (least
believable).  These grades incorporate all the available photometric
and spectroscopic information for each object; the remaining 7 do not
have grades because they are galaxies based on the follow-up
spectroscopy.   We have spectroscopy for 114 of the 148 objects, 90 of
which are optical spectra that have been analyzed.  We can report
stellar spectral types for 83 ($\sim$90\%) of those (again, the
remaining 7 are galaxies).   Although we have biased our spectroscopic
followup towards the brighter objects, such a low rate of finding
galaxies gives us confidence that our YSO selection process is
reasonably successful in that it finds many more stars than galaxies. 
Note that, while all the ``new members'' are also grade A objects, not
all grade A objects are ``new members'' because of the need for
additional data in many cases.  The 83 stellar objects include newly
confirmed YSOs and objects that will most likely turn out to be
background giants when more data are acquired.

There are 34 new members, 3 probable new members, 10 possible new
members, 7 extragalactic objects, 1 other object (Be star), 60 stars
needing additional follow-up observations, and 33 pending any
follow-up observations. All of the individual SEDs appear in Appendix
\ref{sec:seds}.   Combining the 34 new members of Taurus, 3 probable
new members, and 10 possible new members yields a total of 47 new
objects of various shades of confidence.  This represents an increase
of $\sim$20\% (by number) over the previously-identified members
covered by our map.    

Of the remaining 148 objects less the 47 new Taurus members (new,
probable new, and possible new), 60 more objects have optical spectra
resembling stars, but we need additional data to distinguish these
potentially interesting (often transition disk-type SED) objects from
background giants or foreground objects (using, e.g., lithium
abundances, radial velocities, proper motions, etc.; these are in the
``needs additional followup'' category).  If they turn out to be
non-members, the IR excess we observe needs to be explained. Finally,
33 await additional data beyond the Spitzer photometry to confirm or
refute their new Taurus membership status (``pending followup''). 
Thus, more new objects could still be in this data set. For
completeness, the remaining 148-47 objects are also listed in
Table~\ref{tab:newmembers}, sorted by their category,  such that the
objects we grade as most likely to be new members appear at the top of
the list.

\subsection{Detection at other bands}

Since we have a wealth of data at other wavelengths, it is possible
that we can find some additional evidence for youth for our potential
new Taurus members among them.  For example, young stars
are known to be bright in X-rays and ultraviolet.  We did in fact
include the information below in our ultimate ranking of the objects
(see \S\ref{sec:gradations}).

The XEST fields were optimized to cover the previously identified YSO
population, and do not cover our entire Spitzer field. Out of the
previously identified YSOs, 107 of 215 are detected in X-rays, and 50
are detected by the XMM-Newton OM (in ultraviolet). Out of the entire
148 star list of potential new Taurus members, 7 are detected in
X-rays (041940.4+270100=XEST-16-024, 042215.6+265706=XEST-11-078,
042936.0+243555=XEST-13-010,  043326.2+224529=XEST-17-036,
043456.9+225835=XEST-08-003, 043542.0+225222=XEST-08-033, and
044125.7+254349=XEST-07-032), and 2 are also detected by the
XMM-Newton OM (042215.6+265706=XEST-11-078=XEST-11-OM-122 and
042936.0+243555=XEST-13-010=XEST-13-OM-002).  These objects are noted
in Table~\ref{tab:newysot3}.  The two objects detected in both X-rays
and the XMM-Newton OM are already confirmed new members.  Of the
remaining objects detected in X-rays, there are three more confirmed
new members, one probable, one possible, and two pending additional
follow-up; five of them appear in Scelsi \etal\ (2007) as potential
members (see Table~\ref{tab:newysot3}).

The SDSS stripes do not cover our entire field either, but they
provide $u$-band observations for those regions they do cover;
however, extinction strongly affects the numbers of objects detected. 
Out of the 215 previously identified Taurus members, 109 are detected
at SDSS $u$; out of the 148 stars we list as potential new members, 63
are detected at SDSS $u$ (18 of which are confirmed new members).   
Out of those objects, 33 have apparent UV excesses above the locus
formed by {\em all} of the objects in our catalog with Sloan $uriz$
photometry (17 of these 33 are confirmed, probable, or possible new
members). These objects are noted in Table~\ref{tab:newysot3}.

\subsection{Locations of New Candidate Taurus Members}

\begin{figure*}[tbp]
\epsscale{1.0}
\plotone{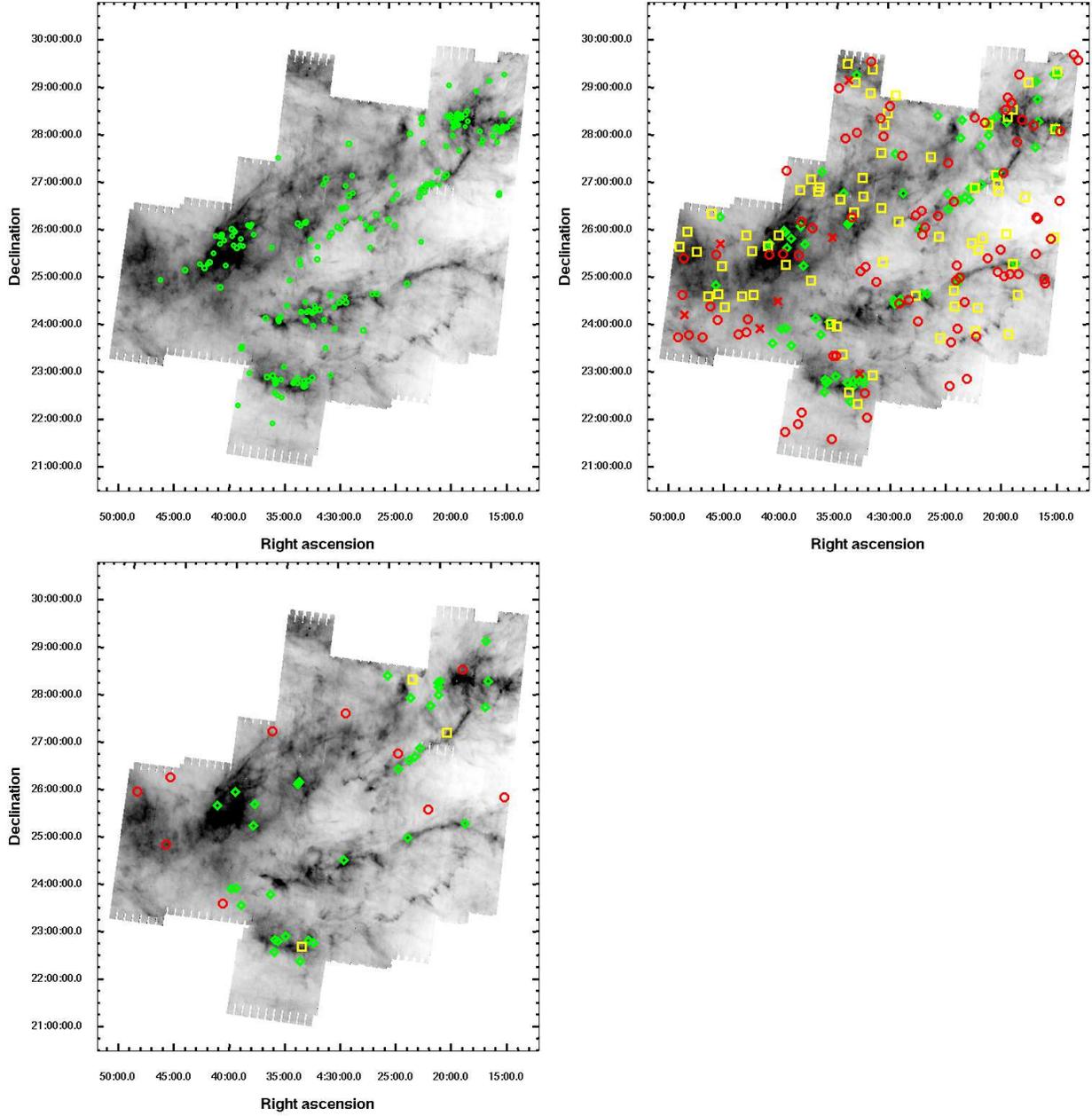}
\caption{Physical location of various samples of Taurus objects
on  the 160 \mum\ image (reverse greyscale): 
{\em (top left)} previously identified Taurus members (green
circles), which tend to follow 160 \mum\ emission;  
{\em (top right)} all new candidate objects: green diamonds = grade A,
yellow boxes = grade B, red circles = grade C, $\times$ =
extragalactic;
{\em (bottom left)} the 47 new Taurus objects: green diamonds =
new members, yellow boxes = probable new members, red circles =
possible new members.
The confirmed new Taurus members thus far 
tend to be found near the previously identified Taurus objects. }
\label{fig:location}
\end{figure*}

As can be seen in Table~\ref{tab:newmembers}, many of the new objects
are quite close to previously-identified Taurus objects.  
Figure~\ref{fig:location} shows the projected location of the sample
of  previously identified YSOs, the new YSO candidates we have
selected, and those new YSO candidates with the results of the
spectroscopy folded in. Because one of the goals of this project is to
look for widely distributed Taurus members, we did not restrict our
search for new objects to the regions already occupied by Taurus
members. While proximity to previously identified YSOs was a component
in our ranking scheme,  it was only one of many criteria
(\S\ref{sec:gradations}).  As can be seen in
Figure~\ref{fig:location}, the new candidate objects are generally
more isotropically distributed than the previously identified members.
However, most of the  new Taurus members tend to be found near the
previously identified Taurus members. There is a new loose grouping of
YSOs found  near ($\alpha,\delta$)=(70,24).  Further spectroscopic
results are needed to complete our search for an extended population
of YSOs in Taurus.

\subsection{Spectral types and YSO classes}

\begin{figure*}[tbp]
\epsscale{0.8}
\plottwo{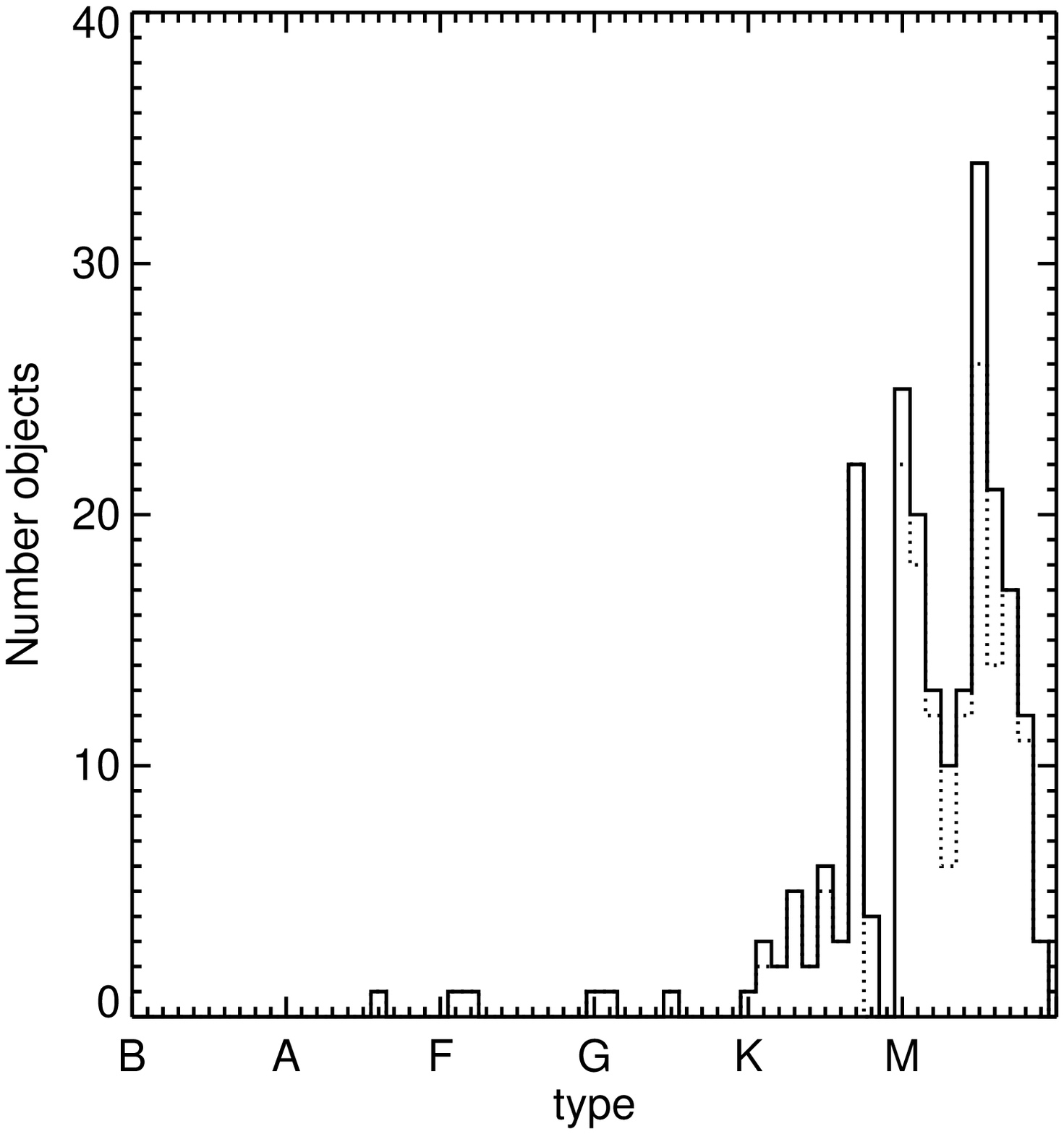}{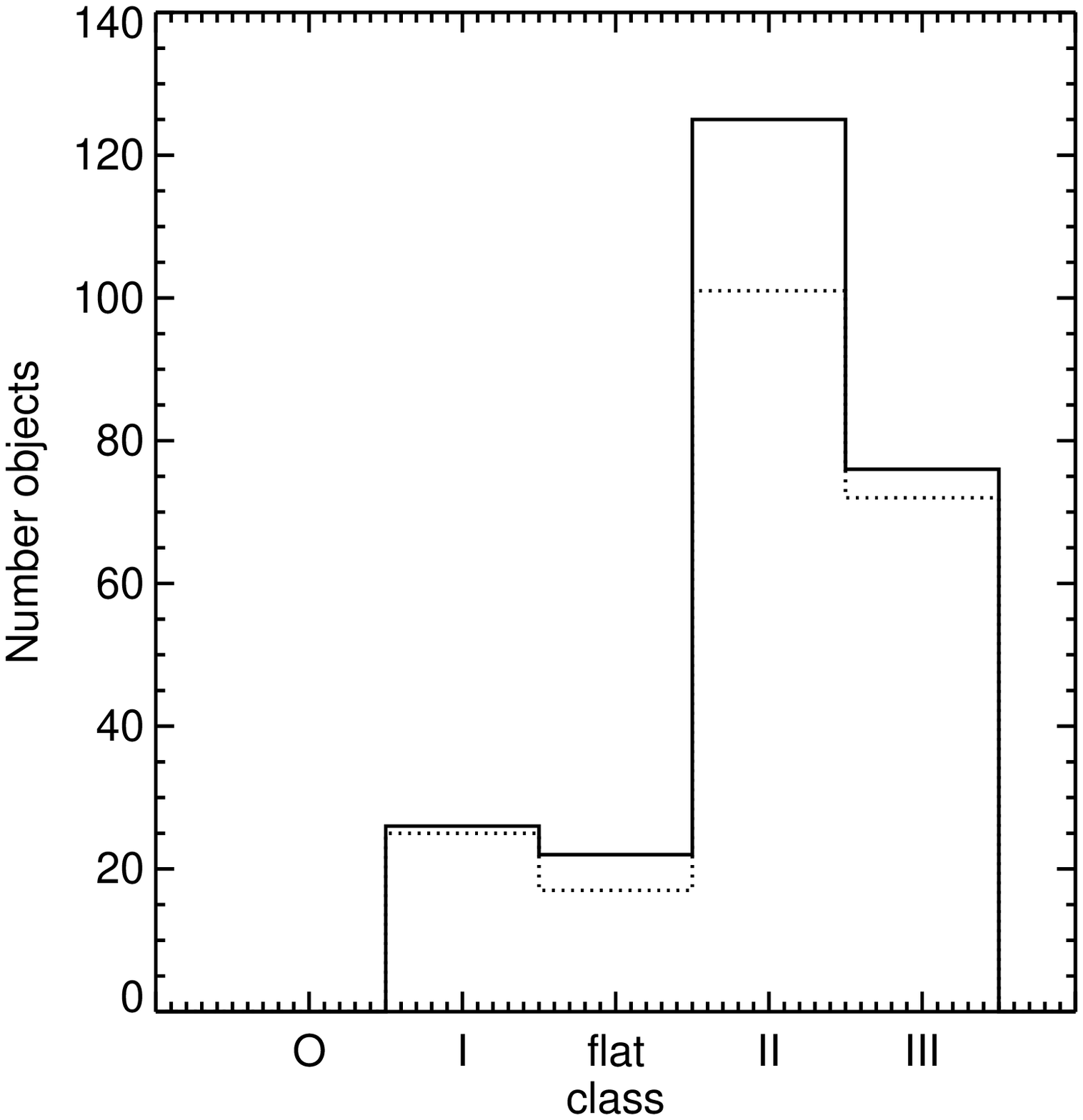}
\caption{LEFT: Histogram of types for all previously-identified
members with spectral types plus new confident members (solid line)
and just the previously-identified members(dotted line). RIGHT:
Histogram of all the previously-identified members plus new confident
members (solid line) and just the previously-identified members(dotted
line)}
\label{fig:imf}
\end{figure*}

Figure~\ref{fig:imf} shows a histogram of the spectral types of the
previously-identified Taurus members plus the 34 confirmed new
objects; most of the new objects are M spectral types -- one is
``early K'', one is a K5e, four are K7-M0, and the rest are M stars. 
Of the  probable new members, there are 1 M star, one A star, and 1
``$<$M0''; of the possible new members, there are 2 M stars, and
the rest are K4 or earlier (1 late F, 2 G stars, 5 K stars). These
earlier types are less secure new members because they have small IR
excesses and/or weak H$\alpha$. 

Most of the 47 new objects (32 of them) are YSO class II.
Figure~\ref{fig:imf} shows a histogram of the types of the
previously-identified objects plus the 32 confirmed new objects. 
There are five ``flat'' YSOs in the 32 new objects, one class I, and 4
class IIIs; the rest are class II. 

\subsection{Ensemble SED Properties of Currently Identified Taurus Members}
\label{sec:netsed}

\begin{figure*}[tbp]
\epsscale{1}
\plotone{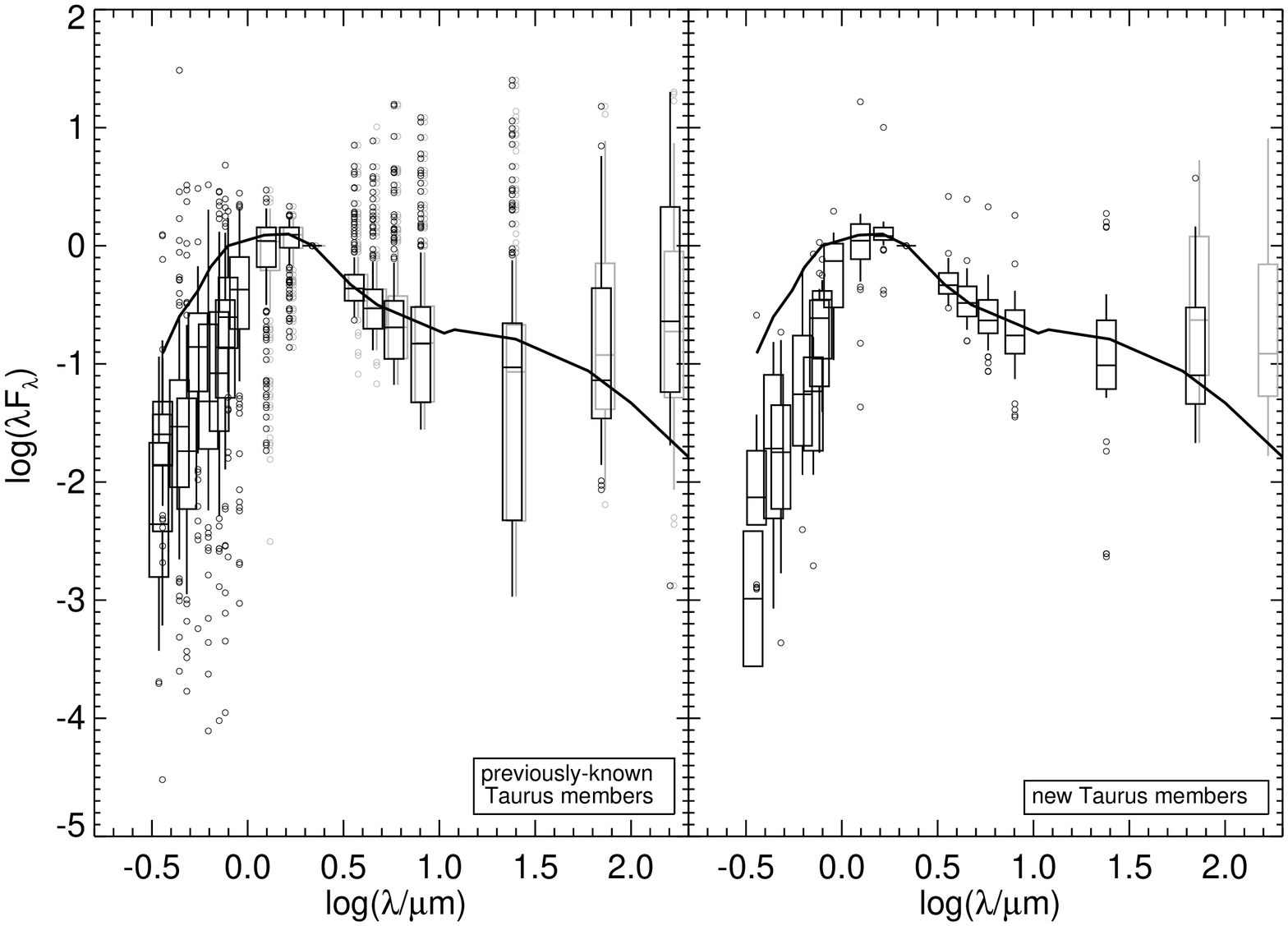}
\caption{
Box plot (see text) of all SEDs, normalized to \ks, for (LEFT) the 211
(out of 215 possible) previously identified Taurus members having a
\ks\ measurement, and (RIGHT) the 34 (confirmed) new members
discovered using the Spitzer data.  The central line in each box
denotes the median; the ends of each box are the first and  third
quartile of the distribution; the lines extend to the most  extreme
values that are not more than 1.5 times the interquartile range; and
the circles are those points outside 1.5 times the interquartile
range. The grey, slightly-offset boxes are the same representation,
but treating all of the upper and lower limits as real detections at
that limit point; see text for more discussion. (Note that none of the
new members have a detection at 160 \mum, so that wavelength has only
a grey box.)  The solid line near the medians is the ``median Taurus
SED'' from D'Alessio \etal\ (1999). 
}
\label{fig:netsed}
\end{figure*}

Having established the location and spectral type distribution of the
new members relative to the established members, we now investigate
the SED properties of the ensemble. Figure~\ref{fig:netsed} shows
representations of the SEDs for all of the 215 previously identified
Taurus members with \ks\ detections (on the left), and the 34
confirmed members discovered using the Spitzer data (on the right). 
Note that (a) these SEDs are normalized to \ks\ (objects missing \ks\
due to saturation are not included, so a total of 211
previously-identified Taurus members are shown), and (b) the SEDs
appear individually in the Appendix below, see \S\ref{sec:seds}.  We
have normalized to \ks\ because we wished to investigate the variation
of the shapes of the SEDs, not their relative intrinsic brightness
and/or extinction; \ks\ provides a value available for most objects
(see Table~\ref{tab:sampleproperties1new}) and is an admittedly
imperfect compromise between disk and photosphere emission. For each
wavelength, there are very broad, non-Gaussian distributions of
points.  In the optical in particular, but also at wavelengths at
least as long as 24 \mum\ (see, e.g., Bary \etal\ 2007), the intrinsic
variability of the objects can contribute significantly to the
scatter.  

The Figure uses box plots; these box plots have been used in other
papers (e.g., Rebull \etal\ 2006; Flaccomio \etal\ 2003) as a
mechanism for interpreting scatter plots.  For each of the wavelength
points, the boxes capture the median and the first and third quartiles
of the distribution in $\lambda F_{\lambda}$ (including measurements
only,  not limits).  The lines extend to the most extreme values that
are not more than 1.5 times the interquartile range, and the circles
are those points outside 1.5 times the interquartile range.  For cases
like those found in, e.g., Rebull \etal\ (2006), there are upper
limits in the distribution, and the Kaplan-Meier (K-M) estimator for
censored data can be used to take into account the upper limits
present in the data. However, in this present case, for each of the
Spitzer points, we have both upper and lower limits in the
distribution, so the K-M estimator fails.  The offset grey boxes in
Figure~\ref{fig:netsed} use all of the upper and lower limits as real
detections at the location of the limit.  The influence of the large
number of limits can particularly be seen at MIPS bands, where the
lower edges of the box are substantially lower with the limits
included as real detections than without.  There are no detections at
160 \mum\ among the 34 new member stars, so only limits can be used.

The solid line near the medians is the ``median Taurus SED'' from
D'Alessio \etal\ (1999).  It is clear that we have not compensated for
reddening in the optical bands, as the D'Alessio SED is significantly
above our medians at blue wavelengths (shorter than 1 \mum).
(D'Alessio \etal\ individually dereddened the SEDs before combining
them to get the median; we would need to apply a reddening of about
$A_{\rm J}$=0.8 to the D'Alessio median SED to make it match our
median SED.)  The D'Alessio SED tracks the rest of our medians
reasonably closely, although it must be noted that our medians include
all of the Taurus members, not just the K5-M2s, and not just the stars
with infrared excesses.  Our medians are slightly below the D'Alessio
SED at 4.5-8 $\mu$m, more below the D'Alessio SED at 24 \mum, then on
the D'Alessio SED at 70 $\mu$m, then above it at 160 \mum; in all
cases, the medians match within the box, e.g., between the first and
third quartiles.  Since there are more lower than upper limits at 24
$\mu$m, the ``true'' median is likely to be lower still, as can be
seen by the location of the grey median line.  Similarly, the ``true''
median at 70 and 160 \mum\ is likely to be lower than our calculated
value. At 160 \mum\ in particular, we are likely seeing the effects of
sensitivity; even our upper limits are reasonably shallow, and the
location of the grey median line is closer to the D'Alessio SED.  The
medians at each band of the 34 new members are not much different than
the medians of the previously-identified members, although they are
slightly brighter.  We are unable to find stars without IR excesses
using Spitzer selection criteria, so we expect our new objects to have
larger excesses on average than the entire ensemble of Taurus objects.
At 70 \mum, the new objects are as a whole much brighter than the
previously identified sample. At 24 \mum, the median is reasonably
comparable but the lower boundary to the box (e.g., location of first
quartile) is much brighter for the new objects than for the previously
identified sample. Both of these effects are undoubtedly a result of
our slight MIPS bias in source selection (see
\S\ref{sec:newsamplesummary}). 

\subsection{Ensemble $L_{\rm IR}/L_{\rm total}$ Properties of Currently Identified
Taurus Members}
\label{sec:ldlstar}

\begin{figure*}[tbp]
\epsscale{1}
\plottwo{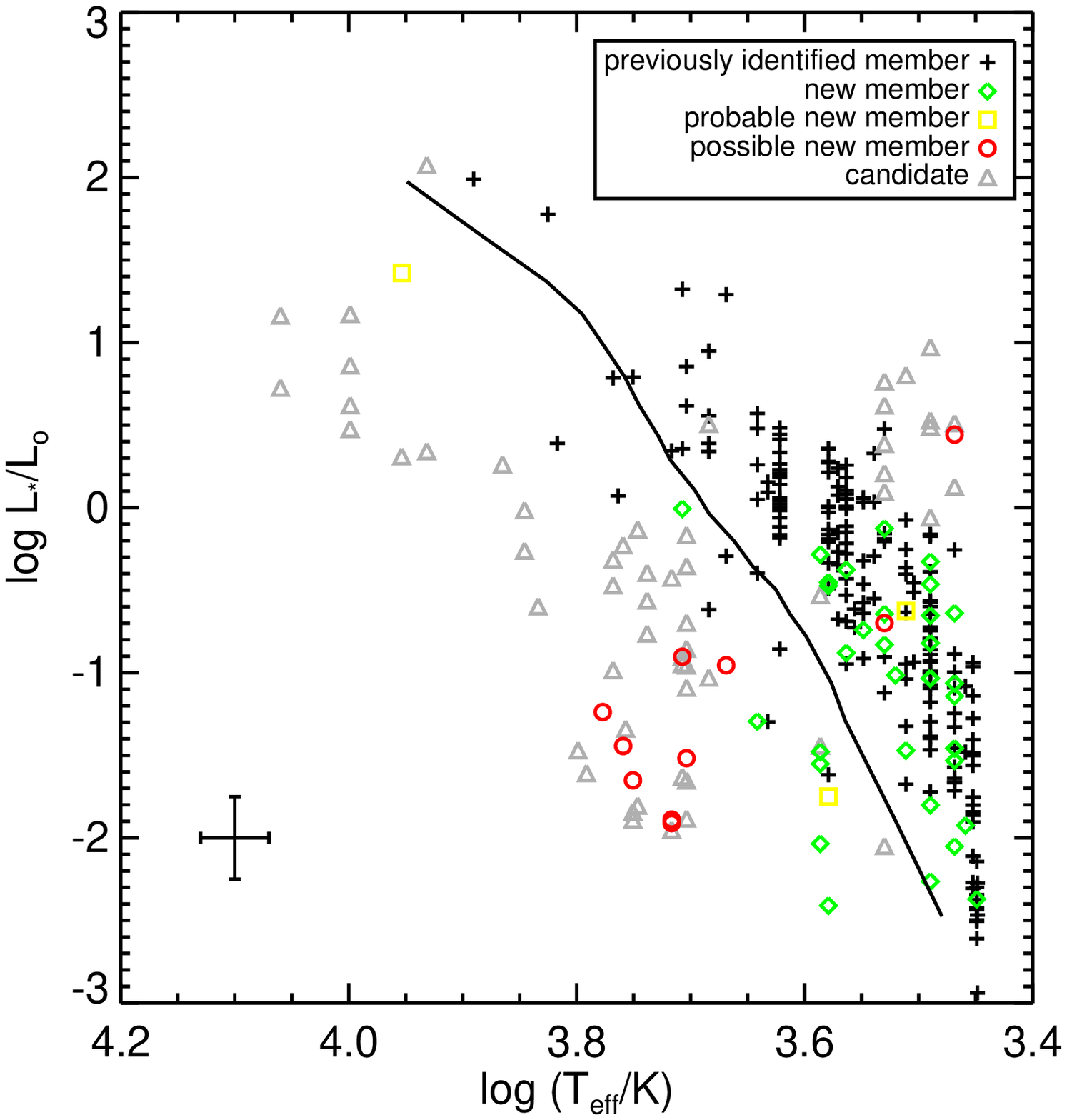}{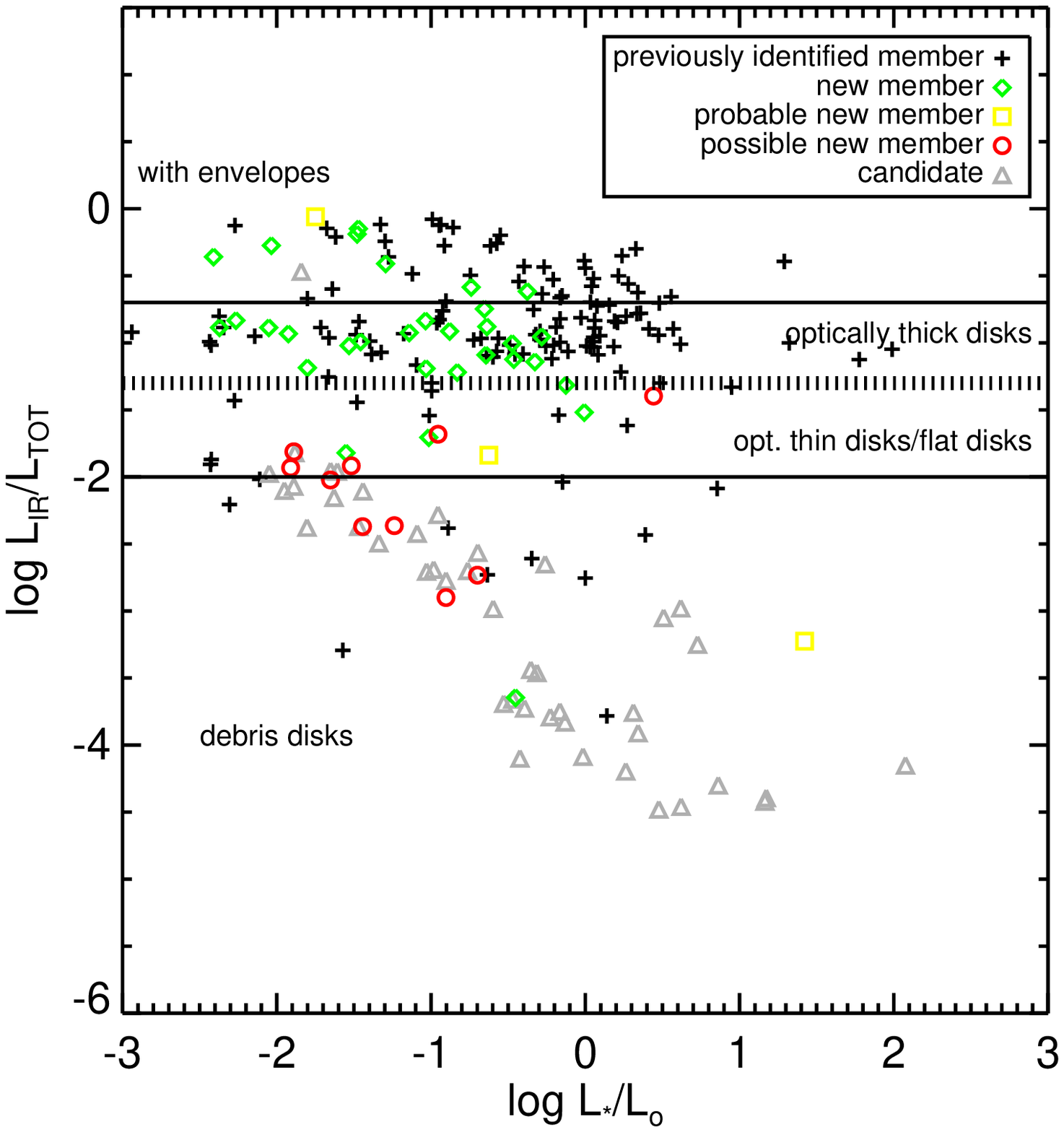}
\caption{LEFT: Hertzprung-Russell (HR) diagram for the
previously-identified members ($+$), the  new members (green
diamonds),  the probable new members (yellow squares), the possible
new members (red circles) and the rest of the candidate members (grey
triangles), using the $L_*$ values calculated (and \teff\ values
assumed) in the process of calculating $L_{\rm IR}/L_{\rm total}$; see
text for caveats.   Typical error bars are indicated in the lower
left; the error in temperature will be larger for earlier types.  The
ZAMS from Siess \etal\ (2000) for the Taurus distance is overplotted
as a solid line for reference. Most of the highest confidence new
members have positions in  this diagram quite consistent with the
positions of previously-identified Taurus members.  All three of the 
probable new members and just two of the possible new members also
have consistent positions.  The remaining possible new members are
apparently under-luminous; see text. Many of the candidates awaiting
further data also have positions consistent with Taurus members, but
many are also apparently sub-luminous.
RIGHT: Plot of $L_{\rm IR}/L_{\rm total}$ against $L_*/$\lsun, with the same
notation as the left figure. Approximate dividing lines for the 
interpretation of $L_{\rm IR}/L_{\rm total}$ are indicated. The
dividing line between optically thick disks and disks that are either
optically thin or flat is hatched to indicate that it is a ``fuzzy''
dividing line.
}
\label{fig:hrd}
\end{figure*}

We calculated $L_{\rm IR}/L_{\rm total}$ for the previously identified
and new candidate Taurus members; see \S\ref{sec:finalverdicts} and
Tables~\ref{tab:knownysot3} and \ref{tab:newysot3}. Since $L_*$ is a
by-product of this calculation, we present a Hertzprung-Russell (HR)
diagram in Figure~\ref{fig:hrd}. The \teff\ values appear quantized
because the \teff\ value for each star was assumed based on its
spectral type. The new members, probable new members, and possible new
members are indicated separately on this diagram, along with the
previously identified and pending samples. Most of the highest
confidence new members have positions in  this diagram quite
consistent with the positions of previously-identified Taurus
members.  All three of the probable new members and just two of the
possible new members also have consistent positions.  Both the
previously identified and new groups have some apparently sub-luminous
members, with additional such objects among the sample awaiting more
data.  These objects could be sub-luminous due to being, e.g., edge-on
disks, or they might not be members.   Further investigation is
warranted.  Many of the candidates awaiting  additional data and/or
analysis fall in the pre-main-sequence regime of this  diagram, but
many do not; these are likely to turn out to be background dusty
giants.  There seems to be a loose clump of new objects near log
\teff\ $\sim$ 3.75, log $L_*$/\lsun\ $\sim-1.8$, and many of the
earlier-type objects still awaiting additional data are also
considerably too faint. These objects are probably unlikely to be
Taurus members -- they could be background debris disk candidates --
and are indicated individually in the data tables. We included the
position of the object in the HR diagram in our individual assessment
and grading of each object (\S\ref{sec:gradations}). 

Figure~\ref{fig:hrd} also shows the $L_{\rm IR}/L_{\rm total}$ vs. log
$L_*$/\lsun\ for the sample.  Due to Spitzer's superior sensitivity,
resolution, and wavelength coverage vs.\ IRAS, the $L_{\rm IR}/L_{\rm
total}$ values here are likely to be more reliable than those values
presented in, e.g., Cohen, Emerson, \& Beichman (1989).  Systems with
$L_{\rm IR}/L_{\rm total} > 0.2$ (e.g., GV Tau AB) are expected to be
disk systems with circumstellar envelopes, and indeed most of the
systems in this range show rising or flat spectral energy
distributions.  Systems with $L_{\rm IR}/L_{\rm total}$ of 0.05--0.2
(e.g., DL Tau, CY Tau) are expected to be optically thick disks
lacking envelopes, with large values indicating more flared disks. 
Most of the known T~Tauri stars in the cloud fall in this range. 
Systems with $L_{\rm IR}/L_{\rm total}$ near 0.01 (e.g., V410 Anon 24)
are either optically thick disks highly flattened by dust settling, or
systems just becoming optically thin as they transition to debris
disks.  Finally, systems with $L_{\rm IR}/L_{\rm total} < 0.01$ (e.g.,
V819 Tau) are similar to classical debris disks. Approximately 15 of
these are found in the survey, among the previously-identified members
combined with the new members.  Most of the previously known and new
objects have substantial disks or envelopes.  Most of the least
confident members (and most of those awaiting more data) appear to
have very tenuous disks.  Certainly some legitimate members appear at
low $L_{\rm IR}/L_{\rm total}$  as well, so some of these candidates
may be legitimate.

\subsection{Comparison to Other Selection Methods}
\label{sec:othermeth}

Thus far, our spectroscopic follow-up suggests very few galaxies are
among our YSO candidates. While we have not obtained spectra of every
candidate, and our spectroscopic follow-up has generally been of the
brighter candidates, it is still important to note that our screening
does seem to successfully weed out galaxies.  

As mentioned above, there are many discussions in the literature also
seeking to identify YSOs using Spitzer color selections (e.g., Allen
\etal\ 2004, Padgett \etal\ 2008b, Rebull \etal\ 2007, Harvey \etal\
2007, Gutermuth \etal\ 2008). In this section, we compare our color
selection to two other popular color selection mechanisms. Harvey
\etal\ (2007) describes the c2d/Gould's Belt method, which works with
large-scale maps of star-forming regions similar to that for Taurus.
With the mitigation of extragalactic contamination as a primary goal,
Harvey \etal\ (2007) apply several color cuts and use shape/fitting
information from the c2d pipeline, combined with manual examination of
the Spitzer images. The Gutermuth \etal\ (2008) criteria are being
used by several different groups, including the IRAC GTO team, and is
designed to find YSOs despite a wide variation of extinction values
and nebulosity found in star-forming clusters within 1 kpc.  We note
that this method has been recently updated in Gutermuth \etal\ (2009),
though the changes to the method are considered relatively small by
the authors, and are not implemented here.

In this section, we have the goal of determining whether we can
directly compare, say, the ``yield'' of this survey with those from
other star-forming regions studied by other groups. We conclude that
it is not at all straightforward, and direct comparison may not
actually be possible because the selection methods are so different. 
Tables~\ref{tab:othermethods} and \ref{tab:othermethods2} make an 
attempt to compare the various methods as we have implemented them, 
with all the caveats discussed below, but just as it is not easy to
compare the ``yield,''  it is not necessarily easy even to compare the
methods on a precisely even footing.  Note that in order to understand
the assumptions that have gone into the numbers in these tables (or
into the classes reported in Tables~\ref{tab:knownysot3} and
\ref{tab:newysot3}), reading the text below is critical.

One of the most significant differences between our {\em data} and
those used to develop both of the other two methods is the depth of
the IRAC Spitzer data that is used as input.  With only two 12-sec
IRAC frames per position, our data is at least a factor of two shorter
integration time (at least square root of 2 less sensitive) than the
other surveys that were used to construct these other selection
methods. On the other hand, Taurus is also much closer than {\em most}
of the other associations considered by the other projects. 
Sensitivity to legitimate YSO cluster members is at least comparable,
but more faint contaminating sources will affect deeper surveys.

An important difference between our {\em selection mechanism} and
both of the other two methods is the use not just of different color
spaces but also of multiple color spaces in serial (the star is a YSO
in this one AND that one AND this other one) vs.\ in parallel (the
star is a YSO in this one OR that one OR this other one).
Table~\ref{tab:othermethods2} explicitly lists the color parameter
spaces used by each method, but the combination of them is complex and
the original paper(s) for each should be consulted. 

To reiterate, our method uses a combination of color cuts in a series
of color spaces (which can be easily applied to other Spitzer data
sets), followed by manual inspection of a variety of properties,
including notably high-spatial-resolution optical imaging (which may
not be easily applied to other Spitzer data sets). We used color
spaces in parallel (this OR that) and manually examined those sources
that met the YSO color criteria in {\em any} space; there are many
thousands of objects that were never inspected manually but are still
very likely to be stars and similarly many thousands more that are
likely to be galaxies based on their colors.  
Table~\ref{tab:othermethods} indicates that we started with 122616
objects, of which 98476 had stellar-like colors in at least one of the
spaces we investigated. The overwhelming majority of those, 99\%, have
colors consistent with stars in {\em all} the spaces we investigated
and were never subject to the scrutiny of the YSO candidates; 513 of
these were also selected as possible YSOs in at least one of the color
spaces we investigated, and most have been discarded. Similarly, there
are 80995 objects that are classified as galaxies in any of the spaces
we investigated, and 99\% of those look like galaxies in ALL the
spaces we investigated; there are 681 that were also selected as
possible YSOs in at least one of the color spaces we investigated, and
most have been discarded. Note too that the numbers of stars/galaxies
reported later in this table for the samples of previously-identified
YSOs and new YSO candidates are again the sample of objects selected
in {\em any} diagram; discussion of the sample selected in {\em all}
diagrams appeared above.

We now compare the other two specific methods with our method in more detail.

\begin{deluxetable}{lrrr}
\tablecaption{Comparison of Selection Methods I.\label{tab:othermethods}}
\tabletypesize{\tiny}
\tablewidth{0pt}
\tablehead{
\colhead{property} & \colhead{This Work} & \colhead{Harvey \etal\
(2007) criteria} &
\colhead{Gutermuth \etal\ (2008) criteria} }
\startdata
\cutinhead{Of entire catalog...}
Part that can be considered & 122616 (18\%)                  & 7119 (1\%) & $\gtrsim$89003 (13\%)\tablenotemark{b}\\
Number of stars             &  98476\tablenotemark{a}        & 965      & 74254 \\
Number of galaxies or other contaminants&  80995\tablenotemark{a} & 4644      & 14176 \\
Number of YSOs              &    870\tablenotemark{c} & 1510       &  573 \\
\cutinhead{Of previously-identified members...}
Part that can be considered &    206 (96\%)         & 154 (72\%) & $\gtrsim$187 (87\%)\\
Number of stars             &   172\tablenotemark{a}& 6          & 82 \\
Number of galaxies or other contaminants&    88\tablenotemark{a}& 7          & 2 \\
Number of YSOs              &   144                 & 141        & 103 \\
\cutinhead{Of sample newly-identified here...}
Part that can be considered &  148 (100\%)        & 128 (86\%) & $\gtrsim$141 (95\%) \\
Number of stars             &  87\tablenotemark{a}& 0          & 76 \\
Number of galaxies or other contaminants&  116\tablenotemark{a}& 28        & 5 \\
Number of YSOs              &  148                 & 100        & 60 \\
\enddata
\tablenotetext{a}{This number of objects meets the star or galaxy
criteria in {\em any} of the color-color or color-magnitude spaces
used. Thus, total number of objects is not equal to number
stars+number galaxies+number YSOs.  See text.} 
\tablenotetext{b}{This method starts with a sample detected at all 4
IRAC bands (89003 objects) and semi-manually adds in additional
objects (see text).  The largest sample of those potential additional
objects are those seen at $JH$\ks[3.6][4.5],  which in this Taurus
dataset is 76522 more objects, so $\gtrsim$165525=89003+76522 might be
reported here instead to indicate this semi-manual addition of
objects. However, there is no easy method to take those additional
objects and break them into stars and galaxies, so the rest of this
section of the table is taken out of 89003 objects. }
\tablenotetext{c}{This number objects meets the YSO criteria in {\em
any} of the color spaces investigated.  After the imaging, etc. tests,
this 870 number is reduced to 148 new candidate YSOs + 144 previously
identified YSOs = 292.}
\end{deluxetable}

\begin{deluxetable}{lccc}
\tablecaption{Comparison of Selection Methods II.\label{tab:othermethods2}}
\tablewidth{0pt}
\tablehead{
\colhead{parameter space} & \colhead{Used by us} & \colhead{Used by
Harvey \etal\ (2007) } & \colhead{Used by Gutermuth \etal\ (2008) } }
\startdata
$[24]/[24]-[70]$  & yes & \nodata & \nodata \\
\ks/\ks $-$[24] & yes & \nodata  & \nodata  \\ 
$[8]/[8]-[24]$ & yes & \nodata  & \nodata  \\ 
$[4.5]/[4.5]-[8]$ & yes & yes & yes \\ 
$[3.6]-[4.5]/[5.8]-[8]$  & yes &  \nodata & yes \\ 
$[24]/[8]-[24]$ &  \nodata & yes & \nodata \\ 
$[24]/[4.5]-[8]$ &  \nodata & yes & \nodata \\ 
$H-$\ks/\ks$-$[4.5] &  \nodata & yes & \nodata \\ 
$[4.5]-[5.8]/[5.8]-[8]$ &  \nodata  & \nodata  & yes \\ 
$[3.6]-[5.8]/[4.5]-[8]$     &  \nodata  & \nodata  & yes \\ 
$[3.6]-[4.5]/[4.5]-[5.8]$   &  \nodata  & \nodata  & yes \\ 
\ks$-[3.6]/[3.6]-[4.5]$     &  \nodata  &  \nodata & yes \\ 
$J-H/H-$\ks         &  \nodata  &  \nodata & yes \\ 
%$H-$\ks/$[3.6]-[4.5]$ & \nodata & \nodata & for measuring $E(H-K_s)$ when no $J$\\
$[4.5]-[5.8]/[5.8]-[24]$    & \nodata   & \nodata  & yes \\ 
\enddata
\end{deluxetable}

%\clearpage

\subsubsection{Harvey \etal\ (2007; c2d) Criteria}

Harvey \etal\ (2007) describe the criteria used by the c2d team and
subsequently the Gould's Belt team.  Especially since these results
are incorporated into Evans \etal\ (2009), which uses statistics of
the Spitzer-selected sample to determine relative lifetimes of the
Class 0/I/flat/II stages, it would be nice to understand how our
sample selection compares so that our data on Taurus can be compared
to the regions studied by these other programs.

To reiterate, we used the following parameter spaces to find candidate
YSOs: (1) [24] vs.\ [24]$-$[70], {\bf or} (2) \ks\ vs.\ \ks$-$[24],
{\bf or} (3) [8] vs.\ [8]$-$[24], {\bf or} (4) [4.5] vs.\ [4.5]$-$[8],
{\bf or} (5) [3.6]$-$[4.5] vs.\ [5.8]$-$[8] with an additional [3.6]
brightness cutoff, {\bf combined with} (6) optical (SDSS/CFHT) imaging
plus the additional qualitative criteria mentioned in
Section~\ref{sec:gradations} above.  The c2d analysis described in
Harvey \etal\ (2007) used the following parameter spaces to find
candidates, assigning a quantitative probability that the object is a
galaxy or YSO candidate: (1) [4.5] vs.\ [4.5]$-$[8], {\bf and} (2)
[24] vs.\ [8]$-$[24], {\bf and} (3) [24] vs.\ [4.5]$-$[8], {\bf and}
(4) (if there was a 2MASS match) $H-K_{\rm s}$ vs.\ $K_{\rm
s}$$-$[4.5],  {\bf and} (5) fitting (and removal) of stars/reddened
stars by the c2d pipeline, {\bf and} (6) shape information gleaned
from the c2d pipeline, {\bf and} (7) manual inspection of the
2MASS+IRAC+MIPS images, {\bf and} (8) manual addition of
previously-identified YSOs to the list of YSOs.  Differences between
the studies include the following: (1) only one CMD is the same
between the two teams; (2) the c2d team is using the intersection of
all their CMDs, and we are using selection in any one color space; (3)
the c2d team is using information obtainable only from their pipeline
(shape, star fitting); (4) the c2d team does not have optical imaging
(though we both examine the Spitzer imaging); (5) the c2d criteria
{\em require} detection in all four IRAC bands, and MIPS-24 (they use
the 2MASS information if it exists); (6) the c2d pipeline performs
PSF-fitting photometry in IRAC and MIPS, and our pipeline does
aperture photometry on IRAC data and PRF-fitting photometry on MIPS
data.  In order to attempt a comparison using the best possible
criteria, we can of course impose the same color cuts. We can't obtain
the shape information because we reduced our photometry differently. 
We can approximate the star fitting by dropping the objects with very
small colors in all bands.  

Out of the entire $\sim$700,000 object catalog, these criteria can
only be applied to 1\% of the objects (see
Table~\ref{tab:othermethods}) because of the multiband detection
requirement. Out of this 1\% of the catalog, which is $\sim$7100
objects, 21\% are identified as YSO candidates (note that we did not
manually examine the images of each of these objects in Spitzer or any
other bands, whereas the c2d team would have done so), 65\% are
identified as galaxies, and 14\% are identified as stars.  Note that
this does {\em not} mean that a c2d-selected YSO candidate sample has
65\% contamination, but rather that out of the objects in the catalog
to which the c2d criteria can be applied, 65\% of these objects are
immediately categorized as galaxies. Among our set of
previously-identified Taurus members, 154 can be classified, but 61 of
them (29\%) cannot be classified in this scheme because of missing
bands.  Seven of them (3\%) are identified as galaxies or other
contaminants.  Six of them (3\%) are classified as stars (the likely
WTTS out of the sample), and 141 (92\%) are identified as YSOs.  As
intended, this selection mechanism (even as we have implemented it) is
strongly biased towards YSOs.  This sample of previously known YSOs
cannot tell us about the contamination rate, but it can tell us about
the fraction of objects that might be missed; a c2d-based YSO
selection operating on this sample would miss $\sim$5\% of the YSOs
with detections in all the requisite bands and infrared excesses.  The
c2d classifications for these previously identified members are listed
in Table~\ref{tab:knownysot3}; the same information for the new
candidate members is in Table~\ref{tab:newysot3}. Among our new
candidate objects, perhaps unsurprisingly, there is a higher fraction
of contaminants.  The multi-band detection requirement means that 14\%
(20 objects) cannot be classified.  Among the remainder of the sample,
there are 28 (19\%) likely galaxies (or other contaminants), no stars
(which makes sense because we are only selecting objects with IR
excesses), and 100 (68\%) likely YSOs in this attempt to use the c2d
criteria on our sample.  Happily, the YSO candidates compose the
largest fraction of our sample. 

The c2d criteria require selection in each of the Spitzer-based CMDs
or CCDs; our selection requires selection in just one of the CMDs or
CCDs. We have noted above in \S\ref{sec:newsamplesummary} the effects of 
instead requiring selection in each of our CMDs or CCDs.  We note
again here that, aside from the previously identified members, the
number of new potential members is comparable to the contaminant hit
rate. The c2d criteria require a 24 \mum\ detection and we are
strongly biased towards objects with a 24 \mum\ detection, so this is
perhaps not surprising.

\subsubsection{Gutermuth \etal\ (2008) Criteria}

Gutermuth \etal\ (2008), in their Section 4.1 and Appendix, describe a
multi-step weeding process.  They require detection in all four IRAC
bands for most of the process, although they allow for semi-manual
addition of objects missing some bands.  The number of sources that
can be considered as input for this process which appears in 
Table~\ref{tab:othermethods} attempts to represent this semi-manual
addition; there are 89003 objects seen at all 4 IRAC bands, and 76522
that are seen at JH\ks[3.6][4.5] but not [5.8] or [8], and these two
components represent the bulk of the sources that can be considered.
Additional provision is made in this method for bright, red objects 
detected at [24] but not all four IRAC bands, and for stars with 24
\mum\ excess; since this represents addition of a few thousand
sources to the 165525 sources already considered, a $\gtrsim$ symbol
is used in the Table.  The Gutermuth method also works in dereddened
colors for some criteria, requiring a high-spatial-resolution \av\
map, which we do not have and thus cannot implement exactly in a
parallel fashion.  We note too that the Gutermuth \etal\ criteria
relies mostly on the [4.5]$-$[5.8] color to avoid effects of
reddening that may be found at 3.6 $\mu$m.  Since the overall median
\av\ towards Taurus is low (\av$\sim$3, which has a very small effect
on 3.6 \mum), we did not deredden our colors, and simply applied the
Gutermuth \etal\ criteria to our observed colors.  We did not
semi-manually add objects to the list following their section 4.2, and
we added a few criteria to remove some blue objects found in our
catalog; to their ``embedded'' criteria, we added an additional
[5.8]$-$[8]$>$$-$0.5 and [3.6]$-$[4.5]$>$$-$0.5, and to their ``Class
II'' criteria, we added [3.6]$-$[4.5]$>$$-$0.5.  Two of the many color
parameter spaces that Gutermuth \etal\ use are the same as ours; see
Table~\ref{tab:othermethods2}.

Differences between the Gutermuth criteria and ours include the
following: (1) only two CMDs are the same between the methods, (2) the
Gutermuth method sometimes uses the intersection of the color spaces
(like c2d) and sometimes takes selection in just one color space as
sufficient (like us), (3) the Gutermuth method provides for a
semi-manual addition of likely YSO objects, based on color cuts, (4)
the Gutermuth method does not require intensive manual examination of
multi-band images, (5) the Gutermuth method, though multi-step, can be
applied to any Spitzer+2MASS catalog of a star-forming region, and is
not dependent on products of pipelines or the presence of ancillary
data (modulo reddening corrections),  and (6) the Gutermuth method,
as properly applied, needs an \av\ map and needs to work in dereddened
colors.  In order to attempt a comparison using the best possible
criteria, we have imposed the same basic color cuts, and, because of
the low overall \av\ towards Taurus, we have continued to work in the
observed color space, and not dereddened anything. We have not
semi-manually added objects to the YSO candidate list (as per
Gutermuth's Phase 2 or Phase 3).

Because of the multi-band detections necessary, these criteria cannot
be applied to most of the entire $\sim$700,000 object catalog; even
with the semi-manual addition of sources allowed for in this method, 
between 50-75\% of the catalog do not have enough detections to be 
included (see Table~\ref{tab:othermethods}), but the process of
semi-manual addition of sources is likely to catch most legitimate
YSOs in the catalog.  The sample with 4-band IRAC detections is
easiest to handle automatically, and out of this sample,  83\% are
dropped as stars, 16\% are dropped as contaminants of any of a variety
of kinds, 282 (0.3\%) are embedded objects (Class 0/I or flat), and
291 (0.3\%) are Class II objects. Among our set of
previously-identified Taurus members, 28 of them (13\%) cannot be
classified because of missing bands.  Just 2 are dropped as
contaminants and 82 (38\%) are classified as stars; 15 (7\%) are
identified as ``embedded'' and 87 (41\%) are identified as Class II.
Again, as intended, this selection mechanism (even as we have
implemented it) is strongly biased towards YSOs.     

The Gutermuth \etal\ classifications (in more finely-grained detail)
for the  previously known objects are  in Table~\ref{tab:knownysot3}
and those for our new candidate members are listed in
Table~\ref{tab:newysot3}.  Among our new candidate objects, 95\% can
be considered, 76 (51\%) are classified as stars, 5 (3\%) are
contaminants, and 60 (41\%) are YSO candidates (21 objects are
identified as embedded and 39 are identified as Class II). There is a
slightly higher fraction of contaminants in the new sample than among
the previously identified sample.   We note that the overwhelming
majority of objects that we dropped as a result of the review
described in \S\ref{sec:gradations} were classified as contaminants
using the Gutermuth classifications.  We also note that the
classification that we derive from a fit to all available points
between \ks\ and 24 \mum\ is in good agreement with the Gutermuth
\etal\ SED classification except in higher-extinction situations, as
expected.

\subsubsection{Conclusions on Different Criteria}

Comparing the different methods is clearly not at all straightforward.
The differences are much more than simply different data reduction
methods (aperture vs.\ PSF-fitting photometry) or survey depths. The
different color spaces that each study investigates, and the different
assumptions that are made (e.g., ORring vs.\ ANDing the color
selection) can be seen most clearly in the different results when
considering the set of previously identified Taurus members.  Since
not all of the previously identified Taurus members have strong IR
excesses, it is not surprising that no Spitzer-based method retrieves
all of them.  Interestingly, the Gutermuth \etal\ selection mechanism
returns a higher fraction of previously identified members without
infrared excesses than the c2d method.    

Direct comparison of the YSOs retrieved in this paper (or over the
Taurus Molecular Cloud as a whole) to any other association is not
easy because the selection methods are so different.  Unfortunately,
numbers from this study cannot simply be dropped into tables from
these other studies.  Even if we were to re-reduce our data in exactly
the same way as the other studies, a direct comparison would be
difficult. 

Even the contaminant rates are hard to compare. Since all of the
follow-up data has yet to be acquired for our candidate objects, it is
hard to do a final assessment of our contamination rate. Out of the 80
objects with spectra reported here, at least 45\% are new members, and
at least 10\% are contaminants. Over 44 square degrees, including the
previously identified plus the new confirmed objects, we recover at
least 4.3 YSOs  per square degree, and at least 0.18 contaminants per
square degree. The c2d method was derived primarily using the c2d
observations of Serpens (and SWIRE), and Harvey \etal\ (2007)
estimates of order 0-1 contaminants in their 0.85 square degrees. 
Scaling this up to our $\sim$44 square degree map, we then would
expect $\sim$50 contaminants in our map using the c2d method! 
Oliveira \etal\ (2009) carry out follow-up spectroscopy for the c2d
Serpens sample, and find higher contamination (25\%) than was
originally estimated in Harvey \etal\ (2007). However, these
contaminants are primarily AGB stars, which should not be as much of a
contaminant source in the Taurus map, given the lower galactic
latitude of Serpens (5$\arcdeg$) as compared with Taurus
($-15\arcdeg$).  Gutermuth \etal\ (2008) estimate $\sim$1 contaminant
left in their 40$\arcmin\times$30$\arcmin$(=0.33 deg$^2$) map.
Scaling  up to 44 square degrees, then we would expect $\sim$130 
contaminants. However, we expect that the actual contamination is much
less because this method has been tested on several maps, including
larger ones. As more follow-up data are obtained for this and the
other associations, a more direct comparison of the complete YSO
inventories in these clouds will eventually be possible.

\subsection{Stellar rotation}

\begin{figure*}[tbp]
\epsscale{0.8}
\plotone{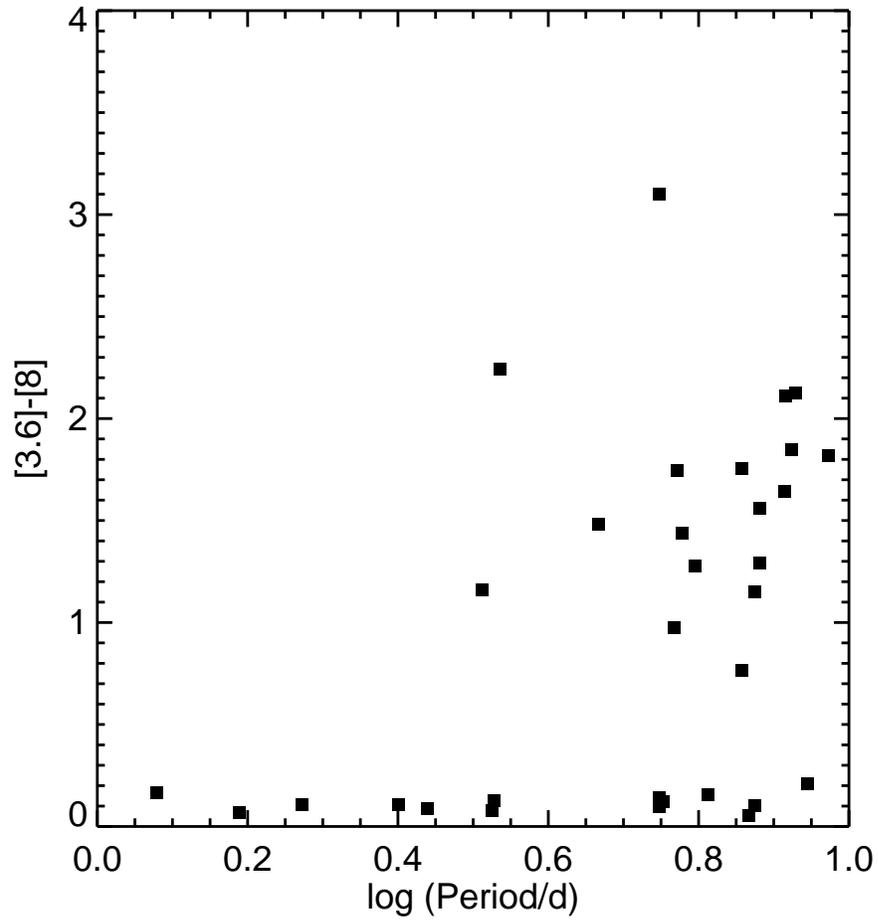}
\caption{Rotation rates of Taurus stars compared to [3.6]$-$[8]
colors. The relationship seen here, where the slowest rotators are
much more likely than the fast rotators to have disks, is consistent
with what has been found in other young associations. }
\label{fig:rotation}
\end{figure*}

Rebull \etal\ (2006) and Cieza \& Baliber (2007) found a correlation
between IRAC excesses and rotation rates in Orion and NGC 2264. 
Taurus is one of the first associations in which rotation of young
stars was studied (see, e.g., Edwards \etal\ 1993), but there are
still only 32 stars with measured periods (see G\"uedel \etal\ 2007,
and references therein, and Grankin \etal\ 2008) and IRAC photometry
in our catalog.  Figure~\ref{fig:rotation} shows the rotation rate
against disk excess for those stars.  The relationship seen here,
where the slowest rotators are much more likely than the fast rotators
to have disks, is consistent with what has been found in the other
associations.

\section{Conclusions}
\label{sec:concl}

We have presented here Spitzer flux densities for 215
previously-identified members of the Taurus Molecular Cloud young
stellar object population.  We constructed Spitzer color-color and
color-magnitude diagrams, investigated where the previously identified
Taurus members were located, and then used those diagrams to select
additional candidate Taurus objects out of a catalog of $\sim$700,000
objects observed with Spitzer over 44 square degrees.  We used a
wealth of supporting data (including ground-based optical imaging) to
winnow that list down to 148 candidate new Taurus members.  We
obtained follow-up optical spectroscopy for about half the sample,
thus far finding 34 new members, 3 probable new members, and 10
possible new members, a potential increase of 15-20\% of young Taurus
members.  Most of the new members are located in close (projected)
proximity to the previously-identified Taurus members; most of them
are Class II M stars. 

In addition to the new members in our sample, there are 60 stars
needing additional follow-up observations, and 33 objects pending any
follow-up observations at all, so more Taurus members may yet be
confirmed out of our list of candidate members.  We also found a
background Be star, a new planetary nebula, a new carbon star, many
background giants, and 100s of galaxies, just 7 of which made it into
our final list of 148 YSO candidates. 

As part of this project's classification as a Spitzer Legacy
Project, enhanced data products have been delivered back to the
Spitzer Science Center (SSC), including the catalogs on which this
present paper is based. 

This study has demonstrated the unique power of Spitzer to efficiently
survey large areas of the sky and provide new information on
membership and YSO properties even in nearby star forming regions such
as Taurus, which has been extensively studied for decades. Even after
many decades of study, our knowledge of membership in Taurus is still
incomplete, but definitely improving, thanks to Spitzer.

\appendix

\section{Spectral Energy Distributions (SEDs) for all of the
Previously-Identified and New Candidate Taurus Members}
\label{sec:seds}

For each of the previously-identified Taurus members, and each of the
new candidate members, we provide an SED here in this on-line-only
Appendix. Notation is as follows: triangles--XMM-Newton OM,
$+$--literature Johnson photometry, *--Sloan photometry,
$\times$--CFHT  photometry, diamonds--2MASS, circles--IRAC,
squares--MIPS. Limits for any  band are indicated by arrows. The
wavelength is in microns; $\lambda F_{\lambda}$ is  in cgs units (erg
s$^{-1}$ cm$^{-2}$). Each plot has the SST Tau catalog number and, if
relevant, a more common name.

For reference, the IRAS PSC all-sky completeness limits at 12, 25, 60,
and 100 \mum\ for a region outside of the Galactic plane are 0.4,
0.5, 0.6, 1.0 Jy, respectively. These limits would appear in the
Figure at log$\lambda F_{\lambda}$=$-$10.00, $-$10.22, $-$10.52, and
$-$10.52.

\begin{deluxetable}{lll}
\tablecaption{Individual miscellaneous non-YSO
objects with YSO-like colors\label{tab:miscobj}\tablenotemark{a}}
\rotate
\tabletypesize{\tiny}
\tablewidth{0pt}
\tablehead{
\colhead{SST Tau number} & \colhead{Other names} & \colhead{Notes} }
\startdata
\cutinhead{Planetary nebulae}
041936.1+271731 & \nodata & New object; CFHT imaging suggests PN\\
043723.4+250242 & PN G174.2-14.6 & Well-studied PN\\
\cutinhead{Carbon stars}
043748.2+254926 & C* 228 & well-studied carbon star\\
042818.5+253140 & V414 Tau & well-studied carbon star\\
043250.3+294239 & IRAS04296+2936 & new (?) carbon star\\
\cutinhead{Background giants from literature}
043724.8+270919 & SV* SVS 1085, HD283751 & new(?) Be background giant\\
043939.9+252034 & JH 225 & B9 with substantial \av \\
042021.2+272102 & phi Tau &  literature giant \\
042136.8+282458 & HD 283570 & literature giant \\
043238.9+235825 & IRAS 04296+2352 & literature giant \\
043858.2+263108 & Elia 3-14 & literature giant \\
043926.9+255259 & Elia 3-15 & literature giant \\
043938.8+261126 & Elia 3-16 & literature giant \\
044057.4+255413 & [TNS87] 8 & literature giant \\
042907.6+244350 & Elias 3-6 & background giant; name incorrectly
associated with IRAS 04260+2437 and HH 414\\
\cutinhead{Background giants based on brightness}
041324.4+290722 &HDS 537& bright, assumed to be giant; our type is M5 \\
041443.5+281708 & [WSB2007] J041443.5+281708 & bright, assumed to be giant\\
041943.4+272056 & \nodata &  bright, assumed to be giant; our type is M3 \\
042115.2+272101 & HD 27482 & bright, assumed to be giant\\
042344.1+225753 & V1142 Tau &  bright, assumed to be giant; our type is M2 \\
042344.1+225757 & \nodata &  bright, assumed to be giant\\
042345.3+234503 & IRAS 04207+2338 &  bright, assumed to be giant\\
042517.3+280440 &IRAS 04221+2757& bright, assumed to be giant; our type is M5 \\
042519.3+261701 &IRAS 04222+2610& bright, assumed to be giant; our type is M4 \\
042630.0+255344 &IRAS 04234+2547& bright, assumed to be giant; our type is M6 \\
042731.3+270958 &IRAS 04244+2703& bright, assumed to be giant; our type is M9 \\
042805.4+284433 & \nodata & bright, assumed to be giant; our type is K6\\ %new
042955.3+225857 & IRAS 04269+2252= 2MASS 04295531+2258579 & bright, assumed to be giant (also Shenoy \etal\ object); our type is M8\\ %new
043121.1+265842 & \nodata & bright, assumed to be giant (also Shenoy \etal\ object); our type is M8\\ %new
043248.0+223952 & GSC 01829-01009 = 2MASS J04324806+2239523&  bright, assumed to be giant; our type is M9 \\
043453.4+270534 & 2MASS J04345345+2705346& bright, assumed to be giant; our type is K7 with substantial \av\ \\
043706.6+214241 & \nodata & bright, assumed to be giant; our type is K2-3 \\
043856.3+271642 & HD 283753 & bright, assumed to be giant; literature type is F6III \\ %new
044118.4+240157 & \nodata & bright, assumed to be giant (also Shenoy \etal\ object); our type is M5\\ %new
044401.6+252014 & DO 10700& bright, assumed to be giant; our type is M5\\
044435.0+250108 & \nodata & bright, assumed to be giant (also Shenoy \etal\ object); our type is M9\\ %new
044510.7+244156 &IRC +20091& bright, assumed to be giant; our type is M8 \\
044659.2+253657 & IRC +30094 = 2MASS J04465929+2536575 & bright, assumed to be giant \\
\cutinhead{Insignificant IR excess}
041501.1+250432&\nodata& our type is G8-K0 \\
041539.2+235344&\nodata& (unobserved)  \\
041643.3+244717&NLTT 12897& (unobserved)  \\
041702.9+242425&\nodata& our type is K0-K2 \\
041707.0+293104&\nodata& our type is K3 \\
041734.0+240153&\nodata& our type is G6 \\
041934.0+252302&\nodata& our type is K2 \\
042216.3+262635&\nodata& our type is G3-G5 \\
042359.7+251452&\nodata& our type is M4 \\
042414.2+252350&\nodata& our type is F5 \\
042635.1+254223&\nodata& our type is G6 \\
042739.6+254208&\nodata& our type is F0 \\
042834.4+262104&\nodata& (unobserved)  \\
042859.4+273625&\nodata& our type is B9-A0 w/ substantial \av  \\
043040.8+235034&\nodata& our type is F6 \\
043203.7+241223&\nodata& our type is F0 \\
043223.4+230059&\nodata& our type is K2-K3 w/ substantial \av  \\
043237.5+292556&\nodata& our type is F0 \\
043254.0+294538&\nodata& our type is G6-K0 \\
043304.9+253705&\nodata& our type is K0-K2 \\
043335.6+242800&2MASS J04333567+2428004& our type is K2 w/ substantial \av  \\
043348.2+274400&\nodata& our type is F0 \\
043354.3+285413&\nodata& (unobserved)  \\
043433.3+244312&IRAS 04315+2436& our type is M5 \\
043458.8+240958&2MASS J04345881+2409587& our type is K2 \\
043544.2+215743&\nodata& our type is F0 \\
043822.9+221048&\nodata& our type is K2-K3 \\
043853.4+251909&\nodata& our type is F8 \\
044018.8+243234&\nodata& our type is G5 \\
044146.7+253824&[THL2004] 2MASS TMRS23& reported as potential Taurus member in literature  \\
044216.7+263917&\nodata& our type is K2 \\
044329.3+260818&\nodata& our type is K1 \\
044334.9+263505&\nodata& our type is G8-K2 \\
044348.0+242723&\nodata& our type is K0-K2 \\
044415.9+255300&\nodata& our type is F1 \\
044453.1+261257&\nodata& our type is A9-F0 \\
044700.1+242745&\nodata& our type is G8-K2 \\
044734.9+234528&\nodata& our type is M2 \\
044827.1+263006&\nodata& our type is F0-F1 \\
044931.5+244935&\nodata& our type is K5 \\
\cutinhead{8 \mum\ pop-ups}
041136.6+293522&\nodata& likely galaxy  \\
041351.6+263819&2MASX 04135166+2638195& likely galaxy  \\
041447.7+254956&2MASX 04144776+2549564& likely galaxy  \\
041525.4+245339&2MASX 04152538+2453388& likely galaxy  \\
041529.1+245942&\nodata& likely galaxy  \\
041556.9+261559&\nodata& likely galaxy  \\
041613.6+253134&2MASX 04161364+2531351& likely galaxy  \\
041711.4+282157&\nodata& likely galaxy  \\
041718.0+291940&\nodata& likely galaxy  \\
041742.5+275427&\nodata& likely galaxy  \\
041756.2+250622&\nodata& likely galaxy  \\
041829.1+285127&\nodata& likely galaxy  \\
041842.1+283543&\nodata& likely galaxy  \\
041843.7+250715&\nodata& likely galaxy  \\
041903.8+271552&\nodata& likely galaxy  \\
041916.1+250413&\nodata& likely galaxy  \\
041929.7+253802&\nodata& likely galaxy  \\
041950.1+251009&\nodata& likely galaxy  \\
042043.7+252715&\nodata& likely galaxy  \\
042047.0+282000&\nodata& likely galaxy  \\
042140.8+282611&2MASX 04214085+2826119& likely galaxy; spectroscopically confirmed xgal \\
042159.7+234810&\nodata& likely galaxy  \\
042252.9+225507&2MASX 04225293+2255071& likely galaxy; spectroscopically confirmed xgal \\
042257.6+243206&\nodata& likely galaxy  \\
042342.1+235903&\nodata& likely galaxy  \\
042343.7+245945&\nodata& likely galaxy  \\
042417.0+234144&\nodata& likely galaxy  \\
042417.9+272942&\nodata& likely galaxy  \\
042429.2+224613&\nodata& likely galaxy  \\
042521.7+262249&2MASX 04252174+2622493& likely galaxy  \\
042634.9+260816&2MASS J04263497+2608161& likely galaxy; spectroscopically confirmed xgal \\
042654.0+262921&\nodata& likely galaxy  \\
042843.2+274001&\nodata& likely galaxy  \\
043028.2+280420&\nodata& likely galaxy  \\
043044.5+282640&\nodata& likely galaxy  \\
043114.1+245838&\nodata& likely galaxy  \\
043205.6+220627&\nodata& likely galaxy  \\
043241.0+251308&\nodata& likely galaxy  \\
043303.9+280846&\nodata& likely galaxy  \\
043329.4+262052&\nodata& likely galaxy  \\
043407.2+280154&\nodata& likely galaxy; very noisy spectrum, could be consistent with low-z galaxy or late G/early K \\
043446.1+290451&2MASX 04344611+2904516& identified by both Gutermuth
and c2d methods as likely YSO, but SED suggests likely galaxy; spectroscopically confirmed xgal \\
043456.7+232501&2MASX 04345670+2325016& likely galaxy  \\
043513.3+232449&\nodata& likely galaxy  \\
043516.1+213943&\nodata& likely galaxy  \\
043712.3+260733&\nodata& likely galaxy  \\
043820.5+215748&2MASX 04382057+2157489& likely galaxy  \\
043828.3+253223&\nodata& likely galaxy  \\
043931.8+214742&\nodata& likely galaxy  \\
043958.0+253354&\nodata& likely galaxy  \\
044114.8+253242&\nodata& likely galaxy  \\
044308.1+240957&\nodata& likely galaxy  \\
044554.8+240843&IRAS04428+2403& galaxy; listed by Kenyon \etal\ (2008) as confirmed Taurus member  \\
044614.5+253100&\nodata& likely galaxy  \\
044717.7+234529&2MASX 04471766+2345296& likely galaxy  \\
044933.6+234518&\nodata& likely galaxy  \\
\enddata
\tablenotetext{a}{These objects were selected at various points in our
selection process, but we have rejected these as YSOs.  See associated
text for much more information.}
\end{deluxetable}

\begin{deluxetable}{lll}
\tablecaption{Literature non-members regarded as potential new members
\label{tab:miscnewobj}}
\rotate
\tabletypesize{\tiny}
\tablewidth{0pt}
\tablehead{
\colhead{SST Tau number} & \colhead{Other names} & \colhead{Notes} }
\startdata
\cutinhead{}
041803.3+244009 & 2MASS 04180338+2440096 & needs additional follow-up; our type is A9\\
041810.7+251957 & [GBM90] L1506 1 & probable new member, our type is K8-M0\\
041823.2+251928 & 2MASS 04182321+2519280 & needs additional follow-up; our type is G:\\
042212.9+254659 & 2MASS 04221295+2546598 & pending follow-up \\
042518.6+255535 & 2MASS 04251866+2555359 &  Shenoy \etal\ object; needs additional follow-up; our type is M5.\\
%042955.3+225857 & IRAS 04269+2252 = 2MASS 04295531+2258579 &  Shenoy \etal\ object; our type is M8 \\
042920.8+274207 & 2MASS 04292083+2742074 & possible new member; our type is M6.\\
043024.1+281916 & 2MASS 04302414+2819165& Shenoy \etal\ object; needs additional follow-up; our type is M5\\
043042.8+274329 & 2MASS 04304284+2743299& Shenoy \etal\ object; needs additional follow-up; our type is M6 \\
%043121.1+265842 & 2MASS 04312113+2658422& Shenoy \etal\ object; our type is M8 \\
043228.1+271122 & 2MASS 04322815+2711228& Shenoy \etal\ object; needs additional follow-up; our type is M6 \\
043344.6+261500 & 2MASS 04334465+2615005& new member; Shenoy \etal\ object; our type is M6e\\
043435.4+264406 & 2MASS 04343549+2644062& Shenoy \etal\ object; needs additional follow-up; our type is M3 \\
%044118.4+240157 & 2MASS 04411845+2401576& Shenoy \etal\ object; our type is M5 \\
044125.7+254349 & 2MASS 04412575+2543492& pending follow-up\\
%044435.0+250108 & 2MASS 04443504+2501082& Shenoy \etal\ object; our type is M9 \\
044539.8+251704 & 2MASS 04453986+2517045& Shenoy \etal\ object; needs additional follow-up; our type is M5 \\
044557.0+244042 & 2MASS 04455704+2440423& needs additional follow-up; our type is K2 \\
044639.8+242526 & 2MASS 04463986+2425260& Shenoy \etal\ object; needs additional follow-up; our type is M5 \\
\enddata
\tablenotetext{a}{These objects were noted in the literature as likely
non-members, but we have promoted them to be candidate members based
on our Spitzer observations.  See text for more information.}
\end{deluxetable}

\clearpage

\section{Comments on individual miscellaneous non-YSO objects with YSO-like
colors}
\label{sec:miscobj}

In the process of conducting our detailed source-by-source
examination, we encountered many objects with YSO-like colors that
were not YSOs, such as plantary nebulae and carbon stars. In some
cases, these objects are new discoveries, or our observations shed new
light on the nature of the object.

\subsection{Planetary nebulae}

Based on imaging, we may have discovered a planetary nebula: SST Tau
041936.1+271731.  It appears bright at 24 and 70 $\mu$m, but CFHT
imaging reveals a circularly symmetric structure that strongly
suggests a planetary nebula.  We have not yet obtained follow-up
spectroscopy.

PN G174.2-14.6 (=043723.4+250242) is a well-studied planetary nebula
(55 references are given in SIMBAD).  It appears to have YSO-like
colors in [24] vs.\ [24]$-$[70], [8] vs.\ [8]$-$[24], [4.5] vs.\
[4.5]$-$[8] (although it appears as faint enough to likely be a galaxy
here), and the IRAC color-color diagram.  We ruled it out as a YSO
candidate based on the literature early on in our weeding process.

\subsection{Carbon stars}

C* 228 (043748.2+254926) is a previously identified carbon star.  It
has YSO-like colors in [24] vs.\ [24]$-$[70] and [8] vs.\ [8]$-$[24]. 
We ruled it out as a YSO candidate based on the literature early on in
our weeding process.

Two of the objects we selected initially as YSO candidates turned out
to be carbon stars when we obtained spectra: 043250.3+294239 (=IRAS
04296+2936) and 042818.5+253140 (=V414 Tau).   IRAS 04296+2936 appears
to have no references in SIMBAD and may therefore be a newly
discovered carbon star; V414 Tau appears in a handful of publications
as a carbon star, including Alksnis \etal\ (2001).  IRAS 04296+2936
has a significantly reddened  spectrum, at least in comparison to V414
Tau.   

The 2MASS NIR $JH$\ks\ photometry for these objects are 8.02, 6.59,
and 5.64 for IRAS 04296+2936, 7.28, 6.11, and  5.49 for V414 Tau; C*
228 is only unsaturated at $H$ and is  3.38 mag.  These first two
objects both have colors considerably redder than those for carbon
giants or dwarfs as appearing in Lowrance \etal\ (2003), and all three
are very much brighter than those in Lowrance \etal\ (2003).   If they
are carbon dwarfs, they would have to be considerably closer than
Taurus, as in $\sim$10 pc.  However, they are so red in $J-H$ that
they are most likely giants; they are so bright that they are most
likely background objects, behind the Taurus molecular cloud.  

Thus far, we have found 3 carbon stars with YSO-like colors in this
$\sim$44 square degree survey.  It is possible, indeed likely, that
more carbon stars are in our survey, but have not been identified as
such.

\subsection{Be background giant(s)}
\label{sec:backgroundBs}

One of the stars we selected as a YSO candidate based on three of our
color spaces, 043724.8+270919 (also known as SV* SVS 1085=HD 283751),
has H$\alpha$ strongly in emission, and we classify it as a B8e. We do
not have any optical photometry for it, but its 2MASS $JH$\ks\ are
9.79, 9.55, and 9.30 mags, respectively.  Using its \ks\ mag and assuming no
reddening (as a worst-case but clearly incorrect scenario), and
comparing the observed \ks\ to that expected for young stars, we
calculate a distance of $\sim$700 pc. It seems to be a Be star well
behind the Taurus cloud. 

043939.9+252034 (JH 225) is listed in the literature as likely
non-member based solely on high proper motions.  We find that it is a
B9 with substantial \av. Its 2MASS $JH$\ks\ are 9.48, 9.01, and 8.73
mags, respectively. As we did above, we calculate a distance of 600
pc. It is also unlikely to be truly a member of Taurus, but additional
follow-up is needed. It does not have emission lines.

Since these objects were selected as YSO candidates, flux densities appear in
the Tables above, and SEDs appear in Appendix~\ref{sec:seds}.

\subsection{Other very bright objects and giants}
\label{sec:giants}

Table~\ref{tab:miscobj} includes several bright objects either assumed
or confirmed to be background giants.

Several giants are confirmed in the literature (via spectroscopy) as
being giants, but are selected in at least one of our color spaces as
being YSO candidates. We ruled these out as YSO candidates based on
the literature early on in our weeding process.

Elias 3-6 (SST Tau J042907.6+244350) has often appeared in lists of
Taurus cloud members. There has been confusion in the literature about
its IRAS association, with a mistaken identification with IRAS
04264+2433 (the HH 414 jet source) often quoted (e.g., Motte and Andre
2001).  Elias 3-6 is actually IRAS 04260+2437.  At \ks=2.85, it is
much brighter than any known Taurus member, is saturated in the
Spitzer IRAC bands, and shows a [24]$-$[70] color of 0.04 mag that is
consistent with a bright photosphere.  Elias \etal\ (1978) classify it
as a field M8 III giant with 6 mag of visual extinction.  The Spitzer
results are fully consistent with Elias' original result that this
object is not a Taurus member. We drop it from the list as a
background giant. 

Several objects listed in Table~\ref{tab:miscobj} are very, very
bright in our survey, and we have assumed primarily based on
brightness (and the fact that they seem to have little legitimate IR
excess as far as we can determine) that they are background giants.
Several of them were selected in at least one of our color spaces as
potential YSOs (given the available photometry points; in many cases
later inspection of these images suggested that they were probably
saturated, or at least in the non-linear regime).  In some of those
cases, since we wondered if they could be YSOs, we obtained spectra of
them; the resultant spectral type is listed in Table~\ref{tab:miscobj}
for use by the community in future studies.

\subsection{Objects with insignificant IR excesses for which we have
spectra}
\label{sec:insignifirx}

Several objects appeared upon early inspection to have significant IR
excesses at Spitzer bands, and thus we obtained spectra of them.
However, upon re-examination, we determined that the IR excesses we
detect were not significantly above the expected photosphere.  Since
these objects are no longer YSO candidates, they do not appear in
tables in the main body of the text above nor in SEDs below. However,
since we obtained spectra for many of them, we still report them here
for use by the community. They are listed in Table~\ref{tab:miscobj}.

\subsection{Extragalactic objects from Luhman \etal}

Eight objects show up as having YSO-like colors in at least one of the
color spaces we investigated and are reported as spectroscopically
confirmed non-members by Luhman \etal\ (2006, 2009b).  Each of these objects
are ones that we had independently discarded as likely galaxies based
on inspection of our optical imaging:  
041916.1+275048,
042754.4+242414,
042336.9+252628, 
042535.5+245739,
043027.1+280707, 
043401.8+231906,
043502.0+233141, and
044554.8+240843.  
This further demonstrates the critical importance of
high-spatial-resolution optical imaging in refining a YSO candidate
list selected based on Spitzer data, before using valuable
spectroscopic telescope time to refute or confirm stellar nature and
youth.

\subsection{IRAS 04428+2403 and other 8 \mum\ pop-up objects}
\label{sec:popups}

\begin{figure*}[tbp]
\epsscale{0.9}
\plotone{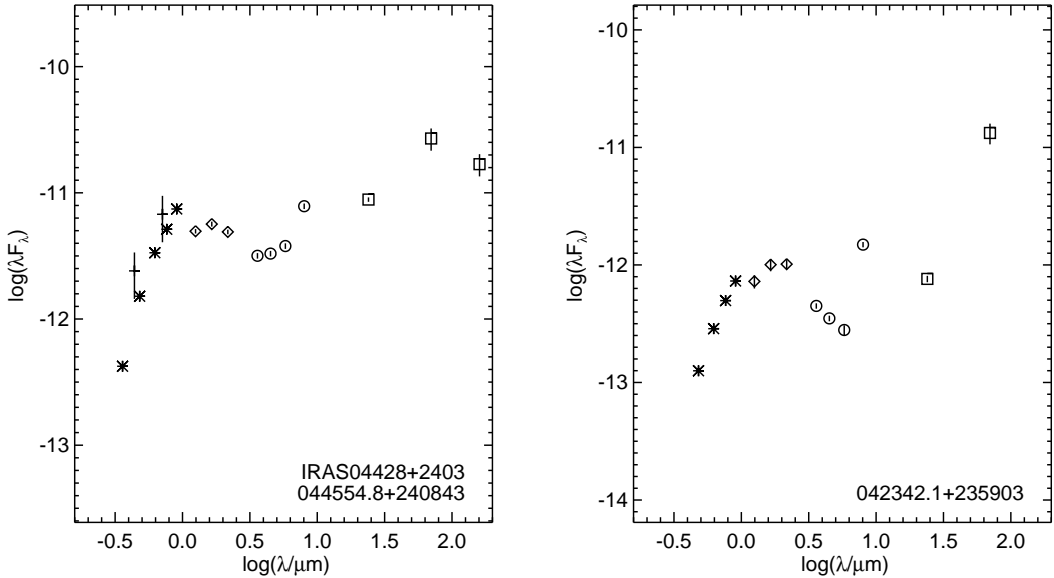}
\caption{Example 8 \mum\ pop-up SEDs for two objects, 
044554.8+240843=IRAS 04428+2403 and 042342.1+235903. Notation is as it
is for all of the other SEDs in this paper. Both of these objects are
likely to be galaxies.}
\label{fig:popupexample}
\end{figure*}

IRAS 04428+2403 (SSTTau 044554.8+240843) appears in Kenyon \etal\
(2008) listed as a confirmed member of Taurus. However, it appears in
two prior  papers in the literature, one which identifies it as a
galaxy (Chamaraux \etal\ 1995), and the other which (Kenyon \etal\
1994) identifies it as only an unconfirmed Taurus member.   By
inspection of all our available imaging (prior to identifying it with
a literature object), we classified it as a galaxy based on its
appearance in the optical images.  The SDSS imaging pipeline
classified this object as extended.  SDSS also obtained a spectrum of
this object, and it is a very reddened (due to the Taurus cloud)
star-forming galaxy with bright H$\alpha$/\ion{N}{2} and \ion{S}{2}
at a redshift of 0.077.   Therefore, we drop this object from our list
of previously-identified members.  

The SED for IRAS 04428+2403 (seen in Figure~\ref{fig:popupexample})
resembles those for a class of objects we found in the process of
looking for candidate YSOs.  These objects seemed to meet all of our
criteria for YSO candidacy (see \S\ref{sec:gradations}).  This
population has SEDs that looked like they could be YSO candidates, but
where the 8 $\mu$m point in the SED appears significantly above a line
connecting the 5.8 and 24 \mum\ points.  Several of these objects were
targeted for optical spectroscopy, and four of them were actually
observed.  All of them turned out to be extragalactic objects.  Tens
if not hundreds of the spatially resolved galaxies in our survey (such
as IRAS 04428+2403, but also more obvious resolved galaxies) also have
this kind of SED.  So, we conclude that even though some of these
objects are point sources as far as we can tell, they are extremely
likely to be galaxies. An example SED for one of the point-like
objects is given in Figure~\ref{fig:popupexample}.  All of the objects
that were apparent point sources (e.g., the ones that fooled us)
appear in Table~\ref{tab:miscobj}.  The ones that are confirmed
extragalactic objects are indicated.

The most likely origin of the 8 micron popup is the polycyclic
aromatic hydrocarbon (PAH) emission feature near 8 \mum.  Geers \etal\
(2009) have found that low-mass young stars almost always lack PAH
emission features at 11 \mum\ in Spitzer spectra or at 3.3 \mum\ in
ground-based spectra. Thus we do not expect faint cloud members to show
this feature. However, these features are common in lower luminosity
galaxies (Weedman \& Houck 2009).  PAH features are seen around
luminous young stellar objects such as the Ae star HD 100546 (Malfait
\etal\ 1998), where strong stellar ultraviolet fluxes can excite PAH
emission.  All of the 8 micron pop-up sources detected in the Taurus
survey, if located at the cloud distance of 137 pc, would have low
luminosities strongly inconsistent with an early spectral type.  Thus
we conclude that none of the 8 micron pop-up sources found in the
Taurus survey field are consistent with low-luminosity Taurus members,
and we have excluded them from the list of candidates.

%042342.1+235903
%041525.4+245339 2massx src
%044554.8+240843 'previously known'

% \section{Possible location for individual discussion of YSO
% candidates}
% \label{sec:option2}
% 
% {\em consensus on one telecon in which i personally was rather less
% than coherent was that individual paragraphs on each YSO candidate
% were desirable.  There could be 176 paragraphs (one per each new obj)
% or 70 paragraphs on each new obj which has a previous (usually SIMBAD)
% identification -- see table for each of the prior names. Taking
% volunteers on this; either way, please see and use the individual
% object information currently encoded in tables, table notes, main
% text, and the appendix.  }

\section{Literature non-members regarded as potential new members}

Several objects listed in the literature as non-members (e.g., assumed
but not spectroscopically confirmed background objects) appear as
having colors consistent with YSOs. In several cases, the literature
regarded them as non-members not because of spectroscopic
confirmation, but because of optical properties.  Since we
investigated their infrared properties for the first time, and found
them to have infrared excesses, we have at least for the time being
promoted them back to being YSO candidates rather than non-member
candidates. These objects are all identified in the relevant tables
above (including Table~\ref{tab:miscobj}), and since they are still YSO
candidates, their SEDs appear in Appendix~\ref{sec:seds}.  

Shenoy \etal\ (2008) use the data from our Taurus-1 survey (as
reported by Luhman \etal\ 2006) combined with 2MASS to create a
catalog of objects they believe to be background objects.  Several of
the objects they report as candidate background objects are based on
2MASS-only measurements (no IRAC).  Now, with the addition of the
MIPS-24 data (as well as the IRAC data for the rest of the survey),
many of them show up as likely YSO candidates in at least one of our
color spaces we investigated.  We have promoted several of the objects
reported in Shenoy \etal\ (2008) to potential members; they are listed
and identified in the relevant tables in the main body of the paper,
but listed in Table~\ref{tab:miscobj} as just ``Shenoy \etal\
object."  Most of these objects do not have obvious signs of youth in
the spectra we have, so additional data are needed to determine if
they are members.

A few objects merit special discussion.  The first, 041810.7+251957
([GBM90] L1506 1),  is listed by Goodman \etal\ (1990) as one of the
objects for which the authors measured polarimetry, and therefore they
assumed was a background object, but is not called out as anything
remarkable in any other sense.  It appears as a strong YSO candidate
in all of the color-magnitude spaces we investigated.  We have it
listed in the tables above as a new member, with a spectral type that
we determined to be K8-M0, substantial emission lines in the spectrum,
and a YSO SED class of II. 

The second object of note is 042920.8+274207 (2MASS 04292083+2742074),
which appears in Luhman \etal\ (2009b) as a background giant, M5III.
We report it as a possible new member based on all the information we
have; our type is M6 with low gravity.  It shares many characteristics
of young stars, but also those of some post-main-sequence objects. 
Further investigation is needed.

\acknowledgements 

We wish to thank the Palomar Observatory, Sloan Telescope, CFHT,
XMM-Newton, and of course Spitzer staff for their assistance using the
telescopes. 

We wish to thank the anonymous referee for thoughful and thorough comments.

This work is based in part on observations made with the Spitzer
Space Telescope, which is operated by the Jet Propulsion Laboratory,
California Institute of Technology under a contract with NASA. Support
for this work was provided by NASA through an award issued by
JPL/Caltech.

This research has made use of NASA's Astrophysics Data System (ADS)
Abstract Service, and of the SIMBAD database, operated at CDS,
Strasbourg, France.  This research has made use of data products from
the Two Micron All-Sky Survey (2MASS), which is a joint project of
the University of Massachusetts and the Infrared Processing and
Analysis Center, funded by the National Aeronautics and Space
Administration and the National Science Foundation.  These data are
served by the NASA/IPAC Infrared Science Archive, which is operated
by the Jet Propulsion Laboratory, California Institute of Technology,
under contract with the National Aeronautics and Space
Administration.  This research has made use of the Digitized Sky
Surveys, which were produced at the Space Telescope Science Institute
under U.S. Government grant NAG W-2166. The images of these surveys
are based on photographic data obtained using the Oschin Schmidt
Telescope on Palomar Mountain and the UK Schmidt Telescope. The
plates were processed into the present compressed digital form with
the permission of these institutions.  This research has made use of
observations obtained with XMM-Newton, an ESA science mission with
instruments and contributions directly funded by ESA member states
and the USA (NASA).

The research described in this paper was partially carried out at the
Jet Propulsion Laboratory, California Institute of Technology, under
contract with the National Aeronautics and Space Administration.  

M.~Audard and C.~Baldovin-Saavedra acknowledge support from a Swiss
National Science Foundation grant (PP002--110504).

\end{document}